\documentclass{aa}

\usepackage{graphicx}
\usepackage{empheq}
\usepackage{amsmath}	% Advanced maths commands
\usepackage{amssymb}	% Extra maths symbols
\usepackage{bm}
\usepackage{xcolor}
\usepackage{nccmath}
\usepackage[capitalize]{cleveref}
\crefname{section}{Sect.}{Sects.}
\Crefname{section}{Section}{Sections}
\usepackage{import}
\usepackage{transparent}

\expandafter\def\expandafter\UrlBreaks\expandafter{\UrlBreaks%  save the current one
  \do\a\do\b\do\c\do\d\do\e\do\f\do\g\do\h\do\i\do\j%
  \do\k\do\l\do\m\do\n\do\o\do\p\do\q\do\r\do\s\do\t%
  \do\u\do\v\do\w\do\x\do\y\do\z\do\A\do\B\do\C\do\D%
  \do\E\do\F\do\G\do\H\do\I\do\J\do\K\do\L\do\M\do\N%
  \do\O\do\P\do\Q\do\R\do\S\do\T\do\U\do\V\do\W\do\X%
  \do\Y\do\Z}
\usepackage{txfonts}
\newcommand{\delimiters}[4][]{
\ifthenelse{ \equal{#1}{1} }{  #2 #3 #4  }
					{ \ifthenelse{\equal{#1}{2}}{ \big#2 #3 \big#4 }
						{ \ifthenelse{\equal{#1}{3}}{ \Big#2 #3 \Big#4 }
							{ \ifthenelse{\equal{#1}{4}}{ \bigg#2 #3 \bigg#4 }
								{ \ifthenelse{\equal{#1}{5}}{ \Bigg#2 #3 \Bigg#4 }
									{ \left#2 #3 \right#4 }
								}
							}
						}
					}
													}
\newcommand{\ev}[2][]{\delimiters[#1]{\langle}{#2}{\rangle}}
\newcommand{\abs}[2][]{\delimiters[#1]{|}{#2}{|}}
													
\newcommand{\e}[1]{_{\text{#1}}}
\newcommand{\h}[1]{^{\text{#1}}}
\newcommand{\dd}{\mathrm{d}}
\newcommand{\ph}{\varphi}
\newcommand{\cst}{\mathrm{cst}}
\newcommand{\ex}[1]{\mathrm{e}^{#1}}
\newcommand{\ii}{\mathrm{i}}
\newcommand{\vect}[1]{\boldsymbol{#1}}
\newcommand{\cplx}[1]{\underline{#1}}
\newcommand{\unit}[1]{\hat{\boldsymbol{#1}}}
\newcommand{\eps}{\varepsilon}

\newcommand{\lightcone}{\mathcal{C}}

\begin{document} 

   \title{Theoretical and numerical perspectives\\ on cosmic distance averages}
   \titlerunning{Cosmic distance averages}

   \author{Michel-Andrès Breton\inst{1} and Pierre Fleury\inst{2}}

   \institute{\inst{1}Aix Marseille Univ, CNRS, CNES, LAM, Marseille, France\\
                \email{michel-andres.breton@lam.fr}\\
              \inst{2}Instituto de Física Teórica UAM-CSIC, Universidad Autónoma de Madrid, Cantoblanco, 28049 Madrid, Spain\\
              \email{pierre.fleury@uam.es}
             }

   \date{Pre-print number: IFT-UAM/CSIC-20-183}

% \abstract{}{}{}{}{} 
% 5 {} token are mandatory
\abstract{      % context heading (optional)
% Why?
The interpretation of cosmological observations relies on a notion of an average Universe, which is usually considered as the homogeneous and isotropic Friedmann-Lemaître-Robertson-Walker (FLRW) model.
% What?
However, inhomogeneities may statistically bias the observational averages with respect to FLRW, notably for distance measurements, due to a number of effects such as gravitational lensing and redshift perturbations.
% How?
In this article, we review the main known theoretical results on average distance measures in cosmology, based on second-order perturbation theory, and we fill in some of their gaps. We then comprehensively test these theoretical predictions against ray tracing in a high-resolution dark-matter $N$-body simulation. This method allows us to describe the effect of small-scale inhomogeneities deep into the non-linear regime of structure formation on light propagation up to $z=10$.
% So what?
We find that numerical results are in remarkably good agreement with theoretical predictions in the limit of super-sample variance. No unexpectedly large bias originates from very small scales, whose effect is fully encoded in the non-linear power spectrum. Specifically, the directional average of the inverse amplification and the source-averaged amplification are compatible with unity; the change in area of surfaces of constant cosmic time is compatible with zero; the biases on other distance measures, which can reach slightly less than $1\%$ at high redshift, are well understood. As a side product, we also confront the predictions of the recent finite-beam formalism with numerical data and find excellent agreement.
}


   \maketitle
%
%________________________________________________________________

\section{Introduction}

On very large scales, our Universe seems to be well described by a spatially homogeneous and isotropic Friedmann-Lemaître-Robertson-Walker (FLRW) model~\citep{green2014how}. This model allows us to predict the dynamics of cosmic expansion as a function of the Universe's content, and of the laws of gravitation. Furthermore, the FLRW model constitutes a rather efficient framework to interpret the observation of remote light sources; in particular, it provides the relation between their redshift~$z$ and their angular or luminosity distance~$D$.

The distance-redshift relation~$D(z)$ is prominent in cosmology, as it is involved in the interpretation of various observables. Its first derivative today defines the Hubble-Lemaître constant, $\dd D/\dd z|_{0}=c/H_0$, whose exact value is still subject to a lively debate~\citep{planck2018cosmological, riess2019large, wong2019holicow, freedman2019CCHP}. More generally, $D(z)$ constitutes the essence of the Hubble diagram of type-Ia supernovae~\citep[SNe,][]{2018ApJ...859..101S,Abbott:2018wog}, which historically revealed the acceleration of cosmic expansion~\citep{1998Natur.391...51P,1998AJ....116.1009R}, as well as the Hubble diagram of gravitational-wave standard sirens in the near future~\citep{2005ApJ...629...15H,2016JCAP...10..006C}. The $D(z)$ relation is also essential in the analysis of the anisotropies of the cosmic microwave background~\citep[CMB,][]{planck2018cosmological}, or in the baryon-acoustic oscillation signal observed in galaxy, Lyman-$\alpha$ or quasar surveys~\citep{Alam:2020sor}, because it converts the observed angular size of the sound horizon~$\theta_*$ into a physical distance $r\e{s}=D(z_*)\theta_*$ that may be predicted by theory.

In the actual inhomogeneous Universe, however, the $D(z)$ relation is affected by various effects, such as gravitational lensing~\citep{schneider1992gravitational} which tends to focus and distort light beams, thereby changing the apparent size and brightness of light sources; it is also affected by the peculiar velocities of the sources and the observer, which correct the observed redshift via the Doppler effect~\citep{hui2006correlated, davis2011effect}. Such effects make $D(z)$ line-of-sight dependent, but it is generally assumed that the FLRW prediction is recovered on average.

The fundamental question of whether the average $D(z)$ is the same as the $D(z)$ of the average Universe goes back more than 50 years, when \cite{zeldovich1964observations} and Feynman (in a colloquium given at Caltech the same year)\footnote{This talk was mentioned in the introduction of \cite{gunn1967propagation}.} suggested the following: if the Universe is lumpy, then a typical light beam should mostly propagate through under-dense regions, and thereby be de-focussed with respect to FLRW; this should imply that $D(z)$ is actually biased up. Many developments and counter-arguments followed from that seminal idea; we refer the interested reader to the introduction of \citet[][hereafter KP16]{kaiser2016bias} and the comprehensive review by \cite{helbig2019calculation} for details.

In that debate, a significant step was made by \citet{weinberg1976apparent}, who showed that in a Universe sparsely filled with point masses, the average flux $\propto\ev[1]{1/D^2(z)}$ is the same as if the matter in those lenses were homogeneously distributed in space. Importantly, Weinberg's calculation was made at first order in the small projected density of the lenses.\footnote{In a more modern language, we may say that the calculation was made at first order in the micro-lensing optical depth~$\tau$, which coincides with the convergence~$\kappa$ if the density of the lenses were smoothed out; see, for instance, \S~II.C of \citet{Fleury:2019xzr}.} As such, it also implies that $\ev{D(z)}$ is unaffected by inhomogeneities at that order, because the difference between $\ev[1]{1/D^2}$ and $1/\ev[1]{D}^2$ only appears at second order. Weinberg nevertheless conjectured, on the basis of flux conservation, that the invariance of $\ev[1]{1/D^2(z)}$ may be exact and hold for any matter distribution. As noted by \cite{ellis1998lensing}, this general flux-conservation argument is, in fact, incomplete because it implicitly assumes that the area of surfaces of constant redshift are unaffected by inhomogeneities, which is a mere reformulation of the whole problem.

For a long period of time, all this discussion remained mostly centred on the observation of individual sources, with the aim of predicting possible biases on the Hubble diagram; it became a somewhat marginal topic from the end of the 1980s, presumably because the precision of cosmological measurements was not sufficient to be sensitive to the expected biases on $\ev{D(z)}$. Interest in that matter was nevertheless revived by \citet{clarkson2014what}, who made the rather surprising claim that lensing affects the distance to the last-scattering surface (LSS) at percent level, which would be dramatic for the standard interpretation of the CMB. This claim was then retracted by (almost) the same team in \cite{bonvin2015do}, who with KP16 clarified that: (i) the average distance to the LSS is not relevant to the standard CMB analysis; and (ii) one must distinguish between the concepts of directional averaging, source-averaging, or ensemble-averaging, which may yield different results~\citep{bonvin2015cosmological}. Such considerations on cosmological averages were actually elaborating on an earlier work by \cite{kibble2005average}.

In the end, for the CMB just as for the Hubble diagram, the whole problem boils down to the validity of Weinberg's conjecture which states that the area of LSS, $A_*$, or the area of constant-redshift surfaces, $A(z)$, are not significantly affected by inhomogeneities. KP16 undertook the difficult task to explicitly check this conjecture in the framework of cosmological perturbations at second order. With a rather intuitive approach, KP16 identified several key effects such as the shortening of the radius reached by rays due to their deflection, or the increase in $A_*$ due to its wrinkles, and eventually reached the conclusion that $A_*$ cannot be biased by more than a part in a million. They identified that the relevant structures responsible for such a bias are rather large in size, of the order of $50\,h^{-1}~\mathrm{Mpc}$.

% From theory to simulations
Most of the theoretical work depicted above was done using cosmological perturbation theory on an FLRW background \citep[see also][]{sasaki1987magnitude, bonvin2006fluctuations, bendayan2012second, 2014CQGra..31t2001U, yoo2016unified}. However, this theoretical framework is not guaranteed to provide a good representation of the Universe, as it does not access the highly non-linear regime of structure formation. That is why one may prefer to rely on numerical simulations and ray-tracing methods, in order to accurately describe the propagation of light in a realistic picture of the cosmos.

% Previous Numerical works
As a first step, a significant research endeavour was dedicated to ray tracing and distance measurements in cosmological toy-models, such as Swiss-cheese models~\citep{2007JCAP...02..013B, Brouzakis:2007zi, 2008PhRvD..77b3003M, Biswas:2007gi, Vanderveld:2008vi, Valkenburg:2009iw, Clifton:2009nv, 2009GReGr..41.1737B, 2010PhRvD..82j3510B, Szybka:2010ky, 2011JCAP...02..025B, Flanagan:2012yv, Fleury:2013sna, Fleury:2014gha, 2014JCAP...03..040T, 2014PhRvD..90l3536P, Lavinto:2015iba, Koksbang:2017arw, Koksbang:2019cen, Koksbang:2019wfg, Koksbang:2020qry}, plane-parallel Universes~\citep{DiDio:2011gf}, or lattice models~\citep{2009PhRvD..80j3503C, 2009JCAP...10..026C, 2011PhRvD..84j9902C, 2012PhRvD..85b3502C, 2015PhRvD..92f3529L, Bruneton:2012ru, Liu:2015bya, Bentivegna:2016fls, sanghai2017raytracing, Koksbang:2020zej}. These works generally agreed with the relevant theoretical predictions. Using $N$-body simulations, \citet{odderskov2016local} showed that at low redshift ($z<0.1$), the averaged luminosity distance is very close from its value in an FLRW background. Within the field of numerical relativity, \citet{giblin2016observable} showed that $\ev{\log D}$ (or averaged magnitude) was not affected by inhmogeneities, at least until $z = 1.5$. However, both studies used simulations with rather low resolution, which might be subject to large variance and therefore could not highlight second-order effects. More recently, \citet{adamek2019bias} used the general-relativistic simulation \texttt{gevolution} \citep{adamek2016gevolution}, and accurate ray tracing to find null geodesics between sources and observer and produce realistic halo catalogues. They found that when averaging over sources, $\ev[1]{1/D^2(z)}$ is very close to its value from a homogeneous Universe, until $z = 3$, thereby confirming Weinberg's conjecture, while $\ev{D(z)}$ is slightly biased as expected. Albeit a high-resolution, \texttt{gevolution} remains a particle-mesh code without adaptive-mesh refinement, which thus cannot access very small scales.

% Presentation of this paper
In the present article, we propose a short theoretical review, and a detailed numerical study of the bias in the distance-redshift relation with respect to the standard FLRW prediction. The theory part builds upon KP16 and fills minor conceptual gaps therein. In the main, numerical, part we use a high-resolution $N$-body simulation part of the `Raygal' suite and propagate photons on null geodesics to infer distance measures, accounting for gravitational lensing and redshift perturbations. Taking advantage of very large statistics and wide redshift range (up to $z = 10$), we investigate the different averaging procedures and study the statistics to the related observables. Furthermore, we numerically estimate the area bias depending on the choice of light-cone slicing.

The article is organised as follows. \Cref{sec:theory} presents the formalism for light propagation and the bias on statistical quantities with respect to the homogeneous case, depending on the averaging procedure; we connect these notions to the area of slices of the light cone. The numerical simulation, ray-tracing methods, and analysis techniques, are presented in \cref{sec:numerical_methods}, while the results are exposed in \cref{sec:results}. We conclude in \cref{sec:conclusion}.  

\paragraph{Notation and conventions} Greek indices $(\mu,\nu, \ldots)$ run from $0$ to $3$ and Latin indices $(i,j,\ldots)$ from $1$ to $3$. Bold symbols denote Euclidean two-dimensional or three-dimensional vectors, and matrices. Over-barred symbols denote quantities computed in a homogeneous-isotropic FLRW model. We adopt units in which the speed of light is unity, $c=1$.

%__________________________________________________________________

\section{Theory}
\label{sec:theory}

This section gathers a number of already-established theoretical results about light propagation in the inhomogeneous Universe, as well as a few novel elements, such as the distinction between lensing magnification and amplification and its interpretation. We shall focus on the statistical averages of distance measures, and how they relate to the area of light-cone slices.

\subsection{Light propagation in a perturbed FLRW Universe}
\label{subsec:light_propagation}

We consider a cosmological space-time described by a spatially flat FLRW model with scalar perturbations. The associated line element reads, in the Newtonian gauge
\begin{equation}
\dd s^2 = a^2(\eta)\left[-(1+2\phi)\dd\eta^2 + (1-2\phi)\dd\vect{x}^2\right] ,
\label{eq:perturbed_FLRW_metric}
\end{equation}
where $\eta$ denotes the conformal time (hereafter simply referred to as time), $x^i$ are comoving coordinates, $a(\eta)$ the scale factor describing cosmic expansion, and $\phi$ the Bardeen potential \citep{bardeen1980gauge} caused by inhomogeneities in the matter density field. We assume that anisotropic stress is negligible so that this potential is unique. Except in the vicinity of compact objects, $\phi\ll 1$ can be treated as a perturbation. The time at the observation event (here and now) is denoted $\eta_0$, where the scale factor is conventionally set to unity, $a_0=a(\eta_0)=1$.

Light propagates along null geodesics of the space-time geometry. In the absence of perturbations (that is, for $\phi=0$), such geodesics are straight lines in comoving coordinates, travelled with unit coordinate speed. In the presence of perturbations, light rays are bent and the coordinate speed of light effectively varies \citep{schneider1992gravitational}. These effects are encoded in the null geodesic equation $k^\nu\nabla_\nu k^\mu=0$, with $k^\mu\equiv \dd x^\mu/\dd\lambda$ and $\lambda$ denotes a past-oriented affine parameter for the light ray. The temporal and spatial components of the geodesic equation read
\begin{align}
\label{eq:geodesic_equation1}
\frac{\dd k^0}{\dd\lambda}
&= -2\mathcal{H} \left(k^0\right)^2 
    - 2\frac{\dd\phi}{\dd\lambda}k^0 
    + 2\frac{\partial\phi}{\partial\eta}\left(k^0\right)^2 \ ,
\\
\label{eq:geodesic_equation2}
\frac{\dd k^i}{\dd\lambda}
&= -2\mathcal{H}k^0 k^i 
    + 2\frac{\textrm{d}\phi}{\dd\lambda}k^i
    - 2\frac{\partial\phi}{\partial x^i}\left(k^0\right)^2 \ ,
\end{align}
where $\mathcal{H}\equiv a^{-1}\dd a/\dd\eta$ is the conformal expansion rate. \Cref{eq:geodesic_equation1} rules the evolution of light's frequency in the cosmic frame; combined with the latter, \cref{eq:geodesic_equation2} describes light bending.

\subsection{Gravitational lensing}
\label{subsec:lensing}

Light bending implies that the images of light sources are displaced and distorted when seen through the inhomogeneous Universe. Let $\vect{\theta}$ denote the position of such an image of a point source, and $\vect{\beta}$ its FLRW counterpart, that is, the position where the image would be seen in the absence of cosmological perturbations. It is customary to refer to $\vect{\beta}$ as the source position.

\subsubsection{Geometric distortions of infinitesimal images}
\label{subsubsec:distortions_infinitesimal_images}

The distortions of an infinitesimal image are then fully encoded in the Jacobi matrix of the mapping $\vect{\theta}\mapsto\vect{\beta}$, also called distortion matrix. This matrix may be parameterised as
\begin{equation}
\label{eq:distortion_matrix}
\bm{\mathcal{A}}
\equiv
\frac{\partial\vect{\beta}}{\partial\vect{\theta}}
=
\begin{pmatrix}
    \cos\omega & -\sin\omega \\
    \sin\omega & \cos\omega  
\end{pmatrix} 
\begin{pmatrix}
    1-\kappa-\gamma_1 & -\gamma_2 \\
    -\gamma_2    & 1-\kappa + \gamma_1  
\end{pmatrix} ,
\end{equation}
with $\kappa$, $\gamma=\gamma_1+\ii\gamma_2$, and $\omega$ are respectively called the convergence, complex shear, and rotation. As a rule of thumb, $\kappa, \gamma$ are typically first order in cosmological perturbations, while $\omega$ is second order~\citep[see for example.][\S~2.3.2]{Fleury:2015hgz}. 
%In the following, we shall therefore neglect the latter.

We define the signed geometric magnification of an image as
\begin{equation}
\label{eq:magnification_signed}
\mu
= \frac{1}{\det\bm{\mathcal{A}}} 
= \frac{1}{(1-\kappa)^2-|\gamma|^2}\ .
\end{equation}
By definition of the determinant of a matrix, its absolute value $|\mu|=\dd^2\vect{\theta}/\dd^2\vect{\beta}$ is the ratio of the angular size of an infinitesimal image, $\dd^2\vect{\theta}$, and the angular size of the underlying source, $\dd^2\vect{\beta}$.

As indicated by its name and definition, the signed magnification of an image can be either positive or negative, which indicates its orientation relative to the source. An image at $\vect{\theta}$ is said to have positive parity if $\mu(\vect{\theta})>0$, and negative parity otherwise. In a Universe made of transparent lenses, any source has an odd total number~$2n+1$ of images, with $n\geq 0$ images of negative parity and $n+1$ images of positive parity~\citep{1981ApJ...244L...1B, schneider1992gravitational}.

The total geometric magnification of a source~$\vect{\beta}$ is the sum of the absolute magnifications of its $2n+1$ images~$\vect{\theta}_i(\vect{\beta})$,
\begin{equation}
\label{eq:magnification_total}
\mu\e{tot}(\vect{\beta})
\equiv \sum_{i=1}^{2n+1} \abs{\mu[\vect{\theta}_i(\vect{\beta})]} .
\end{equation}
It represents the total increase in apparent size of a source relative to is unlensed counterpart.

\subsubsection{Geometric-magnification integrals}

In a transparent Universe, the map~$\vect{\theta}\mapsto\vect{\beta}(\vect{\theta})$, which to an image associates its source, is a well-defined surjective function of $\mathbb{S}^2$ onto $\mathbb{S}^2$. In other words, any image has one and only one source, and every source has at least one image. These properties imply
\begin{align}
\label{eq:integral_signed_magnification}
\int_{\mathbb{S}^2} \dd^2\vect{\theta} \; \mu^{-1}(\vect{\theta}) &= 4\pi \ ,
\\
\label{eq:integral_total_magnification}
\int_{\mathbb{S}^2} \dd^2\vect{\beta} \; \mu\e{tot}(\vect{\beta}) &= 4\pi \ ,
\end{align}
which we refer to as the geometric-magnification integrals.

We note that in the absence of multiple imaging, $\vect{\theta}\mapsto\vect{\beta}(\vect{\theta})$ is a diffeomorphism of $\mathbb{S}^2$, so that \cref{eq:integral_signed_magnification,eq:integral_total_magnification} are merely changes of variables in an integral. The true interest of the magnification integrals is that they hold even in the presence of strong lensing and multiple images.

The total magnification integral~\eqref{eq:integral_total_magnification} is the full generalisation of the result found by \citet{weinberg1976apparent} at linear order and with point lenses. To the best of our knowledge, it was first formulated by \citet{2008MNRAS.386..230W}. The proof goes as follows. For each source element~$\dd^2\vect{\beta}$, $\dd^2\vect{
\theta}\e{tot}=\mu\e{tot}(\vect{\beta})\,\dd^2\vect{\beta}$ is the total solid angle occupied by the associated images. As one sums over $\dd^2\vect{\beta}$, the image sphere gets progressively covered. On the one hand, the whole sphere is eventually covered, because any image has a source --- for any $\vect{\theta}$, there is always a corresponding $\vect{\beta}$. On the other hand, every image point $\vect{\theta}$ is covered only once, because an image cannot have more than one source.

The inverse-magnification integral~\eqref{eq:integral_signed_magnification} can be found in \citet{kibble2005average}. Its proof relies on the relative number of positive- and negative-parity images, mentioned in \cref{subsubsec:distortions_infinitesimal_images}. For each element~$\dd^2\vect{\theta}$ of the image sphere, $\dd^2\vect{\beta}=|\mu^{-1}(\vect{\theta})|\,\dd^2\vect{\theta}$ is the corresponding solid angle in the source sphere. As one sums over $\dd^2\vect{\theta}$, the entire source sphere is covered, again because every source has at least one image. Multiple imaging implies, however, that some regions of the source sphere may be covered several times. When this occurs, since a source~$\vect{\beta}$ always has $2n+1$ images~$\vect{\theta}_i(\vect{\beta})$, $n$ of which having negative parity, their contributions cancel two by two but one,
\begin{equation}
\sum_{i=1}^{2n+1} \mu^{-1}[\vect{\theta}_i(\vect{\beta})] \, \dd^2\vect{\theta}_i
= \sum_{i=1}^{2n+1} (-1)^i \, \dd^2\vect{\beta}
= \dd^2\vect{\beta} \ .
\end{equation}
Therefore, each source element is eventually covered once and only once, which leads to \cref{eq:integral_signed_magnification}. As pointed out by KP16, albeit correct the inverse-magnification integral has little practical interest, because it is difficult to observe the parity of an image.

\subsubsection{Observable magnification: shift and tilt corrections}
\label{subsubsec:shift/tilt}

The geometric magnification~$\mu=\pm \dd^2\vect{\theta}/\dd^2\vect{\beta}$ is a well-defined theoretical notion, but it is not the most observationally relevant one. This is because $\dd^2\vect{\beta}$ represents the coordinate solid angle associated with an image, rather than the unlensed apparent size~$\dd^2\bar{\vect{\theta}}$ of its source. There are two reasons why these quantities differ, namely the `shift' and `tilt', which we elaborate on below.

\begin{figure}[t]
    \centering
    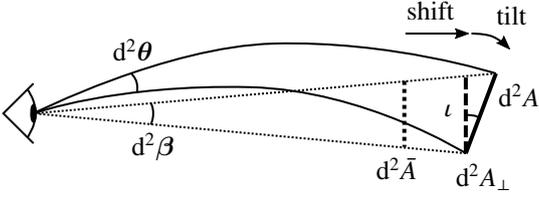
    \caption{Illustrating the difference between the geometrical magnification~$\mu=\dd^2\vect{\theta}/\dd^2\vect{\beta}$ and the observable magnification~$\tilde{\mu}(z)=[\bar{D}\e{A}(z)/D\e{A}(z)]^2=(\dd^2\bar{A}/\dd^2 A)\times(\dd^2\vect{\theta}/\dd^2\vect{\beta})$. The causes of discrepancy are: (i) the position of a source at $z$ may be shifted from the background to the perturbed case, $\dd^2\bar{A}\neq\dd^2 A_\perp$; (ii) the beam's cross section may be tilted due to lensing, $\dd^2 A_\perp=\dd^2 A_z \cos\iota$. This illustration equally applies if $z$ is replaced with $\lambda, \eta, \ldots$}
    \label{fig:shift_tilt_area}
\end{figure}

\paragraph{Observable magnification} We consider an infinitesimal source at redshift $z$ with physical area $\dd^2 A$. Let $\dd^2\vect{\theta}$ be the apparent size of an image of that source, and $\dd^2\bar{\vect{\theta}}$ its unlensed counterpart, that is the solid angle under which $\dd^2 A$ would be seen at the same redshift in FLRW. The observable magnification is defined as
\begin{equation}
\tilde{\mu}(z,\vect{\theta}) \equiv \pm \frac{\dd^2\vect{\theta}}{\dd^2\bar{\vect{\theta}}} \ ,
\end{equation}
where the $\pm$ sign indicates the image parity. By definition, the absolute observable magnification thus quantifies the change of the area distance~$D\e{A}$ to an image due to cosmological perturbations,
\begin{equation}
\label{eq:magnification_DA}
\abs{\tilde{\mu}(z,\vect{\theta})}
= \left[ \frac{\bar{D}\e{A}(z)}{D\e{A}(z,\vect{\theta})}\right]^2\ .
\end{equation}
We note that the above relies on a notion of area distance associated with individual images~$\vect{\theta}$.

\paragraph{Shift and tilt} We now relate the observational magnification~$\tilde{\mu}$ to the geometric magnification~$\mu$. For simplicity, we identify the source with an infinitesimal patch of the surface of constant redshift. However, the results obtained in this paragraph are much more general; in particular, we refer the reader to \cref{sec:shift/tilt_spherical} for an alternative approach based on a spherical source.

Let $\dd^2\vect{\beta}$ be the coordinate solid angle covered by the source. We may multiply and divide the expression~\eqref{eq:magnification_DA} of $|\tilde{\mu}|$ by $\dd^2\vect{\theta}/\dd^2\vect{\beta}$ to get
\begin{equation}
|\tilde{\mu}(z,\vect{\theta})|
=
\frac{\dd^2\vect{\theta}}{\dd^2\vect{\beta}}
\times
\frac{\bar{D}\e{A}^2(z)\,\dd^2\vect{\beta}}
    {D\e{A}^2(z,\vect{\theta})\,\dd^2\vect{\theta}}
=
|\mu(z,\vect{\theta})|
\times
\frac{\dd^2\bar{A}}{\dd^2 A} \ ,
\end{equation}
where $\dd^2\bar{A}=\bar{D}^2\e{A}(z)\,\dd^2\vect{\beta}$ is the physical area sub-tended by the coordinate solid angle $\dd^2\vect{\beta}$ in the absence of perturbations.

As illustrated in \cref{fig:shift_tilt_area}, $\dd^2\bar{A}$ differs from $\dd^2 A$ for two reasons. First, for a given redshift~$z$, the time and radial position of the source event are not necessarily the same in the background~$(\bar{\eta},\bar{r})$ as in the perturbed Universe~$(\eta, r)$; the coordinates of that event are shifted. We call $\dd^2 A_\perp$ the area sub-tended by $\dd^2\vect{\beta}$ at the shifted event; we have
\begin{equation}
\dd^2 A_\perp
= a^2[\eta(z)]\, r^2(z)\,\dd^2\vect{\beta}
\neq a^2[\bar{\eta}(z)]\, \bar{r}^2(z)\,\dd^2\vect{\beta}
= \dd^2\bar{A} \ .
\end{equation}

Second, because of light deflection, the orientation of the source is tilted by an angle $\iota$ with respect to how it would be seen in FLRW. Because they are sub-tended by the same solid angle~$\dd^2\vect{\beta}$, the tilted area~$\dd^2 A$ is larger than its untilted counterpart~$\dd^2 A_\perp=\dd^2 A \times \cos\iota$.

Summarising, the observable and geometrical magnifications are related as
\begin{equation}
\label{eq:geometric_vs_observable_magnification}
\frac
{\tilde{\mu}(z,\vect{\theta})}
{\mu(z,\vect{\theta})}
= \underbrace{
    \frac
    {\dd^2 \bar{A}}
    {\dd^2 A_\perp}
                }_{\text{shift}}
    \underbrace{
    \frac
    {\dd^2 A_\perp}
    {\dd^2 A}
                }_{\text{tilt}}
=
\frac
{a^2[\bar{\eta}(z)] \, \bar{r}^2(z)}
{a^2[\eta(z)]\, r^2(z)} \,
\cos\iota \ .
\end{equation}
We generally expect the shift to be the main driver of the difference between $\mu$ and $\tilde{\mu}$, because the tilt $\cos\iota\approx 1-\iota^2/2$ is a second-order quantity. Specifically, in the numerical results discussed in \cref{subsec:results_averaging}, the effect of tilt will always be sub-dominant; it will be precisely quantified in \cref{subsec:results_tilt}.

\paragraph{Physical origin of the shift} While $\mu$ is a pure-lensing quantity, $\tilde{\mu}$ depends on other phenomena, such as time delays, Sachs-Wolfe (SW) and integrated Sachs-Wolfe (ISW) effects, or peculiar velocities. The latter in particular may lead to significant differences between $\mu(z)$ and $\tilde{\mu}(z)$ at low redshift. If a source has, for instance, a centripetal peculiar velocity with respect to the observer, then its redshift is smaller compared to a comoving source at the same position. Thus, for a given redshift~$z$ its comoving distance must be slightly larger than the one that it would have if it were comoving, $r(z)>\bar{r}(z)$. Because the source event belongs to the observer's past light cone, this also means that it happens slightly earlier, $\eta(z)<\bar{\eta}(z)$. At low $z$, this typically results in $\dd^2 A_\perp>\dd^2\bar{A}$, implying that $\tilde\mu(z)<\mu(z)$. The conclusion would be opposite if the peculiar velocity were centrifugal.

To be more specific, at first order in the peculiar velocities of the source, $\vect{v}\e{s}$ and of the observer, $\vect{v}\e{s}$, the shift\footnote{In fact, \cref{eq:Doppler_magnification} not only allows for the shift of the iso-$z$ surface, but also for the optical aberration effect due to the observer's velocity. If the observer moves towards the source ($\vect{v}\e{o}\cdot\unit{r}>0$), then the source appears smaller to them, $\dd^2\vect{\theta}_\perp<\dd^2\bar{\vect{\theta}}_z$.} reads \citep{kaiser1987clustering,sasaki1987magnitude}
\begin{equation}
\label{eq:Doppler_magnification}
\left(
 \frac{\dd^2 A_\perp}{\dd^2 \bar{A}}
\right)_v
\equiv
1+2\tilde{\kappa}_v \ ,
\quad
\tilde{\kappa}_v
\equiv
\left(\frac{1}{\mathcal{H}r} - 1 \right)
        (\bm{v}_{\rm o} - \bm{v}_{\rm s})\cdot\vect{\beta}
+ \bm{v}_{\rm o}\cdot\vect{\beta} ,
\end{equation}
where $\vect{\beta}$ is the unit vector in the background direction of the source. The $1/(\mathcal{H}r)$ term in \cref{eq:Doppler_magnification} shows that for sources at small distances, $\tilde{\kappa}_v$ may reach large values.

The quantity~$\tilde{\kappa}_v$ is sometimes called `Doppler convergence' \citep{bonvin2008effect, bolejko2013antilensing, bacon2014cosmology}, although it is unrelated to lensing. This expression and notation originate from the fact that we may define an observable distortion matrix~$\tilde{\vect{\mathcal{A}}}$, which is to the distortion matrix~$\vect{\mathcal{A}}$ what $\tilde{\mu}$ is to $\mu$. Namely, if
\begin{equation}
\tilde{\vect{\mathcal{A}}}(z, \vect{\theta})
\equiv \bar{D}\e{A}^{-1}(z)\,\vect{\mathcal{D}}(z,\vect{\theta}) \ ,
\end{equation}
where $\vect{\mathcal{D}}$ is the Jacobi matrix of the Sachs formalism~\citep[for example][\S~2.2]{Fleury:2015hgz}, then $\tilde{\mu}=1/\det\tilde{\vect{\mathcal{A}}}$. We may then introduce a convergence-shear decomposition of $\tilde{\vect{\mathcal{A}}}$ similarly to \cref{eq:distortion_matrix}, thereby defining $\tilde{\kappa}$, to which $\tilde{\kappa}_v$ is an important contribution.

\paragraph{Fixing other parameters} In the above, we have defined the observable magnification~$\tilde{\mu}$ at fixed redshift. This choice was made for concreteness, but the definition of $\tilde{\mu}$ could be adapted if we were to fix another parameter, such as the comoving radius, the emission time, or the affine parameter. A little intellectual challenge would consist in determining which light-cone slicing may ensure $\tilde{\mu}=\mu$. To our knowledge, there is currently no answer to that particular question.

\subsubsection{Amplification and luminosity distance}

Small or remote sources, such as SNe or quasars, are generally unresolved by telescopes. In that context, the key observable is the observed flux, that is the total power received from the source per unit of telescope area, rather than the apparent size of images. We may define the amplification of a source~$\vect{\beta}$ at $z$ as the ratio of the observed flux~$S(z,\vect{\beta})$ with its unlensed counterpart~$\bar{S}(z)$. By virtue of Etherington's reciprocity law~\citep{1933PMag...15..761E}, and assuming a transparent Universe, the amplification is nothing but the total observable magnification,
\begin{equation}
\label{eq:amplification}
\frac{S(z,\vect{\beta})}{\bar{S}(z)}
= \tilde{\mu}\e{tot}(z,\vect{\beta})
\equiv \sum_{i=1}^{2n+1} \abs{\tilde{\mu}[z,\vect{\theta}_i(\vect{\beta})]}
\ .
\end{equation}

By definition of the luminosity distance~$D\e{L}$, we also have
\begin{equation}
\tilde{\mu}\e{tot}(z,\vect{\beta})
=
\left[ \frac{\bar{D}\e{L}(z)}{D\e{L}(z,\vect{\beta})}\right]^2\ .
\end{equation}
This could seem to be at odds with \cref{eq:magnification_DA} and the well-known distance-duality relation $D\e{L}=(1+z)^2 D\e{A}$. This apparent paradox is due to the fact that we have defined $D\e{A}$ for a single image, while $D\e{L}$ accounts for all the images of a given source. The two approaches are reconciled if we consistently distinguish between image-based definitions and source-based definitions. For example, we could define the area distance of a multiply imaged source~$D\e{A}(z,\vect{\beta})$ from the total apparent area occupied by all its images. In that case $[\bar{D}\e{A}(z)/D\e{A}(z,\vect{\beta})]^2=\tilde{\mu}\e{tot}(z,\vect{\beta})$ consistently with distance duality.

Finally, we note that \cref{eq:amplification} is only valid if one compares the background and perturbed fluxes at the same redshift~$z$. Had we compared the two situations, for instance, at fixed affine parameter, the background and perturbed redshift would have differed, which would have affected fluxes through the energy and the reception rate of individual photons.

\subsection{Averaging in cosmology}
\label{subsec:averaging}
The interpretation of cosmological observations, and their confrontation with theoretical predictions, involve various notions of averaging, which are non-trivially related in the presence of gravitational lensing. We review here the relevant definitions and properties of cosmological averages, elaborating on \citet{bonvin2015cosmological,kaiser2016bias,fleury2017how}.

\smallskip

Importantly, from now on we shall neglect multiple imaging, except explicitly stated otherwise. Thus, the lens mapping $\vect{\theta}\mapsto\vect{\beta}$ is assumed to be a diffeomorphism of $\mathbb{S}^2$, and the resulting magnifications are positive. In that context, there is no difference between signed, absolute, and total magnifications any more. We may also treat observable magnification and amplification as synonyms, both denoted $\tilde{\mu}$. This assumption is justified by the relatively rare occurrence of strong lensing from a cosmological perspective, and by the huge gain of simplicity that it brings to the discussions of this section.

\subsubsection{Directional averaging}
\label{subsubsec:directional_averaging_theory}

Let $X(\vect{\theta})$ be an observable in the direction $\vect{\theta}$ on the observer's celestial sphere, such as the temperature anisotropies of the cosmic microwave background, or the apparent surface density of galaxies. Directional averaging~$\ev{\ldots}\e{d}$ corresponds to a statistical average of $X(\vect{\theta})$ where all the observation directions~$\vect{\theta}$ have the same statistical weight; the average is thus weighted by the image solid angle $\dd^2\vect{\theta}$,
\begin{equation}\label{eq:direction_averaging}
\ev{X}\e{d}
\equiv \frac{1}{4\pi} \int_{\mathbb{S}^2} \dd^2\vect{\theta} \; X(\vect{\theta}) \ .
\end{equation}
One may ask how lensing affects directional averages, in particular for distance measurements. We first note that, by virtue of the \cref{eq:integral_signed_magnification}, the directional average of the inverse geometric magnification is unity,
\begin{equation}
\label{eq:direction_average_inverse_geometric_magnification}
\ev{\mu^{-1}}\e{d}
\equiv
\frac{1}{4\pi} \int_{\mathbb{S}^2} \dd^2\vect{\theta}
\; \mu^{-1}(\vect{\theta})
=
1 \ .
\end{equation}
This property is exact and applies to any slicing of the light-cone. However, as pointed out in \cref{subsubsec:shift/tilt}, \cref{eq:direction_average_inverse_geometric_magnification} has only little observational relevance, because the actually observable quantity is $\tilde{\mu}$, which differ from $\mu$ by the shift and tilt described in \cref{fig:shift_tilt_area}. Despite that concern, we may still conclude that
\begin{ceqn}
\begin{equation}
\label{eq:direction_average_inverse_amplification}
    \ev{\tilde{\mu}^{-1}}_{\rm d} \approx 1 \ , 
\end{equation}
\end{ceqn}
in the limit where the tilt/shift corrections are sub-dominant compared to the most relevant gravitational-lensing effects.

Unlike \cref{eq:direction_average_inverse_geometric_magnification}, the accuracy of \cref{eq:direction_average_inverse_amplification} depends on which parameter is fixed in the definition of $\tilde{\mu}$. For instance, \citet{kibble2005average} argued that $\ev[1]{\tilde{\mu}^{-1}(\lambda)}\e{d}=1$ was accurate for sources at fixed affine parameter~$\lambda$; this was checked numerically with ray tracing in post-Newtonian cosmological modelling \citep{sanghai2017raytracing}. However, we shall see in \cref{subsubsec:shift_cst_lambda} that the use of the affine parameter is quite risky at very high redshift. If instead the redshift is kept fixed, then significant departures from $\ev[1]{\tilde{\mu}^{-1}(z)}\e{d}=1$ are expected at low $z$ due to peculiar velocities.

\subsubsection{Source-averaging and areal averaging}
\label{subsubsec:source_area_averaging_theory}

We now consider an observable~$Y$ which is associated with a specific population of sources, such as the distance to type-Ia supernovae or the Lyman-$\alpha$ absorption in quasar spectra. The natural averaging procedure associated with such an observable is called source averaging~$\ev{\ldots}\e{s}$, and is defined as
\begin{equation}
\label{eq:source_average_definition}
\ev{Y}\e{s}
= \frac{1}{N}\sum_{s=1}^N Y_s
= \int_{\mathbb{S}^2} \dd^2\vect{\theta} \;
    \frac{1}{N}\frac{\dd^2 N}{\dd^2\vect{\theta}} \, Y(\vect{\theta}) \ ,
\end{equation}
where $N$ denotes the number of observed sources, and in the second equality we took the continuous limit. The difference with directional averaging is that the sky is not necessarily homogeneously sampled. Clearly, if the sources are not homogeneously distributed in the Universe, then their projected density~$N^{-1}\dd^2N/\dd^2\vect{\theta}$ tends to favour some regions of the sky more than others, thereby breaking the apparent statistical isotropy.

But even if sources are homogeneously distributed in space, gravitational lensing implies that they do not evenly sample the observer's sky. Indeed, lensing tends to make light beams `avoid' over-dense regions of the Universe, thereby favouring under-dense regions in source-averages. This specific effect may be captured in the notion of areal averaging. For example, if all the sources are observed at the same redshift~$z$, we may define the areal average of $Y$ as
\begin{equation}
\label{eq:areal_averaging}
\ev{Y(z)}\e{a}
\equiv \frac{1}{A(z)}
    \int_{\Sigma(z)} \dd^2 A_z \; Y(\vect{x}) \ ,
\end{equation}
with $\Sigma(z)$ the surface of constant redshift~$z$ and $A(z)$ its total proper area. The definition must be adapted if the sources are observed on other slices of the light cone, for instance all at the same emission time~$\eta$ or affine parameter~$\lambda$.

Using the area distance, $\dd^2 A_z=D\e{A}^2(z) \, \dd^2\vect{\theta}$, we may convert areal averages in terms of directional averages as follows,
\begin{equation}
\ev{Y(z)}\e{a}
= \frac{\int_{\mathbb{S}^2} \dd^2\vect{\theta} \; D\e{A}^2(z,\vect{\theta}) \, Y(z,\vect{\theta})}
        {\int_{\mathbb{S}^2} \dd^2\vect{\theta} \; D\e{A}^2(z,\vect{\theta})}
= \frac{\ev{\tilde{\mu}^{-1}(z) \, Y(z)}\e{d}}{\ev{\tilde{\mu}^{-1}(z)}\e{d}} \ ,
\label{eq:direction_to_area_averaging}
\end{equation}
from which we immediately conclude, substituting $Y=\tilde{\mu}$, that
\begin{equation}
\label{eq:areal_averaging_amplification}
\ev{\tilde{\mu}(z)}\e{a} = \frac{1}{\ev{\tilde{\mu}^{-1}(z)}\e{d}} 
\approx 1 \ ,
\end{equation}
by virtue of \cref{eq:direction_average_inverse_amplification}. 
Areal averaging exactly coincides with source-averaging if the sources are homogeneously distributed on $\Sigma(z)$, because then the number of observed sources scales as the area that they occupy, so that $N^{-1}\dd^2 N/\dd^2\vect{\theta}=\dd^2 A_z/\dd^2\vect{\theta}=D\e{A}^2(z)$. If not, corrections arise from the correlation between the fluctuations of the density of sources and the amplification; further corrections such as redshift-space distortions, must also be accounted for if the sources are observed in redshift bins~\citep{fleury2017how,Fanizza:2019pfp}. Such discrepancies between source-averaging and areal averaging typically remain below $10^{-5}$, and hence they may be neglected. Combining this approximation with \cref{eq:areal_averaging_amplification} then yields
\begin{ceqn}
\begin{equation}
\label{eq:source-average_amplification}
\ev{\tilde{\mu}(z)}\e{s}
\approx 
1 \ .
\end{equation}
\end{ceqn}
\Cref{eq:source-average_amplification} was shown to be accurate at the $10^{-3}$ level up to $z=3$ by \citet[][]{adamek2019bias}.

\subsubsection{Ensemble averaging and cosmic variance}
\label{subsubsec:ensemble_average}

We shall close this discussion with the notion of ensemble averaging. Within the standard lore, we envisage all cosmological structures as originating from quantum fluctuations in the primordial Universe~\citep{Peter:2013avv}. From that point of view, $\phi(\eta,\vect{x})$ is a particular realisation of an intrinsically stochastic field, which is believed to be initially Gaussian. In that framework, the ensemble average of any field $Z(\eta,\vect{x})$ that depends on $\phi$, which we may simply denote as $\ev{Z(\eta,\vect{x})}$, would be its expectation value over an infinite number of realisations the Universe.

\begin{figure}
    \centering
    %% Creator: Inkscape 1.0.1 (c497b03c, 2020-09-10), www.inkscape.org
%% Accompanies image file 'ensemble_averaging.pdf' (pdf, eps, ps)
\begingroup%
  \makeatletter%
  \providecommand\color[2][]{%
    \errmessage{(Inkscape) Color is used for the text in Inkscape, but the package 'color.sty' is not loaded}%
    \renewcommand\color[2][]{}%
  }%
  \providecommand\transparent[1]{%
    \errmessage{(Inkscape) Transparency is used (non-zero) for the text in Inkscape, but the package 'transparent.sty' is not loaded}%
    \renewcommand\transparent[1]{}%
  }%
  \providecommand\rotatebox[2]{#2}%
  \newcommand*\fsize{\dimexpr\f@size pt\relax}%
  \newcommand*\lineheight[1]{\fontsize{\fsize}{#1\fsize}\selectfont}%
  \ifx\svgwidth\undefined%
    \setlength{\unitlength}{265.02353073bp}%
    \ifx\svgscale\undefined%
      \relax%
    \else%
      \setlength{\unitlength}{\unitlength * \real{\svgscale}}%
    \fi%
  \else%
    \setlength{\unitlength}{\svgwidth}%
  \fi%
  \global\let\svgwidth\undefined%
  \global\let\svgscale\undefined%
  \makeatother%
  \begin{picture}(1,0.38940161)%
    \lineheight{1}%
    \setlength\tabcolsep{0pt}%
    \put(0,0){\includegraphics[width=\unitlength,page=1]{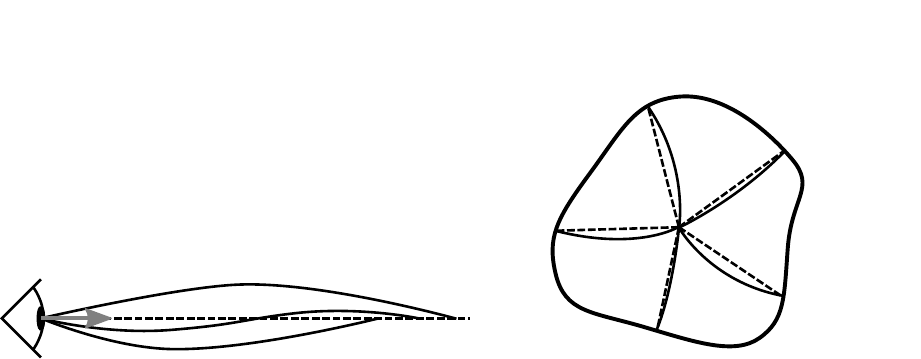}}%
    \put(0.08610256,0.07791024){\color[rgb]{0.50196078,0.50196078,0.50196078}\makebox(0,0)[lt]{\lineheight{1.25}\smash{\begin{tabular}[t]{l}$\vect{\beta}$ \end{tabular}}}}%
    \put(0,0){\includegraphics[width=\unitlength,page=2]{ensemble_averaging.pdf}}%
    \put(0.08530414,0.26124201){\makebox(0,0)[lt]{\lineheight{1.25}\smash{\begin{tabular}[t]{l}$\vect{\theta}$ \end{tabular}}}}%
    \put(0,0){\includegraphics[width=\unitlength,page=3]{ensemble_averaging.pdf}}%
    \put(0.13603851,0.35500022){\makebox(0,0)[lt]{\lineheight{1.25}\smash{\begin{tabular}[t]{l}ensemble\end{tabular}}}}%
    \put(0.61843742,0.35431285){\makebox(0,0)[lt]{\lineheight{1.25}\smash{\begin{tabular}[t]{l}observational\end{tabular}}}}%
    \put(0.82545783,0.28403346){\rotatebox{-41.013201}{\makebox(0,0)[lt]{\lineheight{1.25}\smash{\begin{tabular}[t]{l}$z=\cst$\end{tabular}}}}}%
  \end{picture}%
\endgroup%

    \caption{Correspondence between ensemble averaging and observational averaging procedure depends on which quantity is kept fixed.}
    \label{fig:ensemble_averaging}
\end{figure}

Contrary to directional, areal, or source-averages, ensemble-averaging is thus a strictly theoretical procedure, which is nevertheless used in any cosmological prediction. Ensemble averaging may be connected to other averaging procedures via the ergodicity principle. Which observable averaging is mimicked by ensemble averaging then depends on which quantities are kept fixed when making multiple realisations of the Universe as illustrated in \cref{fig:ensemble_averaging}. For example, if the observed direction of light~$\vect{\theta}$ and redshift~$z$ are kept fixed, then we get directional averaging,
\begin{equation}
\label{eq:ensemble_vs_theta}
\ev{X(z,\vect{\theta})}
= \ev{X(z)}\e{d} \ ,
\end{equation}
because any $\vect{\theta}$ is virtually affected the same statistical weight. In this scenario, the source position~$\vect{\beta}$ may change from one cosmic realisation to another. An alternative scenario would consist, on the contrary, in fixing $\vect{\beta}$ while allowing $\vect{\theta}$ to vary from one realisation to another; this yields
\begin{equation}
\label{eq:ensemble_vs_beta}
\ev{X(z,\vect{\beta})}
= \ev{\mu^{-1}(z)X(z)}\e{d} \ .
\end{equation}
We may divide the above with $\ev[1]{\mu^{-1}}\e{d}$ if directional average is taken on a fraction of the sky only. Other possibilities would consist in fixing another parameter than the redshift, such as time or affine parameter, which would correspond to averaging across other slices of the light cone.

Importantly, ergodicity is sensible only if the region of the Universe over which an observational averaging is performed is statistically homogeneous. In other words, there should not be super-sample inhomogeneity modes. Such an assumption is not satisfied in the standard lore, which predicts inhomogeneity modes at all scales. Thus, any observational average is subject to an irreducible source of uncertainty, called cosmic variance. \Cref{eq:ensemble_vs_theta,eq:ensemble_vs_beta} only hold up to cosmic variance.

\subsection{Biased distance measurements}
\label{subsec:biased_distance}

\Cref{eq:direction_average_inverse_amplification,eq:source-average_amplification} show than only very specific quantities are (almost) unbiased by cosmic inhomogeneities; in particular, most distance measurements happen to be biased. We describe here the nature and amplitude of these biases.

We introduce for convenience the dimension-less distance
\begin{equation}
d(z,\vect{\theta}) \equiv \frac{D\e{A}(z,\vect{\theta})}{\bar{D}\e{A}(z)}
= \abs{\tilde{\mu}(z,\vect{\theta})}^{-1/2} \ .
\end{equation}
Because $d$ is a non-linear function of $\tilde{\mu}$, it exhibits a statistical bias for both directional and source-averaging. For directional averaging we may expand $d$ at second order in $\tilde{\mu}^{-1}-1$ and use \cref{eq:direction_average_inverse_amplification} to get
\begin{equation}
\label{eq:direction_average_distance}
\ev{d(z)}\e{d}-1
\approx -\frac{1}{8} \ev{[\tilde{\mu}^{-1}(z)-1]^2}\e{d}
< 0 \ .
\end{equation}
Similarly, for areal or source-averaging we may expand $d$ in terms of $\tilde{\mu}-1$ which, together with \cref{eq:source-average_amplification} yields
\begin{equation}
\label{eq:source-average_distance}
\ev{d(z)}\e{s}-1
\approx \frac{3}{8} \ev{[\tilde{\mu}(z)-1]^2}\e{s}
> 0 \ .
\end{equation}

The biases appearing in \cref{eq:direction_average_distance,eq:source-average_distance} are not independent, and are usually expressed in terms of the convergence. Indeed, if the amplification is expressed in terms of some convergence and shear similarly to \cref{eq:magnification_signed}, that is, $\tilde{\mu}=[(1-\tilde{\kappa})^2-|\tilde{\gamma}|^2]^{-1}$, then at second order in $\tilde{\kappa},\tilde{\gamma}$,
\begin{equation}
(\tilde{\mu}-1)^2 = (\tilde{\mu}^{-1}-1)^2 = 4\tilde{\kappa}^2 \ .
\end{equation}
Since $\tilde{\kappa}^2$ is a second-order quantity, the difference between its directional, source-, or ensemble-average would be of higher order, and hence it should not matter much which averaging procedure is considered when substituting $\tilde{\kappa}^2$ in  \cref{eq:direction_average_distance,eq:source-average_distance}. Furthermore, if we neglect again the shift and tilt corrections\footnote{This approximation fails at low-$z$, where peculiar velocities generate Malmquist bias~\citep{bendayan2014value,kaiser2015kinematic}.} and write $\tilde{\kappa}=\kappa$, then we simply have
\begin{align}
\ev{d(z)}\e{d} &\approx 1 -\frac{1}{2}\ev{\kappa^2(z)} \ ,
\\
\ev{d(z)}\e{s} &\approx 1 +\frac{3}{2}\ev{\kappa^2(z)} \ .
\label{eq:bias_distance_source}
\end{align}

If \cref{eq:bias_distance_source} is applied at the redshift of the CMB, $z_*\approx 1100$, and $\kappa(z)$ is computed from linear perturbation theory, then the corresponding bias reaches the percent level. This is how \citet{clarkson2014what} concluded that the standard analysis of the CMB, which does not account for such a bias, might be flawed. That conclusion was shown to be incorrect by \citet{bonvin2015do,kaiser2016bias}, because the analysis of the CMB is in fact not sensitive to $\ev{d(z_*)}\e{s}$. 
However, supernova cosmology is. In supernova surveys, it is customary to use the distance modulus rather than the luminosity distance as a distance measure; its perturbation due to inhomogeneities reads $\Delta m= 5\log_{10} d$, and hence its source-averaged bias is
\begin{equation}
\ev{\Delta m(z)}\e{s}
\approx \frac{5}{4\ln 10} \ev{[\tilde{\mu}(z)-1]^2}\e{s}
\approx \frac{5}{\ln 10} \ev{\kappa^2(z)} \ .
\end{equation}
For $z<2$, this bias remains below $10^{-3}$ and hence is negligible in current SN surveys, except for reconstructions of the evolution of the dark-energy equation of state~\citep{fleury2017how}. It would be easily removed if next-generation surveys were using $1/D\e{L}^2(z)$ instead of magnitude as a distance indicator.

We finally note that all the above only holds in a transparent Universe. If this assumption is relaxed, then distance measurements may be further biased by selection effects. For instance, in a Universe made of opaque matter lumps, observed light beams do not evenly sample the density field -- they experience an effectively under-dense Universe. This result in an effective de-focussing of light as originally described by \citet{zeldovich1964observations}, later generalised by \citet{1966SvA.....9..671D} and \citet{dyer1974observations} on the basis of Einstein-Straus Swiss-cheese models~\citep{kantowski1969corrections,Fleury:2014gha}. The resulting bias on luminosity distance measurements typically reaches $10\%$ at $z=1$ for very lumpy models~\citep{Fleury:2013sna}. In \citet{Okamura_2009}, the authors made an attempt to determine the fraction of such opaque lumps based on the halo model of \citet{sheth1999large}. To date, however, there is no compelling evidence of any large effect of opaque lumps on distance measurements in our Universe~\citep[][]{helbig2019calculation}.

\subsection{Reformulation: the area of light-cone slices}
\label{subsec:area_light-cone_slices}

We may now rephrase the average-amplification rules \eqref{eq:direction_average_inverse_amplification} and \eqref{eq:source-average_amplification} in terms of the area of light-cone slices, such as surfaces of constant redshift. We consider for instance the directional average of the inverse amplification:
\begin{equation}
\label{eq:equivalence_inverse_amplification_area}
\ev{\tilde{\mu}^{-1}(z)}\e{d}
= \frac{1}{4\pi \bar{D}\e{A}^2(z)} \int_{\mathbb{S}^2} \dd^2\vect{\theta} \; D\e{A}^2(z,\vect{\theta})
= \frac{A(z)}{\bar{A}(z)} \ ,
\end{equation}
where in the last equality we introduced the area~$A(z)$ of the surface of constant redshift, $\Sigma(z)$, as well as its background counterpart~$\bar{A}(z) = 4\pi \bar{D}\e{A}^2(z)$. \Cref{eq:equivalence_inverse_amplification_area} thus tells us that $\ev{\tilde{\mu}}\e{d}\approx 1$ would be equivalent to $A(z)\approx \bar{A}(z)$; meaning that the area of iso-$z$ surfaces is mostly unaffected by inhomogeneities.

Although it may seem quite natural, the last equality of \cref{eq:equivalence_inverse_amplification_area} is, in fact, not obvious. In the background FLRW space-time, $\bar{\Sigma}(z)$ is a sphere (in comoving coordinates) at constant cosmic time; hence its proper area is clearly $\bar{A}(z)=4\pi \bar{r}^2(z)/(1+z)=4\pi \bar{D}\e{A}^2(z)$. But things are less clear in the inhomogeneous Universe, where $\Sigma(z)$ is wrinkly and is not limited to a constant-time hypersurface. The definition of its proper area~$A(z)$ is then subject to several questions about its uniqueness, if it is frame-dependent and how it relates to the angular distance. We propose to clarify these subtleties below.

\subsubsection{Surfaces of constant redshift and their area}
\label{subsubsec:iso-z_surfaces_area}

\begin{figure}[t]
    \centering
     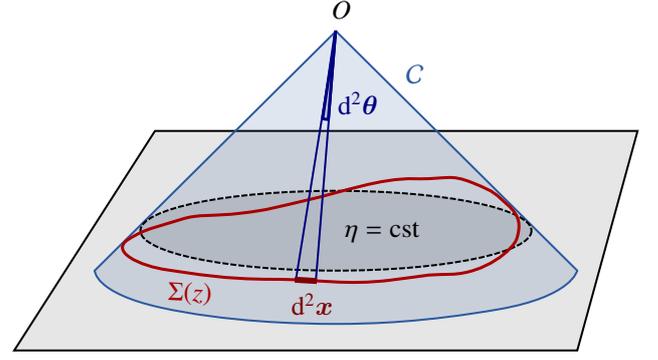
    \caption{Surface of constant redshift, $\Sigma(z)$ (red), is a particular slicing of the light cone $\lightcone$ (blue) that is not included in constant-time hyper-surfaces (grey). We represented light rays as straight lines for simplicity.}
    \label{fig:slicing_z}
\end{figure}

Surfaces of constant redshift, $\Sigma(z)$, correspond to a particular slicing of the light cone~$\lightcone$ of the observation event $O$, as illustrated in \cref{fig:slicing_z}. We note that this slicing is generally not performed at constant time, $\eta(z)\neq\cst$. This last property raises the question of how to actually define the proper area of $\Sigma(z)$.

Let $\dd^2\vect{x}$ be the element of $\Sigma(z)$ subtended by the solid angle $\dd^2\vect{\theta}$ at $O$. For causality reasons, $\dd^2\vect{x}$ must be space-like; thus, there exists a frame such that $\dd^2\vect{x}$ is strictly spatial. We shall call it the `natural frame'\footnote{The natural frame is not unique; there is in fact a class of natural frames which are all related by Lorentz boosts in the local direction of light propagation. The area of $\dd^2\vect{x}$ is invariant under such boosts, as long as they go from one natural frame to another one.} of $\dd^2\vect{x}$, and define the area $\dd^2 A_z$ of $\dd^2\vect{x}$ in that frame. Applying that construction to all elements~$\dd^2 \vect{x}$ of $\Sigma(z)$ and integrating over them then defines its total area~$A(z)$.

Now that we have defined the area of an iso-$z$ surface, we shall see how it relates to the angular distance $D\e{A}(z)$. For that purpose, we note that $\dd^2\vect{x}$ is orthogonal to direction of light propagation in its natural frame. We shall now prove this point. We may see $\lightcone$ as the hyper-surface defined by all the events that are in phase with $O$, for a spherical wave converging at the observer. If $w$ denotes the phase of that wave and $k_\mu=\partial_\mu w$ is the associated wave four-vector, then any displacement $\dd x^\mu$ across $\lightcone$ satisfies $k_\mu\dd x^\mu=\dd w=0$. This applies, in particular, to any $\dd x^\mu\in\dd^2\vect{x}\subset\Sigma(z)\subset\lightcone$. In the natural frame of $\dd^2\vect{x}$, this four-dimensional orthogonality becomes three-dimensional because $\dd x^0=0$; in other words, $\vect{k}\cdot\dd\vect{x}=0$, where $\vect{k}$ is the wave-vector in the natural frame.

The spatial orthogonality between $\vect{k}$ and $\dd^2\vect{x}$ implies that $\dd^2\vect{x}$ forms a Sachs screen space in its natural frame. Thus, $\dd^2 A_z$ is not only the proper area of $\dd^2\vect{x}$, but also the cross-sectional area of the light beam subtended by $\dd^2\vect{\theta}$ in that frame. By virtue of Sachs' shadow theorem~[\citet{sachs1961gravitational}, see also \S~2.1.2 of \citet{Fleury:2015hgz}], the area of a beam is independent of the frame in which it is evaluated, as long as it is projected on a Sachs screen. This unique notion of a beam's cross-sectional area then defines the angular distance according to
\begin{equation}
\dd^2 A_z = D\e{A}^2(z) \, \dd^2\vect{\theta} \ .
\end{equation}
This confirms that the area of $\Sigma(z)$ is indeed related to the angular distance as $A(z)=4\pi\ev[2]{D\e{A}^2(z)}\e{d}$, thereby validating the last equality of \cref{eq:equivalence_inverse_amplification_area}.

We finally note that the above reasoning actually applies to any slice of the light cone. In other words, for any parameter~$p$ such that the iso-$p$ surface~$\Sigma(p)$ is space-like ($p$ may stand, for instance, for the affine parameter~$\lambda$, time~$\eta$, the comoving radius~$r$, etc.) the area of the element~$\dd^2\vect{x}$ subtended by the solid angle $\dd^2\vect{\theta}$ reads $\dd^2 A_p=D\e{A}^2(p)\dd^2\vect{\theta}$ and the total area of the iso-$p$ surface is
\begin{equation}
\label{eq:area_iso-p_surface}
A(p)
= 4\pi\ev{D\e{A}^2(p)}\e{d}
= \ev{\tilde{\mu}^{-1}(p)}\e{d} \bar{A}(p) \ .
\end{equation}

\subsubsection{The photon-flux conservation argument}

In the spirit of the second part of \citet{weinberg1976apparent}, we may also connect the area-averaged amplification to $A(z)$ on the basis of photon conservation. Let $F_0$ be the total number of photons received per unit time by an observer. If the photon number is conserved, then the same photons crossed $\Sigma(z)$ at a rate $F_z=(1+z)F_0$, where the $(1+z)$ factor accounts for time dilation. Importantly, the latter relation holds regardless of the geometry of $\Sigma(z)$; in other words, $F_z=\bar{F}_z$.

Now, the photon flux may be written as
\begin{equation}
F_z
= \int_{\Sigma(z)} \dd^2 \vect{x} \; \vect{n}\cdot \vect{J}(z)
= A(z) \ev{|\vect{J}(z)|}\e{a} \ ,
\end{equation}
where $\vect{n}$ is the outgoing normal to $\Sigma(z)$, $\vect{J}$ is the photon flux density vector, and in the second equality we used that in its natural frame $\vect{n}$ is aligned with $\vect{k}$ and hence to $\vect{J}$. Since $\vect{J}=\vect{\Pi}/(\hbar\omega)$, where $\vect{\Pi}\propto 1/D\e{L}^2(z)$ is the Poynting vector, we conclude that $|\vect{J}(z)|\propto\tilde{\mu}(z)$. Combining this with $F_z=\bar{F}_z$ then yields
\begin{equation}\label{eq:equivalence_mu_area_z}
\ev{\tilde{\mu}(z)}\e{a} = \frac{\bar{A}(z)}{A(z)} \ .
\end{equation}
Therefore, \cref{eq:areal_averaging_amplification} is equivalent to stating that inhomogeneities do not affect the area of iso-$z$ surfaces.
We stress again here that Weinberg's photon-flux-conservation argument does not imply that $\ev{\tilde{\mu}(z)}\e{a}=1$, but rather shows that such an equality is equivalent to $A(z)=\bar{A}(z)$.

\subsubsection{CMB and the area of the last-scattering surface}
\label{subsubsec:LSS}

\begin{figure}[t]
    \centering
     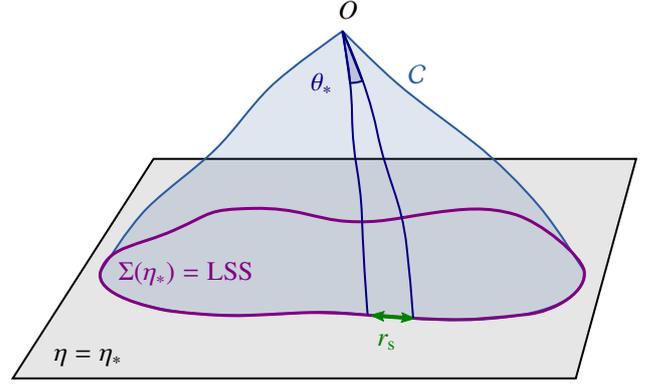
    \caption{Last scattering surface, approximated as a constant-time slice~$\Sigma(\eta_*)$ of the light cone~$\lightcone$. Its area is connected to the number of sound horizons~$r\e{s}$ that appear on the observer's CMB, and hence to its average apparent size~$\theta_*$.}
    \label{fig:slicing_eta}
\end{figure}

Hitherto, our discussion has been focused on surfaces of constant redshift because of their connection with observational averages. The archetypal application would be the analysis of the Hubble diagram in a non-homogeneous Universe, which involves the source-averaged distance modulus. However, shall we be more interested in the analysis of the CMB than in SNe, more relevant would be the last-scattering surface (LSS) and its area~$A_*$.

The area of the LSS is a relevant quantity indeed. As illustrated in \cref{fig:slicing_eta}, $A_*$ essentially drives how many sound horizons~$r\e{s}$ the observer may count in the CMB, and hence their average angular size $\theta_*$ which is one of the main direct CMB observables. The problem is then to estimate to which extent is $A_*$ affected by the inhomogeneities of our Universe. Following KP16, we shall approximate the LSS as a surface of constant time,\footnote{This is of course a gauge-dependent statement, see for example \citet{Ellis:2018led} for a discussion.} $\Sigma(\eta_*)$. By virtue of \cref{eq:area_iso-p_surface} for $p=\eta_*$, we have $A_*\approx \bar{A}_*$ where the departure from strict equality stems from the difference between $\tilde{\mu}(\eta_*)$ and $\mu(\eta_*)$, that is from the small shift and tilt effects emphasised in \cref{subsubsec:shift/tilt}.

From a pedagogical point of view, surfaces of constant time~$\Sigma(\eta)$ are very attractive because their natural frame (in the sense of \cref{subsubsec:iso-z_surfaces_area}) is the comoving frame. Indeed, for any displacement $\dd x^\mu\in\Sigma(\eta)$ we have by definition $\dd x^0=0$ in comoving coordinates. Because of this, the shift and tilt corrections to the area~$A_*$ of the LSS can be made particularly explicit. Expressing $A_*$ as the area of a polar surface, we have indeed
\begin{align}
\label{eq:area_LSS_beta}
A(\eta_*)
&= a^2(\eta_*)
\int_{\mathbb{S}^2} \dd^2\vect{\beta} \; \frac{r_*^2(\vect{\beta})}{\cos\iota_*(\vect{\beta})} \ ,
\\
\cos\iota_*
&= \left(
        1 + \left|
                \frac{1}{r_*}
                \frac{\partial r_*}{\partial\vect{\beta}}
            \right|^2
    \right)^{-1/2} \ ,
\end{align}
where $r_*(\vect{\beta})\equiv r(\eta_*,\vect{\beta})\neq \bar{r}(\eta_*)$ is the comoving radial coordinate of the point of LSS with angular coordinates~$\vect{\beta}$, which is generally shifted with respect to its background counterpart~$\bar{r}_*(\vect{\beta})$. The angle $\iota_*(\vect{\beta})$ is the tilt between the normal to the LSS and the radial direction; it encodes the wrinkles of the LSS which tend to increase its area. Recall that $\vect{\beta}$ denotes the `true' angular position of a point of the LSS, not to be confused with the direction~$\vect{\theta}$ in which that point would be observed. 

In order to get a theoretical prediction for $\delta A_*/\bar{A}_*$, we may expand \cref{eq:area_LSS_beta} at second order in cosmological perturbations, and assume ergodicity to turn integrations over $\vect{\beta}$ into ensemble averages (see \cref{subsubsec:ensemble_average}). This yields
\begin{equation}
\label{eq:perturbation_area_LSS}
\frac{\delta A_*}{\bar{A}_*}
= \underbrace{
    \ev{
    \frac{2\delta r_*(\vect{\beta})}{\bar{r}_*}
    + \frac{\delta r_*^2(\vect{\beta})}{\bar{r}_*^2}
    }
    }_{\text{shift}}
    +
    \underbrace{
    \frac{1}{2} \ev{\iota^2_*(\vect{\beta})}
                }_{\text{tilt}} \ ,
\end{equation}
with $\delta A_*\equiv A_*-\bar{A}_*$.

KP16 proposed a quite intuitive analysis of the shift term, $\delta r_*$; we shall paraphrase their idea here, while further details and minor corrections are given in \cref{appendix:LSS}. The first effect of inhomogeneities is the presence of the gravitational potential~$\phi$, which changes the effective (coordinate) speed of light as
\begin{equation}
c\e{eff} =
\left|\frac{\dd\vect{x}}{\dd\eta}\right|
= \sqrt{\frac{1+2\phi}{1-2\phi}}
= 1+2\phi + \mathcal{O}(\phi^2)\ .
\end{equation}
As a consequence, during a fixed travel time $\eta_0-\eta$, the comoving distance travelled\footnote{In KP16 that same quantity is denoted~$\lambda$. We adopt $s$ instead in order to avoid confusions with the affine parameter.}~$s$ is slightly changed, $s(\eta)=\eta_0-\eta+\delta s(\eta)$. At the LSS, this reads
\begin{equation}
\delta s_*(\vect{\beta}) = \int_{\eta_*}^{\eta_0} \dd\eta \; 2\phi[\eta, \vect{x}(\eta)] \ ,
\end{equation}
where $\vect{x}(\eta)$ is the photon trajectory connecting the observer to the point $\vect{\beta}$ of the LSS. We note that this is nothing but the usual Shapiro time delay seen from a different point of view.

Second, because of gravitational lensing, light rays are wiggly, and hence the comoving radius~$r$ that they reach after travelling a comoving distance~$s$ is slightly smaller than $s$. At the LSS we may write $r_*=s_*+\delta r\e{geo}$, with
\begin{equation}
\delta r\e{geo}(\vect{\beta}) = \int_0^{s_*} \dd s \; [\cos\iota(s)-1]
\approx -\frac{1}{2}\int_0^{\bar{r}_*} \dd r \; \iota^2(r) \ ,
\end{equation}
at second order in perturbations, and where $\iota$ is the angle made between the instantaneous photon propagation direction and the axis spanned by $\vect{\beta}$. It coincides with $\iota_*$ at the LSS.

When both effects (time delay and wiggles) are taken into account, the radial shift of the LSS with respect to its background counterpart reads
\begin{equation}
\delta r_*(\vect{\beta})
\equiv r_*(\vect{\beta})-\bar{r}_* 
= \delta s_*(\vect{\beta}) + \delta r\e{geo}(\vect{\beta}) \ .
\end{equation}
We note that $\delta s_*$ is first-order, while $\delta r\e{geo}$ is second-order in cosmological perturbations. It is thus essential to go beyond the Born approximation when evaluating $\delta s_*$ for consistency. Because of that hierarchy, $\delta s_*$ may also be considered the main driver of the wrinkles~$\iota_*$ of the LSS. 

Once ensemble average is taken, \cref{eq:perturbation_area_LSS} yields
\begin{align}
\label{eq:perturbation_area_LSS_KP16}
\frac{\delta A_*}{\bar{A}_*}
&= \int_0^{\bar{r}_*} \dd r \; \frac{(2\bar{r}_*-r)r}{\bar{r}_*^2} \, J(r)
\approx 5\times 10^{-7}\ ,
\\
\label{eq:definition_J}
J(r)
&\equiv
2\int_0^{\infty} \frac{\dd k}{2\pi} \; k^3
    P_\phi\left(\eta_0-r, k\right) \ ,
\end{align}
where $P_\phi$ denotes the power spectrum of the gravitational potential. More details can be found in \cref{appendix:LSS}. \Cref{eq:perturbation_area_LSS_KP16} agrees with Eq.~(A.44) of KP16, albeit obtained via a slightly different path.

\subsection*{Summary and goal of the remainder of this article}

Inhomogeneities may bias cosmological observations, notably via the effect of gravitational lensing on distance measurements. Biases depend on the notion of averaging that is involved. By virtue of the inverse-magnification integral $\ev[1]{\mu^{-1}}\e{d}=1$, some specific observables are expected to be almost unbiased: $\ev[1]{d^2(z)}\e{d} \approx 1/\ev[1]{d^{-2}(z)}\e{s} \approx 1$. Other combinations of $d$, such as the magnitude, generally exhibit a potentially much larger bias, on the order of $\ev[1]{\kappa^2}$. Departures from the exact $\ev[1]{d^2(z)}\e{d} = 1$ stem from $\tilde{\mu}\neq \mu$ and may be interpreted as being due to shifts and tilts of iso-$z$ surfaces with respect to their background counterpart. Apart from these shift and tilt effects, the area of iso-$z$ surfaces is unaffected by inhomogeneities. An equivalent reasoning may be applied to other slices of the light cone, such as surfaces of constant time whose area is relevant for CMB observations.

In the remainder of this article, we propose to numerically evaluate: (i) the accuracy of $\ev[1]{d^2(z)}\e{d} \approx 1/\ev[1]{d^{-2}(z)}\e{s} \approx 1$; (ii) the amplitude of the $\mathcal{O}(\ev[1]{\kappa^2})$ bias on other observables; (iii) the performance of the prediction~\eqref{eq:perturbation_area_LSS_KP16} for the area of iso-$\eta$ surfaces. Our investigation will be based on accurate ray tracing in a high-resolution $N$-body simulation, so as to fully capture non-linear effects which are difficult to control in a pure-theory approach.
  
\section{Numerical methods}
\label{sec:numerical_methods}

In this section, we present the numerical set-up and the various tools that are used to obtain the results reported in \cref{sec:results}.

\subsection{Simulation}

We use the $N$-body code \textsc{RAMSES} \citep{teyssier2002cosmological, guillet2011simple} with dark matter (DM) only. \textsc{RAMSES} uses a Particle-Mesh with Adaptive-Mesh-Refinement (PM-AMR) method, which computes the evolution of the gravitational potential and density field from particles and gravity cells. AMR allows one to probe high-density regions and hence the highly non-linear regime of structure formation.

The simulation's box comoving length is $2625~h^{-1}$Mpc with 4096$^3$ particles in a $\Lambda$CDM cosmology with \textsc{WMAP}-7 best-fit parameters \citep{komatsu2011seven}, namely $h = 0.72$, total-matter density $\Omega\e{m} = 0.25733$, baryon density $\Omega\e{b} = 0.04356$, radiation density $\Omega\e{r} = 8.076\times 10^{-5}$, spectral index $n\e{s} = 0.963$ and power-spectrum normalisation $\sigma_8 = 0.801$. The corresponding DM-particle mass is $1.88 \times 10^{10}h^{-1}$M$_\odot$.
The initial power spectrum is computed with \textsc{CAMB}~\citep{lewis2000efficient}. Initial conditions are generated using a 2LPT version of \textsc{Mpgrafic} \citep{prunet2008initial} to avoid transients \citep{scoccimarro1998transients}, which allows us to start the simulation at $z = 46$.

\citet{fidler2015general, fidler2016relativistic} showed that Newtonian $N$-body simulations, such as the one used in this article, yield physical quantities computed in the so-called $N$-body gauge. In principle, a small relativistic correction must be applied to translate such results into the Newtonian gauge~\citep{chisari2011connection}. We choose to neglect these corrections, and hence we identify the coordinates and the gravitational potential computed from the simulation with the coordinates and metric perturbation~$\phi$ in \cref{eq:perturbed_FLRW_metric}.

\subsection{Light cones}
\label{sec:light-cones}

To produce light cones from our simulation we use the onion-shell method \citep{fosalba2008onion, teyssier2009fullsky}. At each synchronisation (coarse) time step of the simulation, we output a thin spherical shell whose mean radius is the comoving distance to a central observer at the snapshot time. The shells contain all the required information about the particles (positions and velocities) and about the grid cells (gravitational potential and acceleration). Furthermore, the shells are produced with a non-zero thickness, in the sense that every spatial cell appears at different times, which allows us to compute time derivatives.

We produce three different light cones for a given observer at the centre of the simulation, which correspond to three different depths and sky coverage. The simulation box size allows us to build a full-sky cone up to a radius equal to half the box length, corresponding to $z\lesssim 0.5$. Going further would imply that some parts of the cone would repeat due to the periodic boundary conditions of the simulation. Such replication effects are suppressed by reducing the angular width of the light cone beyond $z\approx 0.5$. An intermediate narrow cone is built up to $z=2$ with a sky coverage of $2500\;\mathrm{deg}^2$, while our deep narrow cone goes up to $z=10$ covering $400\;\mathrm{deg}^2$. The cones are oriented so that light rays do not cross the same structures at different times.

Haloes on the light cones are identified using the parallel Friend-of-Friend (PFoF) code \citep{roy2014pfof} with linking length $b = 0.2$ and at least 100 DM particles. A halo's position is defined from its centre of mass, while its velocity is defined as the mean velocity of the particles that it contains. The properties of the DM haloes of the present simulation have been studied in the Appendix of \cite{corasaniti2018probing}.

We choose to model neither the haloes' intrinsic luminosity, nor the luminosity threshold for their detection by the observer. Thus, the results presented in this article exploit all the available data within the redshift ranges of interest.

\subsection{3D relativistic ray tracing}

Observables are extracted from the light cones using a fully relativistic ray-tracing procedure based on the \textsc{Magrathea} library \citep{reverdy2014propagation}.
Ray-tracing is performed backwards, that is, towards the past starting from the observation event where $\lambda=0$. Initial conditions are fixed by the observation direction~$\vect{n}$ and by setting $k^0=1$. This means that the affine parameter coincides with conformal time at O. The observer is chosen to be comoving, meaning that its peculiar velocity is set to zero, $\vect{v}\e{o}=\vect{0}$. This implies that $k^i\propto n^i$, where the proportionality factor is such that $k^\mu k_\mu = 0$.

From these initial conditions, the geodesic equations~\eqref{eq:geodesic_equation1} and \eqref{eq:geodesic_equation2} are integrated numerically with a fourth-order Runge-Kutta integrator. Specifically, photon trajectories are computed within the 3D AMR structure with four steps per AMR cell. Since the underlying $N$-body code uses a Triangular Shaped Cloud (TSC) interpolation scheme, we use an inverse TSC to estimate the gravitational potential and acceleration at the exact position of a photon. Using another interpolation method may lead to inconsistencies, such as self-accelerating particles.

The 3D TSC scheme requires 27 cells with the same refinement level to interpolate the value of a field at a position~$\vect{x}$. In practice, we start with the finest level, that is, the level of the smallest cell that contains $\vect{x}$; if there are less than 27 neighbouring cells with the same refinement level, then we try again with the next coarser level, and so on.

We stop the ray tracing if (and only if) that operation is impossible even at the coarse level, which means that 
the ray reaches the limits of our numerical background light-cones (described in \cref{sec:light-cones}) and there is no more data available to pursue its propagation.
Importantly, we save all the information about every integration step of each ray's trajectory. Besides, rays are traced irrespective of the structures that they encounter; in other words, matter is assumed to be transparent.

\subsection{Infinitesimal beams}
\label{sec:infinitesimal_beams}

The most common way to numerically evaluate the distortion matrix~$\bm{\mathcal{A}}$ is based on the multi-plane lensing formalism~\citep{1986ApJ...310..568B}, where the matter distribution near the line of sight (LOS) is projected onto various planes which are then treated as thin lenses~\citep{jain2000raytracing, hilbert2009raytracing}.
Here we want to fully exploit the 3D information of the \textsc{RAMSES} AMR octree. For that purpose, a first option consists in integrating the projected Hessian matrix~$\nabla_a\nabla_a\phi$ of the gravitational potential along the actual trajectories of light rays,
\begin{equation}
\label{eq:infinitesimal_matrix}
\mathcal{A}_{ab} = \delta_{ab} - \frac{2}{c^2}\int^{r\e{s}}_0 \dd r \; \frac{(r\e{s}-r)r}{r\e{s}}\nabla_a\nabla_b\phi[\eta(r),\vect{x}(r)]
\ ,
\end{equation}
where $r\e{s}$ is the comoving distance to the source, $a,b$ take the values $1,2$, and the two-dimensional gradient $\nabla_a$ is transverse to $\vect{n}$ (to the LOS). In practice, the 3D Hessian $\partial_i\partial_j\phi$ is computed on the mesh, and then converted in spherical coordinates to extract its angular (transverse) part $\nabla_a\nabla_b\phi$.

We shall refer to this approach as the `infinitesimal-beam method', because it describes the distortions of an infinitesimal light source. We note that this way of computing $\bm{\mathcal{A}}$ is comparable to the method used in \textsc{Ray-Ramses} \citep{barreira2016ray}, except that here $\nabla_a\nabla_b\phi$ is evaluated on the actual ray trajectory rather than on the background trajectory. In other words, we do not resort to the Born approximation.

\subsection{Ray bundles and finite-beam effect}
\label{sec:ray_bundles}

\begin{figure}
\centering
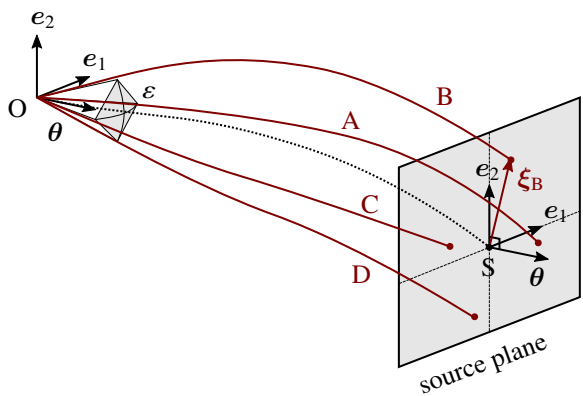
\caption{Ray-bundle method. A central light ray (dotted line) is accompanied with four auxiliary rays A, B, C, D (red solid lines). Each auxiliary ray makes an angle $\eps$ at O with respect to the central ray. The distortion matrix is estimated by comparing the relative positions of the auxiliary rays in a plane orthogonal to the line of sight~$\vect{\theta}$.}
\label{fig:jacobian_with_bundles}
\end{figure}

The second option to compute the distortion matrix~$\bm{\mathcal{A}}$ is based on a bundle of rays (\textit{a minima} three), which may be seen as a finite light beam subtended by an extended light source \citep{fluke1999raybundle, fluke2011shape}. In that `ray-bundle method', each ray is accompanied with four auxiliary rays making an angle $\eps$ with the central one, as depicted in \cref{fig:jacobian_with_bundles}. The components of $\bm{\mathcal{A}}$ are then computed from finite coordinate differences between the rays, rather than from gradients.

More precisely, our method goes as follows:
First, stop the central ray when the relevant parameter (such as redshift, time or comoving distance) has reached the desired value; this defines the fiducial source event~S. 
Second, stop the auxiliary rays at the same affine parameter~$\lambda$ as the central ray's at S. This criterion is arbitrary and other possibilities are implemented in the code.
Third, project the relative positions of the auxiliary rays on some source plane. For simplicity, we chose it to be orthogonal to the LOS\footnote{We have also implemented another prescription where the screen is orthogonal to the photon direction.}~$\vect{\theta}$. This defines the transverse separation~$\vect{\xi}$ between the auxiliary rays and the central one.

Our estimator $\hat{\vect{\mathcal{A}}}$ for the distortion matrix is then motivated by the fact that if two rays separated by a small $\Delta\vect{\theta}$ at O should have angular coordinates differing by $\Delta\vect{\beta}=\Delta\vect{\xi}/r=\hat{\vect{\mathcal{A}}}\,\Delta\vect{\theta}$,
\begin{equation}
\hat{\vect{\mathcal{A}}}
\equiv
\frac{1}{2r\eps}
\begin{bmatrix}
(\vect{\xi}\e{A}-\vect{\xi}\e{C})\cdot\vect{e}_1 & (\vect{\xi}\e{B}-\vect{\xi}\e{D})\cdot\vect{e}_1 \\
(\vect{\xi}\e{A}-\vect{\xi}\e{C})\cdot\vect{e}_2 &
(\vect{\xi}\e{B}-\vect{\xi}\e{D})\cdot\vect{e}_2
\end{bmatrix} \ ,
\end{equation}
where $r$ is the radial position where the central ray was stopped, and $\vect{e}_1, \vect{e}_2$ are unit vectors defining the initial separation of the auxiliary rays with respect to the central ray.

We note that the choices made in step 2 and 3 induce a spurious tilt in our estimate of the distortion matrix. We have checked that the effect of this tilt is negligible in all the results involving $\hat{\vect{\mathcal{A}}}$ in this article. The specific analysis of the tilt in \cref{subsec:wrinkly_surface}, which require a better accuracy, will rely on a different method.

The finite separation of the rays in the bundle method may cause some discrepancies with the infinitesimal-beam approach. Those may be quantified using the finite-beam formalism developed by \cite{fleury2017weak, fleury2019cosmic, fleury2019weak}.\footnote{This formalism was initially developed to tackle the Ricci-Weyl paradox in the gravitational-lensing theory. It was later applied to determining the effect of the finite size of light sources in weak-lensing surveys.} In particular, the finite-beam corrections to the angular power spectrum of convergence, $P_\kappa$, and shear, $P_\gamma$, are found to read
\begin{align}
\frac{P_\kappa(\ell;\eps)}{P_\kappa(\ell;0)}
&= \frac{1+J_2(2\eps\ell)-J_0(2\eps\ell)+2J_2(\sqrt{2}\eps\ell)}{2(\eps\ell)^2}
\label{eq:finite_beam_convergence_powerspectrum} \ ,\\
\frac{P_\gamma(\ell;\eps)}{P_\gamma(\ell;0)}
&= \frac{1-J_0(2\eps\ell)}{(\eps\ell)^2} \ ,
\label{eq:finite_beam_shear_powerspectrum}
\end{align}
where $P_{\kappa,\gamma}(\ell;0)$ denote the power spectra computed with the infinitesimal-beam approach described in \cref{sec:infinitesimal_beams}.

The complete derivation of \cref{eq:finite_beam_convergence_powerspectrum,eq:finite_beam_shear_powerspectrum} is provided in \cref{sec:appendix_finite_beams_calculations}; it relies on the weak-lensing, flat-sky, and Limber approximations. We note that \cref{eq:finite_beam_convergence_powerspectrum,eq:finite_beam_shear_powerspectrum} differ from the results highlighted in \citet[][Eqs.~(120), (121)]{fleury2019cosmic}, because they correspond to different beam geometries. The latter were computed from the distortions of circular beams, while the former correspond to square-shaped beams as depicted in \cref{fig:jacobian_with_bundles}.

\begin{figure}
\includegraphics[width=\columnwidth]{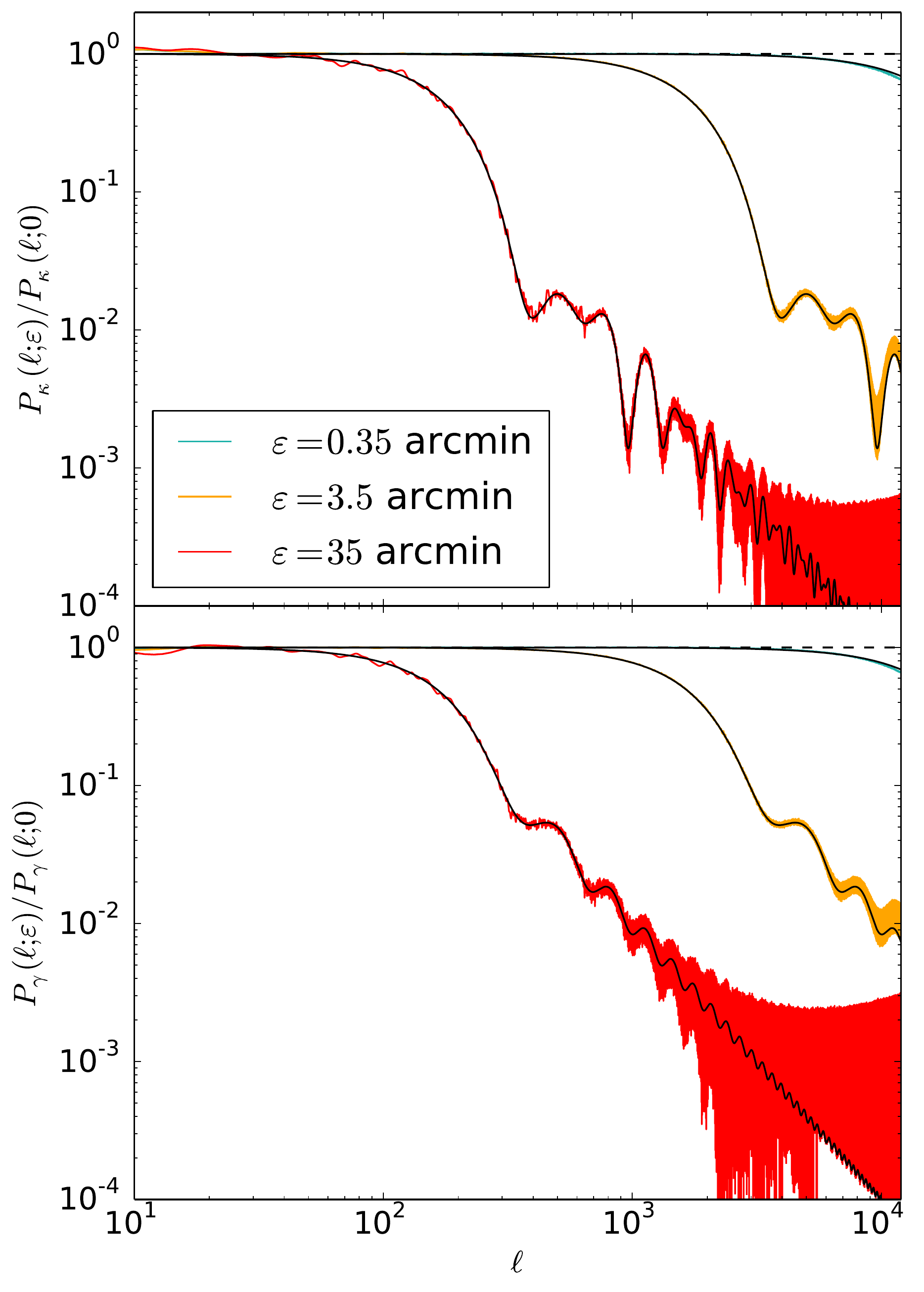} 
    \caption{Finite-beam corrections to the angular power spectrum of convergence (\emph{top panel}) and shear (\emph{bottom panel}) at $z = 1.95$, for different semi-aperture sizes. Black lines indicate the theoretical predictions of \cref{eq:finite_beam_convergence_powerspectrum,eq:finite_beam_shear_powerspectrum}, while coloured lines indicate ray-tracing results.}
    \label{fig:convergence_shear_ps_finite}
\end{figure}

\Cref{fig:convergence_shear_ps_finite} compares the predictions of \cref{eq:finite_beam_convergence_powerspectrum,eq:finite_beam_shear_powerspectrum} with ray tracing. 
Three different beam semi-apertures are considered: $\eps = 35\;\mathrm{arcmin}, 3.5\;\mathrm{arcmin}$ and $0.35\;\mathrm{arcmin}$, to which we may add $\eps=0$ corresponding to infinitesimal beams. For each value of $\eps$ but $0$, we compute the convergence and shear using the ray-bundle method, at $z=1.95$ on the intermediate narrow light cone, and for LOS dictated by \textsc{Healpix} \citep{gorski2005healpix}. Power spectra are extracted using \textsc{PolSpice} \citep{szapudi2001fast, chon2004fast}, so as to correctly allow for the angular selection function associated with the narrow cone's geometry.

Power-spectrum estimates from \textsc{Healpix} turn out to be robust until $\ell \approx {\rm nside}$.\footnote{The total number of pixels in a full-sky \textsc{Healpix} map is given by \\ $N\e{pix} = 12\times {\rm nside}^2$.} Since finite-beam effects typically kick in from $\ell\sim\eps^{-1}$, we set $\mathrm{nside} = 4096$ for $\eps = 35\;\mathrm{arcmin}$ and $\eps=3.5\;\mathrm{arcmin}$, while we set $\mathrm{nside} = 8192$ for the smallest beam size $\eps = 0.35\:\mathrm{arcmin}$, so as to ensure that the power spectra are reliable at the scales of interest.

The excellent agreement between~\cref{eq:finite_beam_convergence_powerspectrum,eq:finite_beam_shear_powerspectrum} and ray tracing, as shown in \cref{fig:convergence_shear_ps_finite}, is the first numerical evidence of the accuracy of the finite-beam formalism. This confirms that the finite-beam corrections that may arise in the present work are well understood and under control. In particular, the damping of $P_\kappa(\ell;\eps)$ $P_\gamma(\ell;\eps)$ is expected to slightly reduce the variance of convergence and shear, which are involved in distance biases. Such effects will not change the conclusions of our analysis.
%However this effect is completely under control and would not qualitatively change the present analysis.

Except otherwise stated, in the remainder of this article we set $\eps = 0.35\;\mathrm{arcmin}$. Smaller beam sizes are excluded because they would exceed the resolution of the simulation. 

\subsection{Producing observables for statistical averages}

We now turn to the generation of observables, for the purpose of computing statistical averages. As seen in \cref{sec:theory}, directional averaging and source averaging are distinct operations for which different numerical techniques must be applied.

\subsubsection{\textsc{Healpix} maps for directional averages}
\label{sec:production_healpix_maps}

Directional averaging consists in affecting equal weights to all directions of the observer's sky. This condition is easily satisfied by dividing the sky into pixels of equal area, which is the purpose of \textsc{Healpix}. In order to estimate the directional average~$\ev{X}\e{d}$ of an observable $X$, we thus shoot a ray bundle in each direction~$\vect{\theta}$ dictated by \textsc{Healpix}, compute $X(\vect{\theta})$, and take their average.

\subsubsection{Halo catalogues for source averaging}
\label{sec:production_catalogs}

Source averaging gives the same statistical weight to each source on the observer's light cone. Thus, computing source averages requires to produce a source catalogue, and to determine the null geodesic that connects each source to the observer.

In this work, sources are identified with the DM haloes, which are extracted from the simulation as described in \cref{sec:light-cones}. Geodesic identification, besides, follows \citet{breton2019imprints} \citep[see also][]{adamek2019bias}. In a nutshell, a photon is shot towards the comoving direction of a source; due to gravitational lensing the photon generally misses the source, so that LOS must be corrected and the operation iterated upon convergence at the source \citep[for an illustration, see Fig. 1 in ][]{breton2019imprints}. This procedure eventually yields the full trajectory of light for each source, as well as its observed position. We note that we do not account for multiple images of the same source, meaning that we stop the geodesic-finding algorithm as soon as one valid ray is found.

From the $N$-body code we also know the gravitational potential and velocity of each source in the catalogue. This data notably allow us to accurately compute the redshift, accounting for all the special- and general-relativistic effects at first order in the metric perturbation.
The quantitative features of the mocks\footnote{The halo catalogues, as well as convergence and magnification \textsc{Healpix} maps are available at \url{http://cosmo.obspm.fr/raygalgroupsims-relativistic-halo-catalogs}} used in this work are summarised in \cref{tab:lightcones}.
\begin{table}[h!]
	\centering
	\caption{Three light cones are used for the present work. This table indicates the type of light cone, the area covered, the maximum redshift, the number of DM haloes in our catalogues and the nside parameter used for \textsc{Healpix} maps (except otherwise stated).}	
	\begin{tabular}{l|c|c|c|c} % four columns, alignment for each
	Cone & Area (deg$^2$) & $z_{\rm max}$ & $N_{\rm haloes}$ & nside \\
	\hline
    Full-sky & - & 0.5 & $1.4 \times 10^7$& 2048\\
    Intermediate narrow  & 2500 & 2 & $1.2 \times 10^7$& 4096 \\
    Deep narrow  & 400 & 10 & $3 \times 10^6$& 8192 \\
	\end{tabular}
	\label{tab:lightcones}
\end{table}
\subsection{Surfaces on the light cone}
\label{subsec:surface_perturbations_numerical}

In this article, we shall consider various ways to slice the observer's past light cone, depending on which parameter is fixed; namely:
% Isochromes
Surfaces of constant redshift (iso-$z$),
% Isochrones
constant time (iso-$\eta$),
% Isopedes
constant comoving distance travelled (iso-$s$), and
% Isoeides
constant affine parameter (iso-$\lambda$).
In the background FLRW model, all these surfaces are spherical and correspond to each other following specific one-to-one relations. These are denoted with an over-bar; for instance, $\bar{z}(\eta)$ is the background redshift on the background iso-$\eta$ surface. In practice, we determine these background relations by shooting a single ray in the simulation with $\phi=0$.

In the inhomogeneous case, the surfaces are determined by shooting rays in directions~$\vect{\theta}$ set by \textsc{Healpix}. Since all the properties of the ray and its location are saved at each integration step, it is straightforward to determine the perturbed surfaces, such as iso-$\eta$ surfaces $\vect{x}(\eta)$, as well as the value of all the other parameters across the surfaces, so that $z(\eta, \vect{\theta})\not= \bar{z}(\eta)$. The comoving distance travelled~$s$ is computed at each integration step according to $s_{i+1}=s_i + |\vect{x}_{i+1}-\vect{x}_i|$.

Subtleties arise in the case of iso-$z$ surfaces. The significant contribution of peculiar velocities to the observed redshift raises two issues: First, since velocities are only defined for particles, interpolation on the grid is necessary to estimate $z$ at each time step. For that purpose, we use a TSC interpolation using all the DM particles in the redshift range of interest with a buffer zone.
    Second, it may happen that a light ray meets the same redshift multiple times during its propagation. In other words, the function $\lambda\mapsto z(\lambda,\vect{\theta})$ is not one-to-one in the inhomogeneous Universe; iso-$z$ surfaces are not uniquely defined. In this work we restrict the analysis to the two extremal iso-$z$ surfaces, namely the closest to the observer $\{\vect{x}[r\e{min}(z,\vect{\theta}),\vect{\theta}]\}$, and the farthest from the observer $\{\vect{x}[r\e{max}(z,\vect{\theta}),\vect{\theta}]\}$. We denote these surfaces $\Sigma_-(z)$ and $\Sigma_+(z)$ respectively.

\subsection{Computing the area of wrinkly iso-$\eta$ surfaces}
\label{subsec:wrinkly_surface}

In order to check the theoretical predictions of \cref{subsubsec:LSS} regarding the area of the LSS, and more generally of the iso-$\eta$ surfaces, we need to numerically evaluate the expression
\begin{equation}
\label{eq:area_iso_eta_surface}
A(\eta)
= a^2(\eta)
\int_{\mathbb{S}^2} \dd^2\vect{\beta} \; r^2(\eta,\vect{\beta}) \, \sqrt{1+\left|\frac{\partial \ln r}{\partial\vect{\beta}}\right|^2}
\ ,
\end{equation}
where $\vect{\beta}$ denotes the `true' position of a point of the iso-$\eta$ surface, as opposed to the direction $\vect{\theta}$ in which it would be observed. Such a computation thus requires the numerical determination of $r(\eta,\vect{\beta})$ and its gradient $\partial r/\partial\vect{\beta}$.

In practice, however, we have a more direct access to $r(\eta,\vect{\theta})$ because the iso-$\eta$ surface is determined by ray shooting (see \cref{subsec:surface_perturbations_numerical}), which yields $r(\eta,\vect{\theta})$ and $\vect{\beta}(\eta,\vect{\theta})$ for each $\vect{\theta}$ of a \textsc{Healpix} map. One could in principle compute $r(\eta,\vect{\beta})$ by finding the null geodesics between the observer and the direction $\vect{\beta}$ at each iso-$\eta$ surface, but this procedure would be computationally expensive. Another option consists in directly building a lower-resolution \textsc{Healpix} $\vect{\beta}$-map, such that in each pixel $r(\eta,\vect{\beta})$ is the average of the $r(\eta,\vect{\theta})$ for which $\vect{\beta}(\eta,\vect{\theta})$ falls into that pixel.

An even cheaper possibility consists in using the fact that the conversion between $\vect{\theta}$ and $\vect{\beta}$ is dictated by lensing quantities, which we do compute for each ray. We shall adopt this method here. Specifically, in terms of $\vect{\theta}$,  \cref{eq:area_iso_eta_surface} reads
\begin{align}
A(\eta)
&= a^2(\eta)
\int_{\mathbb{S}^2} \dd^2\vect{\theta} \; \frac{r^2(\eta,\vect{\theta})}{\mu(\eta,\vect{\theta})} \,
            \sqrt{ 1 +
                    \left|\vect{\mathcal{A}}(\eta,\vect{\theta})\,
                        \frac{\partial \ln r}{\partial\vect{\theta}}
                    \right|^2} \ ,
\\
&= 4\pi a^2(\eta)
    \ev{\frac{r^2(\eta)}{\mu(\eta)}
        \sqrt{ 1 +
                    \left| \vect{\mathcal{A}}(\eta)\,
                        \frac{\partial \ln r}{\partial\vect{\theta}}
                    \right|^2
                }
        }\e{d} \ .
\end{align}
The quantities $r, \vect{\mathcal{A}}, \mu$ are indeed evaluated in each direction~$\vect{\theta}$ of our \textsc{Healpix} maps, thereby making the computation of $A(\eta)$ much easier. In fact, the corrections due to the presence of $\mu$ and $\vect{\mathcal{A}}$, that is, the difference between integrating over $\vect{\theta}$ or $\vect{\beta}$, turn out to be very small -- about $1\%$ of $\delta A(\eta)/\bar{A}(\eta)$; these corrections could thus be neglected in first approximation.

% Computation of gradients
We use two different methods to compute the gradient $\partial r/\partial\vect{\theta}$ from the map $r(\eta,\vect{\theta})$ so as to better control numerical artefacts:
   First, the `finite differences' method, where we estimate derivatives from finite differences between pixels.
    Second, the `spherical harmonics' method, where we first decompose the map into spherical harmonics, $r(\eta,\vect{\theta})=\sum_{\ell, m} r_{\ell m}(\eta) Y_{\ell m}(\vect{\theta})$ and then compute gradients from the gradients of spherical harmonics. The same procedure is applied to the mask (with zero padding), which so as to normalise the gradient of the original map. We use \verb|healpy| routines \citep{zonca2019healpy}.

In practice, the spherical-harmonics method requires a smoothing beforehand to ensure that we recover the initial map through the operation map $\rightarrow r_{\ell m} \rightarrow$ map. The smoothing scale must be as small as possible but larger than pixel size; we thus adopt a Gaussian beam with $\mathrm{FWHM} = 5~\mathrm{arcmin}$. Although smoothing is not needed in the finite-difference method, we apply it as well to ensure that their results are comparable.

\subsection{Uncertainties on numerical averages}
\label{subsec:variance_theory}

When computing the (source or directional) average~$\ev{X}$ of an observable $X$ from mock data, the result generally differs from theoretical predictions; in other words, the ergodicity principle is not exactly satisfied. This may happen for two reasons: First, the number of mock observations in the sample is finite; this leads to a Poisson uncertainty on the estimation of any average quantity. Second, there may be super-sample inhomogeneity modes \citep{hui2006correlated}, which may bias the estimator of $\ev{X}$. This is particularly relevant to the mock data extracted from the narrow cones, which may be, for example, slightly over-dense or under-dense with respect to the simulation box.

We shall account for this uncertainty by adding error bars on any numerical average presented in the next section. The size of such error bars, that is, the uncertainty $\sigma$ on $\ev{X}$, is computed as
\begin{equation}
\label{eq:uncertainty_average_X}
\sigma^2 =
\underbrace{
\frac{1}{N} \sum_{\ell=0}^\infty \frac{2\ell+1}{4\pi} \, C_\ell^X
}_{\text{Poisson}}
+
\underbrace{
\sigma\e{ss}^2
}_{\text{super sample}} \ .
\end{equation}
In \cref{eq:uncertainty_average_X}, the first term represents the Poisson uncertainty due to the finite sample size; $C_\ell^X$ denotes the $\ell$th multipole of $X$, and $N$ is the number of mock observations -- the number of pixels in the map for directional average, or the number of sources for source-averaging. When $N\gg 1$, this first term may be neglected.

The second term in \cref{eq:uncertainty_average_X}, $\sigma\e{ss}^2$, is the super-sample variance. This contribution may be understood as a generalisation of the cosmic variance mentioned in \cref{subsubsec:ensemble_average}. Suppose that we compute the directional average of $X$ within a cone with half angle $\alpha$ at the observer,
\begin{equation}
\ev{X}_\alpha
= \int_0^\alpha \dd\vartheta \, \sin\vartheta
    \int_0^{2\pi} \dd\ph \; X(\vartheta,\ph)
= \int_{\mathbb{S}^2} \dd^2\vect{\theta} \; W(\vect{\theta}) \, X(\vect{\theta}) \ ,
\end{equation}
with $\vect{\theta}=(\vartheta,\ph)$, and where
\begin{equation}
W(\vect{\theta}) \equiv \frac{[\vartheta<\alpha]}{2\pi(1-\cos\alpha)},
\end{equation}
is the cone's window function. We note that $\ev{X}_{\alpha=\pi}=\ev{X}\e{d}$ by definition. $\ev{X}_\alpha$ is a random variable, because it depends on the actual orientation of the cone. Its variance is the super-sample variance~$\sigma\e{ss}^2$ that we are looking for,
\begin{equation}
\sigma\e{ss}^2
= \ev{\ev{X}_\alpha^2}
= \int \dd^2\vect{\theta} \, \dd^2\vect{\theta}' \;
    W(\vect{\theta}) \, W(\vect{\theta}')
   \ev{X(\vect{\theta}) X(\vect{\theta}')} \ .
   \label{eq:configuration_space_sample_variance}
\end{equation}
Decomposing the window function~$X$ and observable~$W$ in spherical harmonics,
% Our conventions
%\begin{align}
%X(\vect{\theta})
%&= \sum_{\ell,m} X_{\ell m} Y_{\ell m}(\vect{\theta}) ,\\
%W(\vect{\theta})
%&= \sum_{\ell,m} W_{\ell m} Y_{\ell m}(\vect{\theta}) \ ,
%\end{align}
%
%\begin{equation}
%    \overline{X_{\ell m} X^*_{\ell' m'}}
%    = \delta_{\ell, \ell'}\,\delta_{m,m'} \, C_\ell^X \ ,
%\end{equation}
for which
\begin{equation}
W_{\ell m}
=
\begin{cases}
\dfrac{\delta_{m0}}{\sqrt{4\pi}}
    & \ell=0, \\[2mm]
\dfrac{\delta_{m0}}{\sqrt{4\pi(2\ell+1)}}
    \dfrac{P_{\ell-1}(\cos\alpha)-P_{\ell+1}(\cos\alpha)}
        {1-\cos\alpha}
    & \ell \geq 1,
\end{cases}
\end{equation}
where $P_\ell$ are Legendre polynomials, we find
\begin{align}
\sigma\e{ss}^2
&= \sum_{\ell,m} |W_{\ell m}|^2\,C_\ell^X \ ,
\\
&= \frac{1}{4\pi}
    \left[
    C_0^X +
    \sum_{\ell=1}^\infty
    \frac{C_\ell^X}{2\ell+1}
     \left|
            \frac{P_{\ell-1}(\cos\alpha)-P_{\ell+1}(\cos\alpha)}
                    {1-\cos\alpha}
        \right|^2
    \right] .
    \label{eq:correlated_variance}
\end{align}
In the full-sky limit ($\alpha=\pi$), the uncertainty on $\ev{X}_\alpha$ is expectedly dictated by the monopole only, $\sigma\e{ss}^2=C_0^X/4\pi$. 

\subsection{Variance within a finite simulation box}
\label{sec:variance_finitebox}
The variance derived in \cref{subsec:variance_theory} depends on $C_\ell^X$, which ultimately depends on the matter density power spectrum~$P(k)$. As such, it would seem natural to use the information from all the wavelengths available in $P(k)$.
However there is a subtlety when estimating the variance from $N$-body simulations: these are usually cubic boxes with periodic boundary conditions, with a mean density equal to zero inside the cubic volume by definition. This means that, unlike the real Universe, there can be no inhomogeneity modes with wavelengths larger that the box itself.

To mimic this effect, \cite{gelb2994cold} imposed a cut-off in the matter power spectrum at $k_{\rm min} = 2\pi/L$ with $L$ the comoving size of the box. This approach has been widely studied either to estimate 3D statistics \citep{bagla2005comments, power2006impact} or 2D weak-lensing analysis \citep{harnois-deraps2015simulations}. 
To go further, one may convolve the power spectrum with the appropriate cubic window function in real space \citep{pen1997generating, sirko2005}; we found that this last correction was negligible because our box is large enough.

Finite-box corrections effectively change the low-$k$ behaviour of the power spectrum of any quantity that depends on the gravitational potential or on the density contrast. As a consequence, the angular power spectrum $C_\ell^X$ of any related observable $\ell$ is modified at low $\ell$ compared to its theoretical predictions in an infinite Universe. Depending on the shape of $C_\ell^X$, this effect may be more or less pronounced; in particular, we expect a strong impact when most of the power is carried by large scales. In that case, it is crucial to carefully account for finite-box effects as well as evaluating power spectra beyond Limber's approximation. See \citet{kilbinger2017precision} and references therein for a review about low-$\ell$ corrections in weak-lensing studies.

We finally mention that, besides the aforementioned low-$k$ corrections, there are high-$k$ corrections due to mass assignment, shot noise and aliasing. These are already well known \citep{hockney1981computer} and negligible in the present study.

\subsection{Constrained Gaussian random field and ensemble averaging}
\label{subsec:constrained_grf_method}

The ergodic principle does not hold when averaging over a small volume. For example, the ensemble average of the gravitational potential vanishes by definition, $\ev{\phi} = 0$; yet, its average over a small spatial region around the observer should be almost equal to $\phi_0$, which in general is non-zero. This mis-match, due to small-scale correlations, may lead to spurious discrepancies between ensemble averages and numerical averages at low redshift.

It is possible to smoothly transition from the constraint at the observer to the expectation from ensemble average through the constrained random field formalism \citep{hoffman1991constrained, vandeweygaert1996peak, mitsou2020general}. Take again the example of the gravitational potential~$\phi$ near the observation event, which is subject to the constraint $\phi(\vect{0}) = \phi_0$. Following \citet{desjacques2020statistics}, the constrained ensemble average of $\phi$ then reads
\begin{equation}
\label{eq:constrained_mean}
    \ev{\phi(\vect{x})|\phi_0} = \phi_0\, \frac{\xi_\phi(r)}{\xi_\phi(0)} \ ,
\end{equation}
where $\xi_\phi(r)$ is the unconstrained two-point correlation function of the gravitational potential and $\xi_\phi(0)=\sigma_\phi^2$ its variance at $z =0$. We implicitly assume that all quantities are evaluated on the light cone to alleviate notation, $\xi_\phi(r) \equiv\xi_\phi(\eta_0-r,r)$.
For numerical applications, $\xi_\phi(r)$ is estimated from the linear power spectrum with an infrared cutoff at $k\e{min} = 2\pi/L$, as per \cref{sec:variance_finitebox}. 

The constraint~$\phi(\vect{0})=\phi_0$ also impacts the two-point correlation function of $\phi$. Indeed, since its value is fixed at $\vect{0}$, we expect the variance of $\phi$ to vanish as we approach that point. The constrained two-point correlation function reads
\begin{align}
\zeta_\phi(\vect{x},\vect{x}')
&\equiv \ev{\phi(\vect{x})\phi(\vect{x}')|\phi_0}
    - \ev{\phi(\vect{x})|\phi_0}\ev{\phi(\vect{x}')|\phi_0} \ ,
\\
&= \xi_\phi(|\bm{x}-\bm{x}'|)
    - \frac{\xi_\phi(r)\xi_\phi(r')}{\xi_\phi(0)} \ .
\label{eq:constrained_variance}
\end{align}
\Cref{eq:constrained_mean,eq:constrained_variance} are particularly useful for fields that are mostly correlated on large scales, such as the gravitational potential whose power spectrum scales as $P(k)/k^4$.

\section{Results}
\label{sec:results}

In this section, we confront the theoretical predictions of \cref{sec:theory} with numerical results obtained with the methods described in \cref{sec:numerical_methods}. Directional averages and source-averages of cosmological distance indicators are considered in \cref{subsec:results_averaging}. In the next sub-sections, we then focus on the rather subtle shift (\cref{subsec:results_shift}) and tilt (\cref{subsec:results_tilt}) corrections to the amplification~$\tilde{\mu}$, that is, to the area of various light-cone slices.

\subsection{Directional and source averages}
\label{subsec:results_averaging}

\subsubsection{Directional averaging}
\label{subsubsec:direction_averaging_results}

We analyse here the statistical properties of the \textsc{Healpix} maps generated as described in \cref{sec:production_healpix_maps} for directional averages. To avoid numerical uncertainties due to short light propagation, we focus on data with $z \geq 0.2$. 

\begin{figure}
\includegraphics[width=\columnwidth]{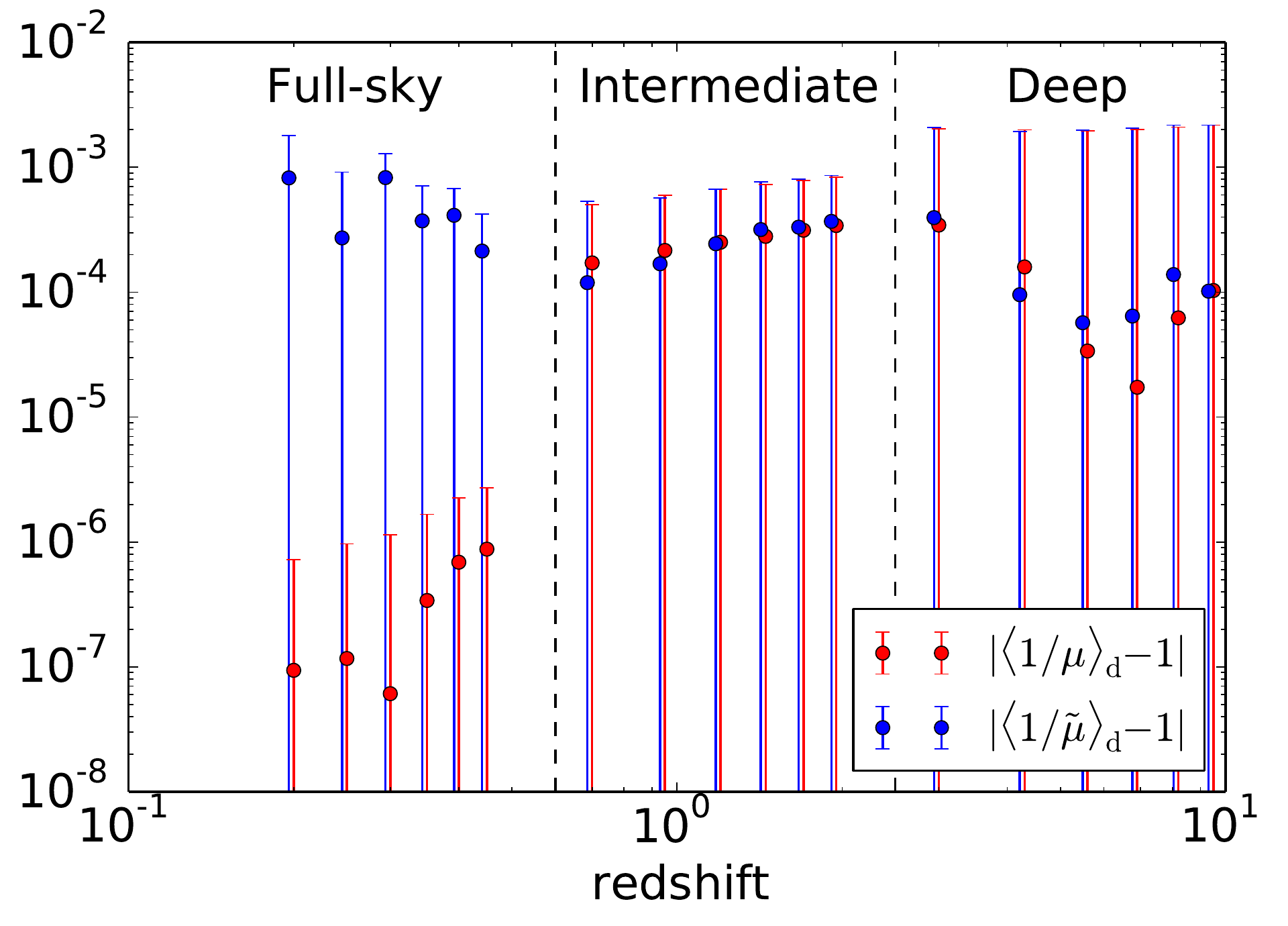}  
    \caption{Departures from 1 of the directional average of the inverse geometric magnification ~$\ev[1]{\mu^{-1}(z)}\e{d}$ and of the inverse observable magnification~$\ev[1]{\tilde{\mu}^{-1}(z)}\e{d}$ at constant redshift~$z$. Dots indicate numerical averages over \textsc{Healpix} maps, while error bars allow for super-sample and Poisson variance; see \cref{sec:appendix_variance_source_direction_averages} for their expressions. The full-sky variance on $\ev[1]{\mu^{-1}(z<0.5)}\e{d}$ does not contain super-sample variance.
    The three vertical dashed lines indicate the limits of our three light cones: full-sky ($z < 0.5$), intermediate narrow ($0.5 < z < 2$), and deep narrow ($2 < z < 10$).}
    \label{fig:direction_averaging_invmu}
\end{figure}

% Test <1/mu> = 1
\Cref{fig:direction_averaging_invmu} shows how much the directional average of the inverse geometric and observable magnifications depart from unity. Dots indicate numerical averages, while error bars account for both Poisson and super-sample variance as described in \cref{subsec:variance_theory}; an exception is the full-sky estimate of $\ev[1]{\mu^{-1}}\e{d}$ for $z<0.5$, which is only affected by Poisson variance. The exact expressions that we use are provided in \cref{sec:appendix_variance_source_direction_averages}.

First of all, we note that we do not recover exactly $\ev[1]{\mu^{-1}(z)}\e{d}=1$, which yet should be exactly satisfied. For the full-sky cone ($z<0.5$) the discrepancy is extremely small $(<10^{-6})$ and is attributed to the discretisation of the full-sky map: averages are performed over a large but finite number of points. This interpretation is supported by the fact that $\ev[1]{\mu^{-1}(z<0.5)}\e{d}$ is consistent with unity within the Poisson uncertainty. For the two other cones ($z>0.5$), the discrepancy is larger $(\sim 10^{-4})$. This should not come as a surprise, because $\ev[1]{\mu^{-1}}\e{d}=1$ is exact on the full sky only. The intermediate and deep narrow cones are simply slightly under-dense or over-dense with respect to the average box. Again, this interpretation is supported by the fact that $\ev[1]{\mu^{-1}(z)}\e{d}-1$ falls well into the error bars accounting for super-sample variance.

The direction-averaged inverse amplification $\ev[1]{\tilde{\mu}^{-1}(z)}\e{d}$ departs from $1$ by almost $10^{-3}$ even for the full-sky data. We may note that, unlike $\mu$, $\tilde{\mu}$ is subject to super-sample variance even on a full sky; however, the main reason for which $\ev[1]{\tilde{\mu}^{-1}(z)}\e{d}\neq 1$ at low $z$ is the shift effect due to peculiar velocities (see \cref{subsubsec:shift/tilt}). For the other two cones $\ev[1]{\tilde{\mu}^{-1}(z)}\e{d} - 1$ is mostly due to the super-sample variance as the one affecting $\ev[1]{\mu^{-1}(z)}\e{d} - 1$.

Our results show that the approximation $\ev[1]{\tilde{\mu}^{-1}(z)}\e{d} \approx 1$ is accurate up to $10^{-3}$ up to $z=10$. They also indicate that incomplete sky coverage in actual observations may in the end be the main cause of any departure from $\ev[1]{\tilde{\mu}^{-1}(z)}\e{d}=1$ at high redshift.

\begin{figure}
\includegraphics[width=\columnwidth]{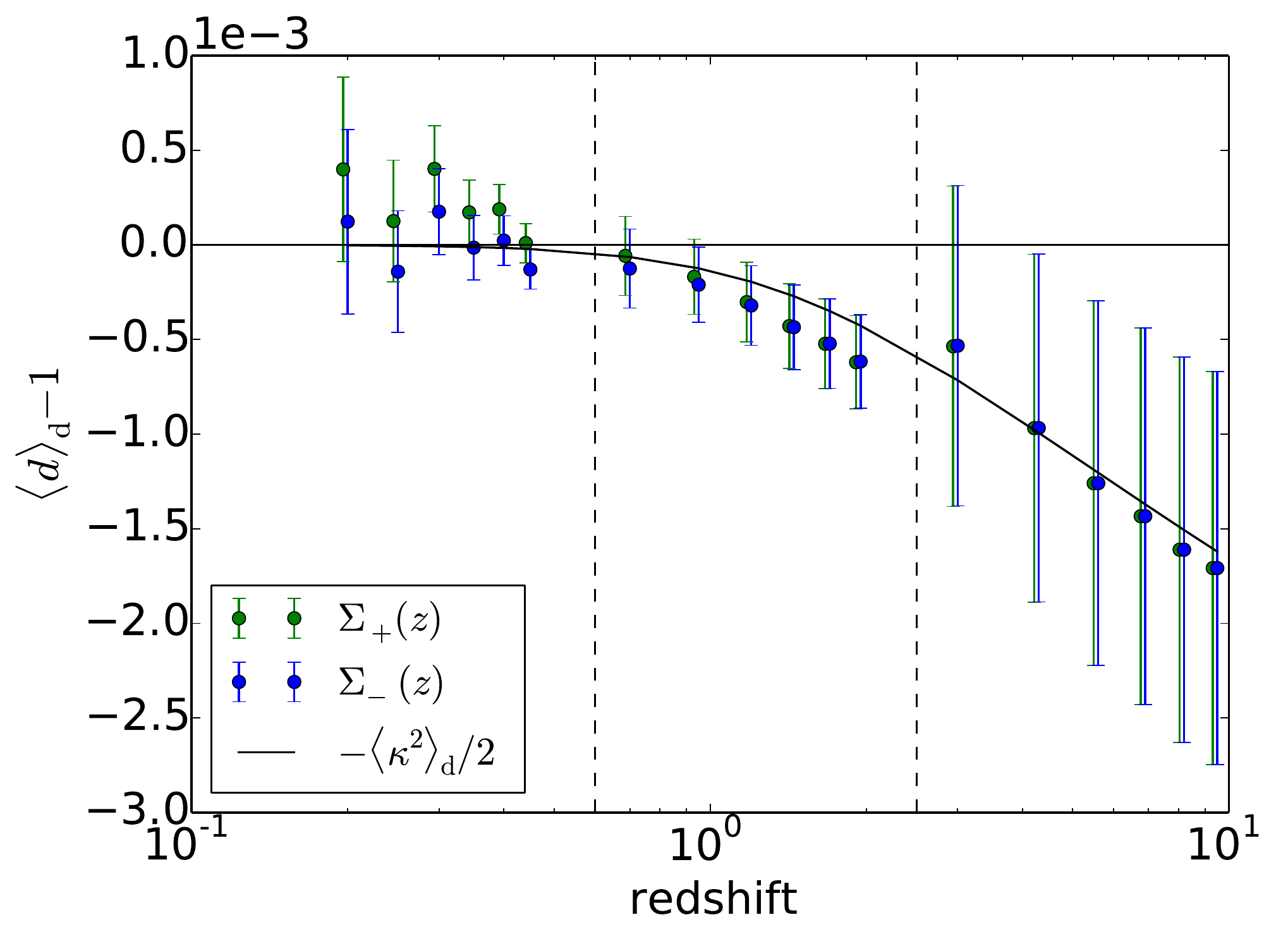}  
    \caption{Bias on the distance-redshift relation when averaged over directions, compared with the theoretical prediction~$\ev{d(z)}\e{d}=1-\ev[1]{\kappa^2(z)}/2$. Since the iso-$z$ surface is not unique, we indicated results for the closest surface~$\Sigma_-(z)$ and farthest surface~$\Sigma_+(z)$ from the observer. Expressions for the error bars may be found in \cref{sec:appendix_variance_source_direction_averages}.}
    \label{fig:direction_averaging_d}
\end{figure}

% Evaluate <d>-1
We now turn to the bias of angular or luminosity distance. For directional averaging, we expect $d(z)$ to be negatively biased according to $\ev{d(z)}\e{d} \approx 1 - \ev[1]{\kappa^2(z)}/2$. 
We evaluate $\ev[1]{\kappa^2}$ from the data directly as  $\ev[1]{\kappa^2}\e{d}$, and we have checked that a pure-theory estimate based on the matter power spectrum gives the same results. As shown in \cref{fig:direction_averaging_d}, that theoretical prediction is in good agreement with numerical results within the error bars dominated by super-sample variance. Again, low-redshift departures are due to peculiar velocities whose shift effect is not accounted for in $\kappa^2$. While $\ev[1]{\mu^{-1}(z)}\e{d}$ and $\ev[1]{\tilde{\mu}^{-1}(z)}\e{d}$ remains unity within error bars, $\ev{d(z)}\e{d}-1$ clearly departs from zero for $z > 1$.
%
% Accuracy of the expansion

\begin{figure}
\includegraphics[width=\columnwidth]{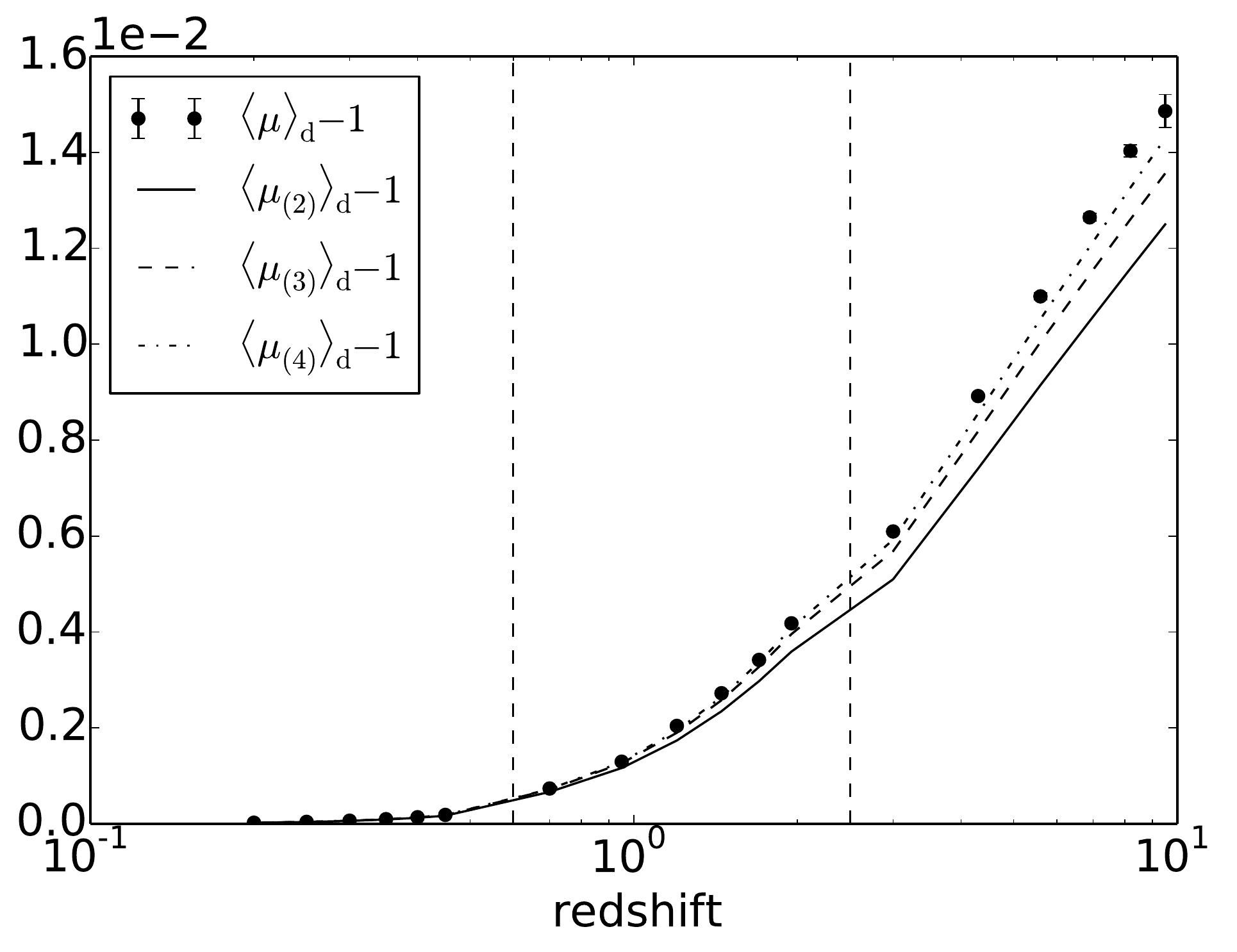}  
    \caption{Accuracy of the second-order expansion for the direction-averaged magnification. The error bars are only due to Poisson variance.}
    \label{fig:direction_averaging_mu}
\end{figure}

We finally evaluate the accuracy of the expansion that allowed us to express all distance biases in terms of $\ev[1]{\kappa^2(z)}$ in \cref{subsec:biased_distance}. We shall take the direction-averaged magnification in order to illustrate that point. Taylor-expanding $\mu$ at order $n$ in $\mu^{-1}-1$ yields
\begin{equation}
\label{eq:Taylor_expansion_mu}
\mu
=
\left[1-(1-\mu^{-1})\right]^{-1}  
=
\underbrace{
            \sum_{k=0}^n (1-\mu^{-1})^k
            }_{\mu_{(n)}}
+ \mathcal{O}(1-\mu^{-1})^{n+1} \ .
\end{equation}
\Cref{fig:direction_averaging_mu} confronts $\ev{\mu}\e{d}$ with its Taylor-expansion~$\ev[1]{\mu_{(n)}}\e{d}$ for $n=2,3,4$, where all these quantities are evaluated numerically. The error bars only allow for Poisson errors, because super-sample variance affects $\mu$ and $\mu_{(n)}$ in the same way. We see that the quadratic expansion~$\ev{\mu_{(2)}}\e{d}$ provides a good approximation of the exact result for $z<1$. Beyond that, we observe discrepancies reaching about $10\%$ at $z=10$. These are due to departures from the weak-lensing regime ($|\mu^{-1}-1|\ll 1$), which are more likely to happen as the redshift increases. \Cref{fig:direction_averaging_mu} also illustrates the convergence of the series expansion of \cref{eq:Taylor_expansion_mu}.

\subsubsection{Source averaging}
\label{subsubsec:results_source_averaging}

We now analyse the mock halo catalogues produced for the three light cones (full sky, intermediate narrow and deep narrow) as described in section~\ref{sec:production_catalogs}. We arrange the haloes in tomographic bins of width $\Delta z = 0.08, 0.2$ and $1.5$ for the full sky, intermediate and deep light cones, respectively. 

\begin{figure}
    \includegraphics[width=\columnwidth]{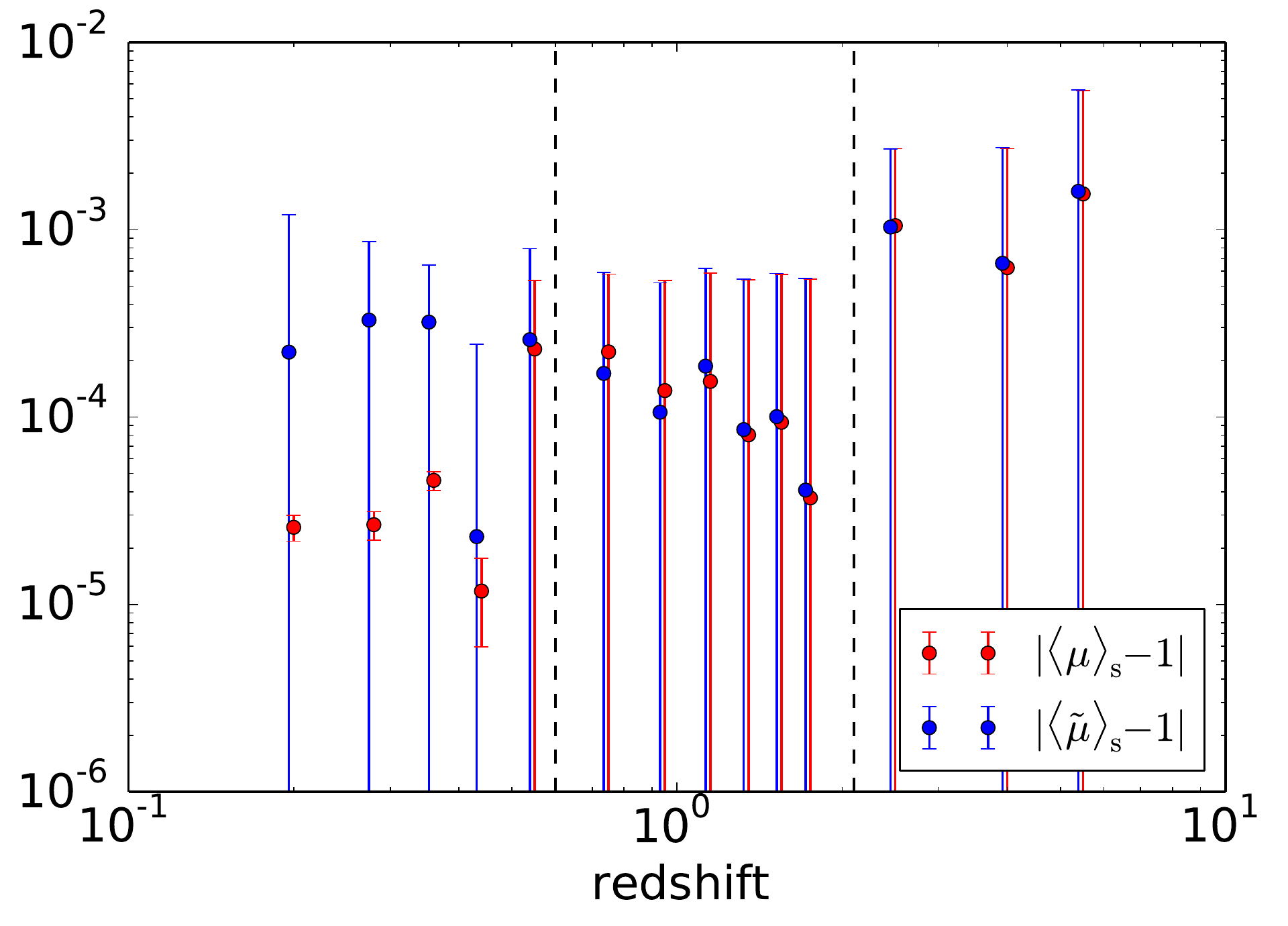} 
 \caption{Departures from 1 of the source-average of the geometric magnification~$\ev[1]{\mu(z)}\e{s}$ and of the observable magnification~$\ev[1]{\tilde{\mu}(z)}\e{d}$ at constant redshift~$z$. Dots indicate numerical averages over binned halo catalogues, while error bars allow for super-sample and Poisson variance; see \cref{sec:appendix_variance_source_direction_averages} for their expressions. The full-sky variance on $\ev[1]{\mu(z)}\e{s}$ does not contain super-sample variance.}
\label{fig:source_averaging_mu}
\end{figure}

We have seen in \cref{subsubsec:source_area_averaging_theory} that the source-average of the geometric and observable magnifications are expected to be almost unity~$\ev{\mu(z)}\e{s}\approx\ev{\tilde{\mu}(z)}\e{s}\approx 1$.\footnote{In principle these (approximate) relations should apply to the total absolute magnification rather than to the signed magnification. As mentioned in \cref{subsec:averaging}, they coincide in the absence of multiple imaging, which is a good approximation here. In fact, no negative-parity image was found in our halo catalogue. 
%This is partly due to the ray-finder procedure described in \cref{sec:production_catalogs}, which is expected to preferably find the main image of a source. 
This is mostly due to the source-averaging procedure, which is expected to give less strong-lensing events that the directional averaging one, coupled with the fact that there are less sources at high redshift.
Another reason is the use of ray bundles, which tend to smooth out the matter inhomogeneities on very small scales and thereby reduce the occurrence of strong lensing.} Unlike directional averaging, departures from equality may be caused by the non-trivial clustering of sources in addition to the shift and tilt corrections responsible for $\tilde{\mu}\neq\mu$. Numerical results for $\ev{\mu(z)}\e{s}, \ev{\tilde{\mu}(z)}\e{s}$ are depicted in \cref{fig:source_averaging_mu}. Again, error bars account for both Poisson and super-sample (cosmic variance) whose expressions are given in \cref{sec:appendix_variance_source_direction_averages}. Since the number of mock haloes per bin is smaller than the number of pixels of the maps used in \cref{subsubsec:direction_averaging_results}, Poisson variance is larger here, especially at high redshift where it exceeds super-sample variance.

For the full-sky data ($z<0.5$), we observe that $|\ev{\mu(z)}\e{s} - 1| \approx 10^{-5}$, which is about 100 times larger than what was found for $|\ev[1]{\mu^{-1}(z)}\e{d}-1|$. This discrepancy, which goes beyond the estimated uncertainty, is due to the fact that the latter does not properly account for the spatial clustering of haloes. Halo clustering is present in the $\dd^2 N/\dd^2\vect{\theta}$ kernel in the definition~\eqref{eq:source_average_definition} of source averaging. The correlation between spatial clustering and lensing convergence was predicted to be on the order of $10^{-5}$ in \citet{fleury2017how,Fanizza:2019pfp}, which agrees with the present results. As for $\ev{\tilde{\mu}}\e{s}$, just as in \cref{fig:direction_averaging_invmu} the numerical results are in agreement with unity within the error bars dominated by peculiar velocities.

The interpretation of the results for the intermediate ($0.5<z<2$) cone is similar to the directional-averaging case. The relative effect of the shift, that is,  the main difference between $\mu$ and $\tilde{\mu}$, reduces as the impact of super-sample variance increases. As for the deep cone, Poisson variance dominates due to the small number of haloes per tomographic bin. In both cases, $\ev{\mu}\e{s}$ and $\ev{\tilde\mu}\e{s}$ are compatible with unity within error bars, so that no unexpected bias arises.

\begin{figure}
    \includegraphics[width=\columnwidth]{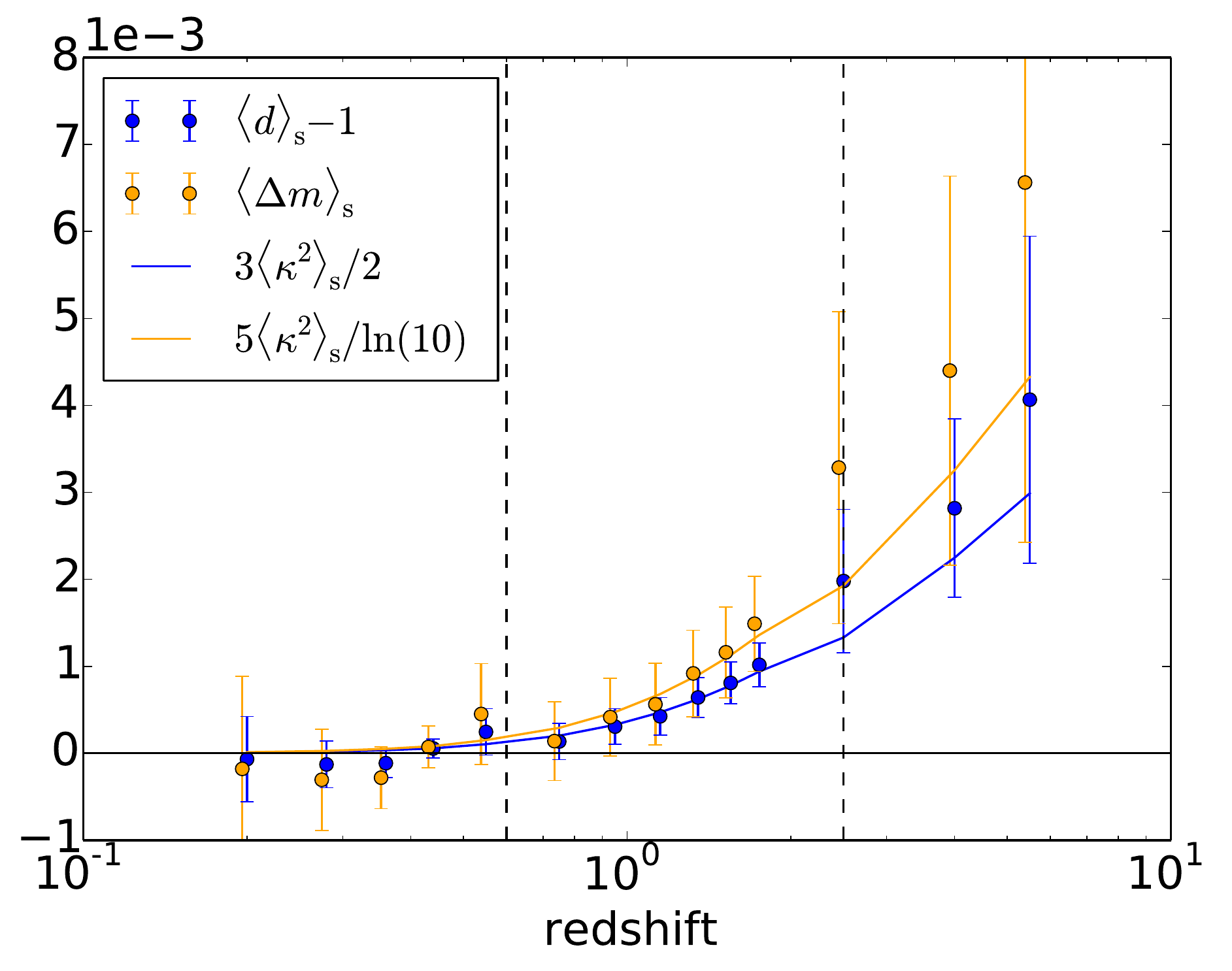} 
 \caption{Bias of the source-averaged distance-redshift~$\ev{d(z)}\e{s}$ (blue) and magnitude-redshift~$\ev{m(z)}\e{s}$ (yellow) relations. The associated solid lines are the theoretical predictions for these quantities as presented in \cref{subsec:biased_distance}. Error bars account for Poisson and super-sample variance; their expressions can be found in \cref{sec:appendix_variance_source_direction_averages}.}
 \label{fig:source_averaging_d_m}
\end{figure}

% Show <d> and <Delta m>
We now turn to more observationally relevant biases. For SN surveys, the common practice consists in fitting the magnitude-redshift relation $m(z)$ with the FLRW prediction~\citep{2018ApJ...859..101S}. Such a method is thus biased by $\ev{\Delta m(z)}\e{s}=5\ev{\log_{10}d(z)}\e{s}$. For standard-siren Hubble diagrams, it may be more common to fit the luminosity distance-redshift relation, which would be biased by~$\ev{d(z)}\e{s}$. Numerical results on these biases are reported in \cref{fig:source_averaging_d_m}, and compared with the theoretical predictions presented in \cref{subsec:biased_distance}. Similarly to \cref{subsubsec:direction_averaging_results}, we estimate the variance of the convergence from the data itself. We note that $\ev[1]{\kappa^2(z)}\e{s}$ slightly differs from $\ev[1]{\kappa^2(z)}\e{d}$; their relation is essentially given by \cref{eq:direction_to_area_averaging}.

Again, our results agree with theoretical predictions within the error bars. We note that $\ev{d(z)}\e{s}$ reaches a few $10^{-3}$ at high redshift, which may become non-negligible for the standard-siren Hubble diagrams of the future LISA mission~\citep{2016JCAP...10..006C}. Such a bias would be easily removed by fitting $D\e{L}^{-2}(z)$ instead of $D\e{L}(z)$, as pointed out by \citet{fleury2017how}.

We finally mention that, since $d$ and $m$ are non-linear functions of $\mu$, theoretical predictions on their bias based on a second-order Taylor expansion are subject to the same small inaccuracies as displayed in \cref{fig:direction_averaging_mu}. These inaccuracies are however much smaller than the uncertainty on those quantities, and hence may be safely neglected.

\subsection{Focus on the shift correction}
\label{subsec:results_shift}

In \cref{subsec:results_averaging}, we have analysed the statistical biases to distance measures for both directional and source averaging. In particular, we have found no unexpected violation of $\ev[1]{\tilde{\mu}^{-1}(z)}\e{d}\approx 1$ within numerical uncertainties. As seen in \cref{subsec:area_light-cone_slices}, this relation may be understood in terms of the area of iso-$z$ surfaces, namely $A(z)\approx \bar{A}(z)$ -- the area is unaffected by inhomogeneities.

We now propose to further focus on the two subtle corrections that are making the above `$\approx$' differ from equality, namely the shift and tilt effects (see \cref{fig:shift_tilt_area}). We start, in the sub-section, with the analysis of the shift, that is, the discrepancy between the mean radius of a given light-cone slice (such as iso-$z$ or iso-$\eta$) and its value in the FLRW background.

\subsubsection{Wiggly ray effect: Mean distance reached at fixed distance travelled}
\label{sec:wiggly_ray_res}

\begin{figure}[h!]
\includegraphics[width=\columnwidth]{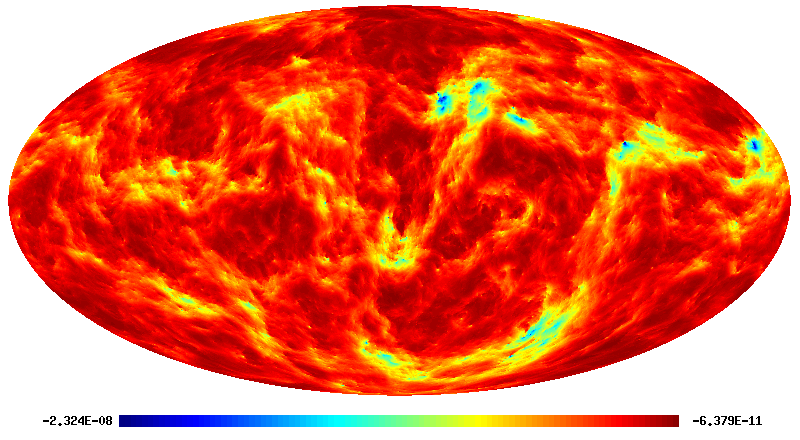} 
    \caption{Map of wiggly-ray effect, that is, the relative fluctuations of the comoving distance reached at fixed distance travelled, $\delta r\e{geo}(s)/\bar{r}(s)=r(s)/\bar{r}(s)-1$, for $s=\bar{s}(z=0.2)$.}
    \label{fig:r_s}
\end{figure}

We first consider the mean comoving distance that is reached by a photon after it travelled over a given comoving distance. As briefly described in \cref{subsubsec:LSS}, because light rays do not travel in straight lines, the radius reached for a distance travelled~$s$ is shorter than $\bar{r}=s$. We may thus write $r(s)=\bar{r}(s)-\delta r\e{geo}(s)$, where $\delta r\e{geo}$ encodes this wiggly-ray effect. This is illustrated with a \textsc{Healpix} map of $\delta r\e{geo}(s)/\bar{r}(s)$ at $z=0.2$ in \cref{fig:r_s}. We see that the fluctuations are very small, on the order of $10^{-8}$, and vary on relatively small angular scales.

On average, the wiggly-ray correction is found to read (see \cref{appendix:ensemble_average_rgeo})
\begin{equation}
\label{eq:approx_eq_comoving_at_constant_travelled}
\frac{\ev[1]{\delta r_{\rm geo}(s)}}{\bar{r}(s)}
= -\int_0^{\bar{r}}\dd r \; \frac{(\bar{r}-r)r}{\bar{r}^2} \, J(r) \ .
\end{equation}
It may be noted, however, that \cref{eq:approx_eq_comoving_at_constant_travelled} was obtained with the help of a few approximations, among which is Limber's. A slightly more accurate computation, in the spirit of Eq.~(A26) in KP16, would yield
\begin{equation}
\label{eq:comoving_at_constant_travelled}
\frac{\ev[1]{\delta r_{\rm geo}(s)}}{\bar{r}(s)} 
= -8\int_0^{\bar{r}}\dd r \; \frac{\bar{r}-r}{\bar{r}^2} \, g(r)
    \int_0^r \dd R \; \frac{R}{r-R} \, g(R) \, \xi_\phi'(r-R) \ ,
\end{equation}
with $g(r) \equiv D_+(\eta_0-r)/a(\eta_0-r)$ where $D_+$ is the linear growth factor, $\xi_\phi$ is the two-point correlation function of the gravitational potential at $z=0$, and $\xi_\phi'$ is its derivative. 

\begin{figure}
\includegraphics[width=\columnwidth]{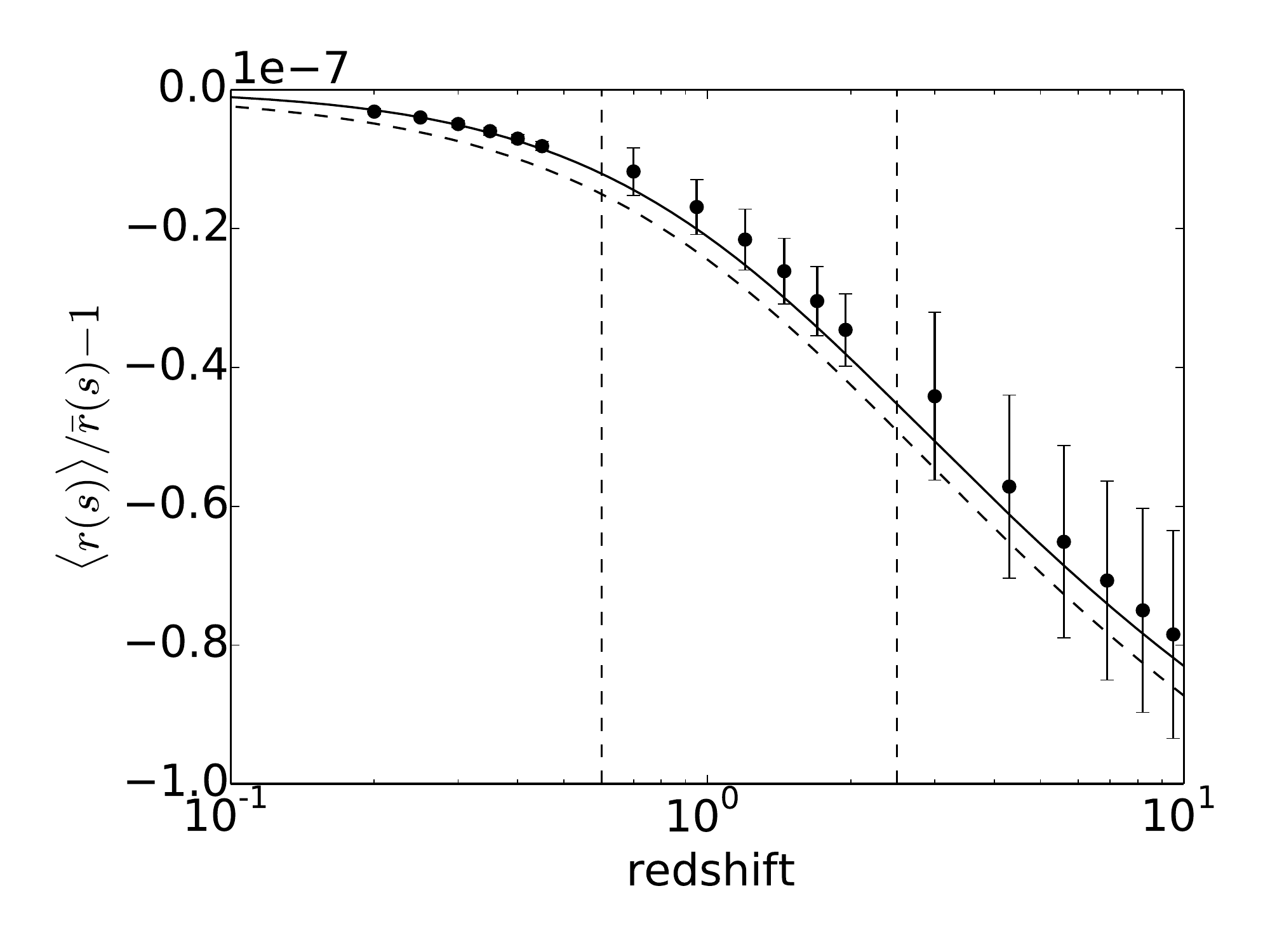} 
    \caption{Wiggly ray effect: Mean fractional reduction~$\ev{r(s)}/\bar{r}(s)-1$ of the comoving distance~$r(s)$ reached after travelling a comoving distance $s$. The $x$-axis indicates the redshift $\bar{z}(s)$ instead of $s$ for better readability. Dots indicate directional averages in the simulation~$\ev{r(s)}\e{d}$; the dashed and solid lines respectively indicate the theoretical predictions of \cref{eq:approx_eq_comoving_at_constant_travelled,eq:comoving_at_constant_travelled}. Error bars are computed following \cref{sec:appendix_variance_rgeo}.}
    \label{fig:mean_comoving_at_cte_travelled}
\end{figure}

These theoretical predictions are successfully confronted with numerical results in \cref{fig:mean_comoving_at_cte_travelled}. Error bars allow for both Poisson and correlated variance, whose expression is given in \cref{sec:appendix_variance_rgeo}. We note that the predictions of \cref{eq:approx_eq_comoving_at_constant_travelled,eq:comoving_at_constant_travelled} are very similar, but only the latter falls within error bars for the full-sky ($z < 0.5$) and intermediate cones ($0.5<z<2$).

\subsubsection{Shapiro effect: Mean distance reached at fixed time}
\label{subsubsec:delta_s_results}

\begin{figure}[h!]
\includegraphics[width=\columnwidth]{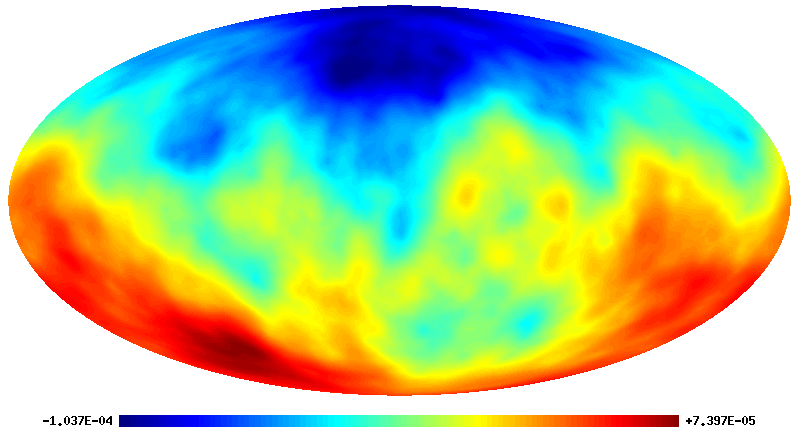} 
    \caption{Map of the relative fluctuations of the comoving distance reached at fixed time, $r(\eta)/\bar{r}(\eta)-1$ for $\eta=\bar{\eta}(z = 0.2)$. These fluctuations are dominated by the first-order Shapiro time-delay effect.}
    \label{fig:map_r_eta}
\end{figure}

We now investigate the mean comoving distance reached at constant time~$\ev{r(\eta)}$; in other words, we consider the shift of iso-$\eta$ surfaces, which would be relevant for CMB-like observations. A \textsc{Healpix} map of $\delta r(\eta)/\bar{r}(\eta)$ is given in \cref{fig:map_r_eta} for illustration. Compared to \cref{fig:r_s}, we note that fluctuations are much larger (of order $10^{-4}$) and take place on much larger angular scales.

As discussed in \cref{subsubsec:LSS}, in theory this shift may be decomposed in two components: (i) the wiggly-ray effect~$\delta r\e{geo}$ considered above; and (ii) the Shapiro time-delay effect. The latter indeed changes the comoving distance travelled~$s(\eta)$ during a given time, compared to its background counterpart~$\bar{s}(\eta)=\eta_0-\eta$, depending on the path-integrated gravitational potential experienced by light. Summarising,
\begin{equation}
r(\eta)
= \bar{r}(\eta) + \delta r(\eta) \ ,
\qquad
\delta r(\eta) = \delta r\e{geo}(\eta) + \delta s(\eta) \ ,
\end{equation}
and as shown in \cref{appendix:LSS} we expect
\begin{equation}
\label{eq:delta_r_eta_unconstrained_prediction}
\ev{\delta r(\eta)}
= - \ev[1]{\delta r\e{geo}(\eta)} > 0 \ ,
\end{equation}
because of the post-Born corrections to $\delta s$ which turn out to be minus twice the geometrical contribution.

\begin{figure}
\includegraphics[width=\columnwidth]{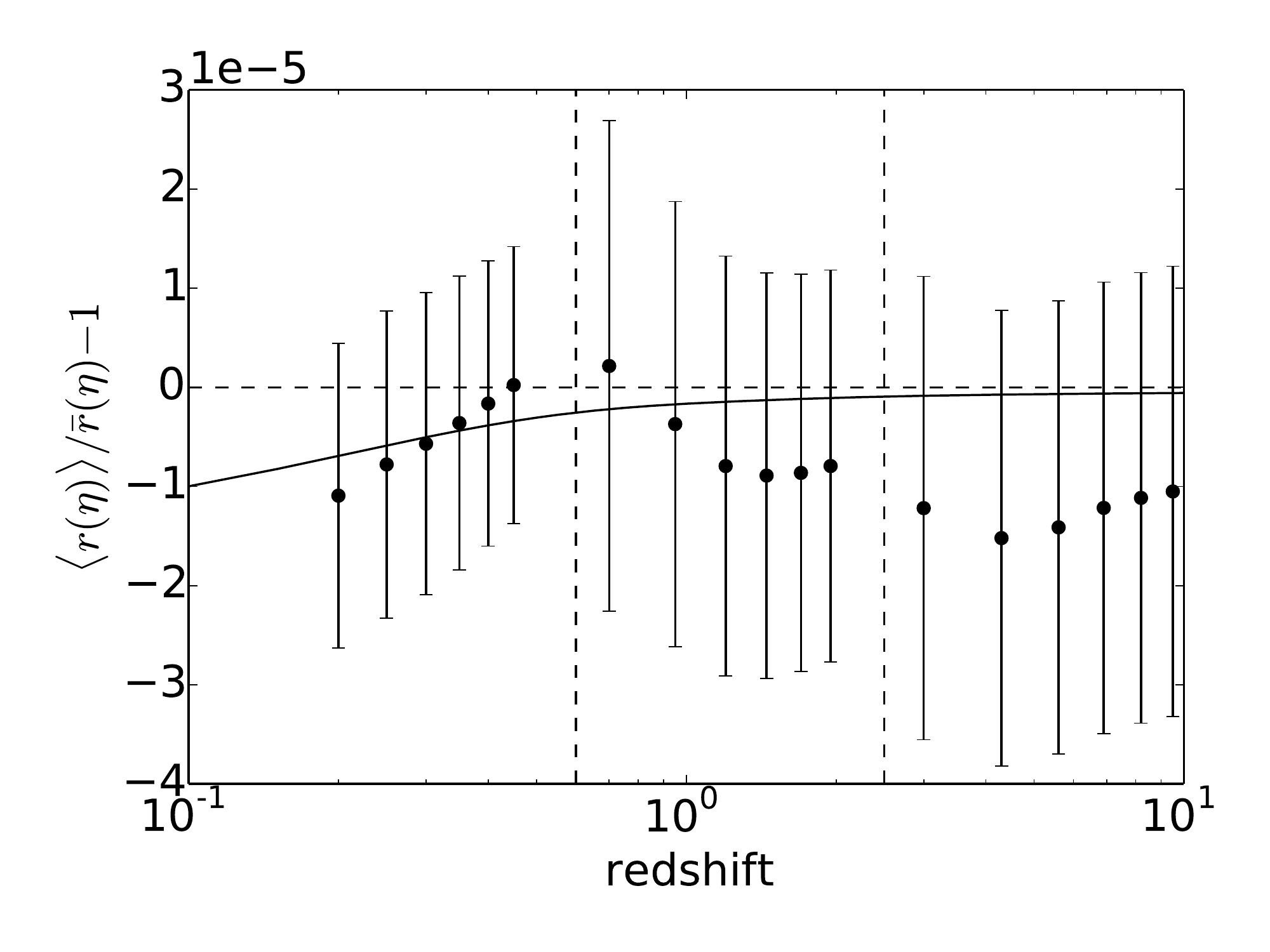} 
    \caption{Mean comoving distance reached at fixed time, $\ev{r(\eta)}$, compared to its background counterpart $\bar{r}(\eta)=\eta_0-\eta$. The difference combines the wiggly-ray and Shapiro time-delay effects. The $x$-axis indicates $\bar{z}(\eta)$ instead of $\eta$ for better readability. Dots indicate numerical results and error bars are computed according to \cref{sec:appendix_variance_constrained}. The dashed line indicates the unconstrained theoretical prediction~\eqref{eq:delta_r_eta_unconstrained_prediction}, while the solid line shows the prediction~\eqref{eq:mean_comoving_cte_t_constrained} that accounts for our knowledge of the gravitational potential~$\phi_0$ at the observer.}
    \label{fig:mean_comoving_at_cte_time}
\end{figure}

Numerical results are confronted with theory in \cref{fig:mean_comoving_at_cte_time}. The first striking feature is perhaps that error bars are about two orders of magnitude larger than the ones of \cref{fig:mean_comoving_at_cte_travelled}. This is because the time-delay contribution $\delta s(\eta)$ is a first-order quantity, while $\delta r\e{geo}(\eta)$ is second-order. Thus, the corresponding super-sample variance is much larger. As such, the prediction~\eqref{eq:delta_r_eta_unconstrained_prediction} naturally falls in the error bars, which is not particularly informative.

We may add an extra layer of refinement since we know the exact value of the gravitational potential in the simulation at the observer, $\phi_0 \approx -7\times10^{-6}$ (which is about $-2$~km/s). This information may be used to improve our prediction of $\delta s(\eta)$ at low redshift. Precisely, following the method outlined in \cref{subsec:constrained_grf_method}, we find that the average Shapiro contribution under the constraint $\phi(0)=\phi_0$ totally overwhelms the geometrical and post-Born terms,\footnote{This finding does not contradict the conclusions of \citet{Hall:2019hkt} that the impact of our local gravitational potential is negligible in current weak-lensing and galaxy-clustering surveys, because they concern very different cosmological quantities.} yielding
\begin{equation}
\label{eq:mean_comoving_cte_t_constrained}
\ev{\delta r(\eta)|\phi_0}
= 2\phi_0 \int_0^{\bar{r}(\eta)} \dd r \; \frac{\xi_\phi(r)}{\xi_\phi(0)}
\gg \ev[1]{\delta r\e{geo}(\eta)}\ .
\end{equation}
The corresponding prediction is indicated by a solid line in \cref{fig:mean_comoving_at_cte_time}, and is observed to reproduce the behaviour of numerical results at low redshift.

\begin{figure}
\includegraphics[width=\columnwidth]{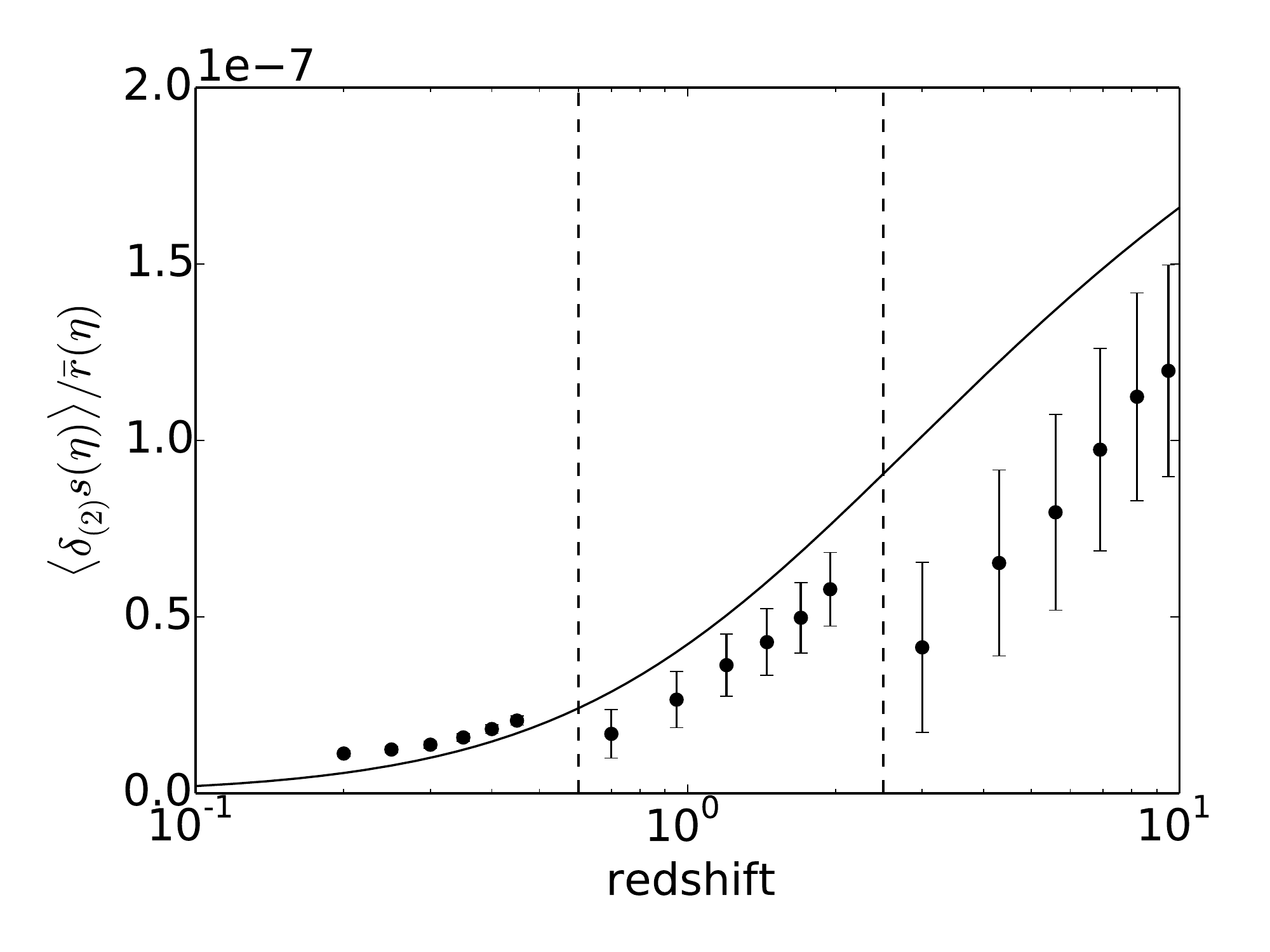} 
    \caption{Second-order (post-Born) contribution to $\delta s(\eta)$. Dots indicate the numerical estimate from \cref{eq:estimator_delta_2_s}, while the solid line is the theoretical prediction $-2\ev[1]{\delta r\e{geo}(\eta)}$. Error bars are twice those of \cref{fig:mean_comoving_at_cte_travelled}.}
    \label{fig:mean_correction_comoving_at_cte_time}
\end{figure}

The fact that \cref{fig:mean_comoving_at_cte_time} is dominated by super-sample variance makes it a priori impossible to check the unconstrained prediction~\eqref{eq:delta_r_eta_unconstrained_prediction}, which yet would be the most relevant at high redshift, where the effect of $\phi_0$ becomes negligible. However, we may apply the following trick to numerically extract the second-order contribution of $\ev{\delta s(\eta)}$. We consider the following estimator
\begin{equation}
\label{eq:estimator_delta_2_s}
\widehat{\delta_{(2)}s}(\eta)
= \frac{\ev[1]{\mu^{-1}(\eta)\delta r(\eta)}\e{d}}{\ev[1]{\mu^{-1}(\eta)}\e{d}}
    - \ev{\delta r(\eta)}\e{d} \ .
\end{equation}
By construction, this combination eliminates the wiggly-ray and first-order Shapiro contributions to $\delta r(\eta)$, while preserving the post-Born term of $\delta s$; see \cref{subsubsec:extracting_post_Born} for details. Thus, we expect $\widehat{\delta_{(2)}s}=-2\ev[1]{\delta r\e{geo}}$. This is indeed roughly what is observed in \cref{fig:mean_correction_comoving_at_cte_time}, although the error bars seem to be under-estimated. This confirms our understanding of the subtle behaviour of $r(\eta)$.

\subsubsection{Doppler effect of peculiar velocities: spatio-temporal shift at fixed observed redshift}

We now turn to observations performed at fixed redshift~$z$. In the inhomogeneous Universe, several phenomena may affect the observed redshift of a source at a given position: Doppler effect due to peculiar velocities, Sachs-Wolfe (SW) and ISW effects. These imply that for a given redshift, light may have been emitted slightly closer to the observer (and later), or slightly further (and earlier) compared to the background FLRW case. In other words, we have $r(z)=\bar{r}(z)+\delta r(z)$ and $\eta(z)=\bar{\eta}(z)+\delta\eta(z)$, with $\delta r=-\delta \eta$ the associated spatio-temporal shift of $\Sigma(z)$.

The Doppler effect of peculiar velocities is expected to dominate, especially at low-$z$. Assuming that the observer is comoving ($\vect{v}\e{o}=\vect{0}$), which is the case in the simulation, the shift due to the source's peculiar velocity~$\vect{v}$ reads
\begin{equation}
\frac{\delta r(\vect{\theta}, z)}{\bar{r}(z)}
= - \frac{\vect{\theta}\cdot\vect{v}(\vect{\theta},z)}{\mathcal{H}(z)\bar{r}(z)} \ ,
\label{eq:delta_z_at_cte_zobs}
\end{equation}
at first order. We note that, combined with the (opposite) temporal shift, we find that the corresponding area perturbation reads $\delta[a^2(\eta) r^2]/[a^2(\bar{\eta})\bar{r}^2]=2\tilde{\kappa}_v$, in agreement with \cref{eq:Doppler_magnification}.

\begin{figure}
\includegraphics[width=\columnwidth]{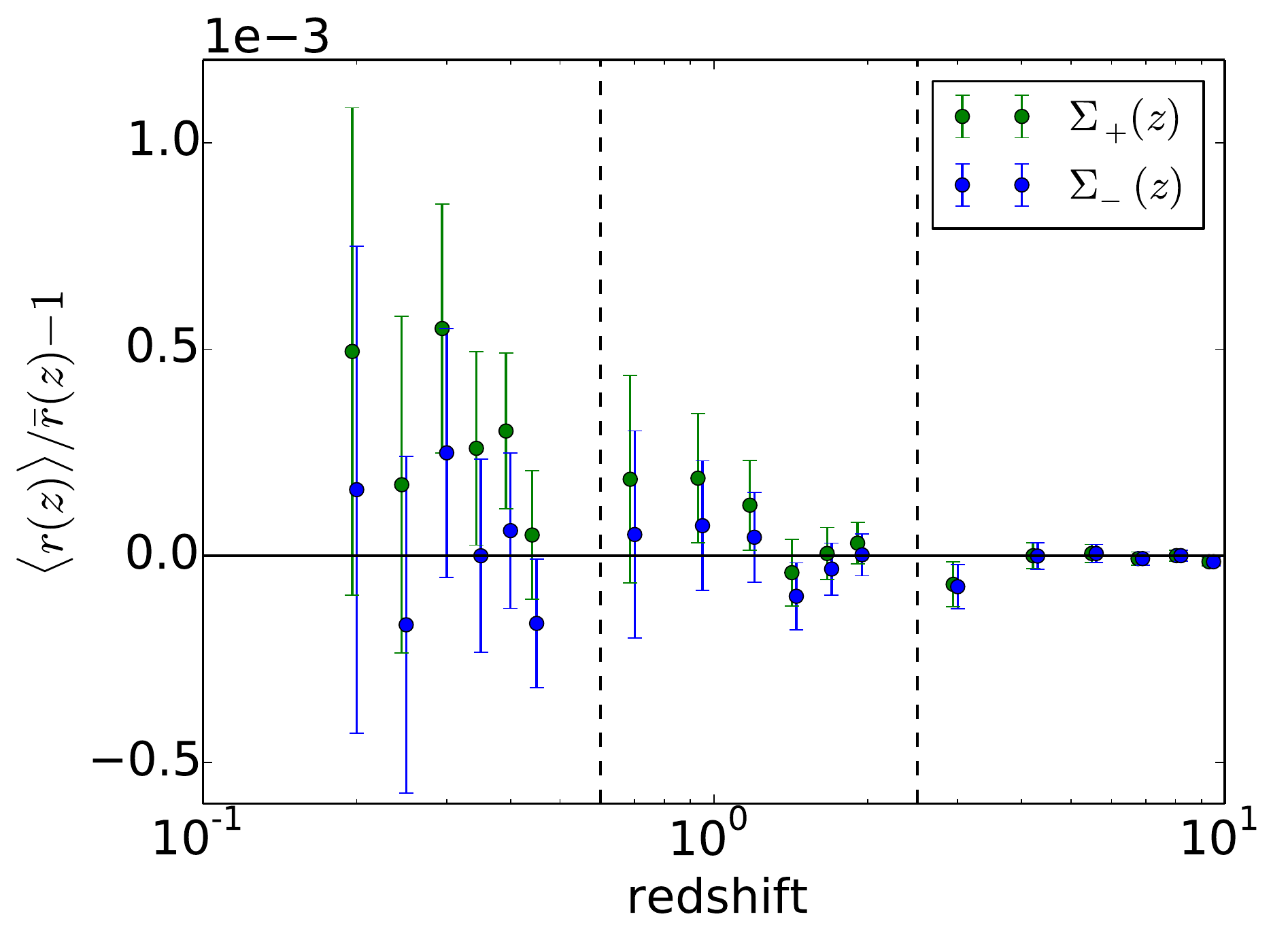} 
    \caption{Radial shift of surfaces of constant redshift. Dots indicate numerical results for $\ev{\delta r(z)}\e{d}/\bar{r}(z)$ while error bars account for Poisson variance and for the super-sample variance due to peculiar velocities only. For each $z$, we provide two values of the mean shift: one across the closest iso-$z$ surface [$\Sigma_-(z)$, blue] and the other across the fathest iso-$z$ surface [$\Sigma_+(z)$, green].}
    \label{fig:mean_z_at_observed_z}
\end{figure}

Numerical results for $\ev{\delta r(z)}\e{d}$ are shown in \cref{fig:mean_z_at_observed_z}. Two values of $\delta r(z)$ are provided for each $z$. This is because the fluctuations of peculiar velocities may be important on the light cone, so that several events may have the same redshift: $\Sigma(z)$ is not unique. We plotted here the smallest one, $\ev{\min\delta r(z)}\e{d}$, and the largest one, $\ev{\max\delta r(z)}\e{d}$. Numerical values for $\ev{\delta r(z)}\e{d}$ account for all the redshift components at first order in metric perturbations; however, the super-sample variance contribution to the error bars only account for peculiar velocities, which are highly dominant. See \cref{sec:appendix_variance_redshift} for their expression.

The mean shift is compatible with zero within the large error bars, which is why we did not work on a more elaborate theoretical prediction for $\ev{\delta r(z)}$. We also note that the two extremal surfaces~$\Sigma_-(z), \Sigma_+(z)$ that we have considered differ more as we get closer to the observer. We shall now explain this point: at high redshift there are few virialised objects, and hence velocity dispersion is small next to such objects; thus, at high $z$, $\Sigma(z)=\Sigma_-(z)=\Sigma_+(z)$ is typically unique. As $z$ decreases, more haloes form, which implies that more regions have high velocity dispersion,\footnote{This is known as the Finger-of-God effect in galaxy clustering and redshift-space distortions analysis.} thereby de-multiplying $\Sigma(z)$ and spreading its occurrences. The $1/(\mathcal{H}\bar{r})$ factor in \cref{eq:delta_z_at_cte_zobs} further enhances that spread.

\subsubsection{Affine parameter at constant time}
\label{subsec:considerations_iso-lambda}

While surfaces of constant redshift~$\Sigma(z)$ or time~$\Sigma(\eta)$ are observationally relevant, we may also consider surfaces of constant affine parameter~$\Sigma(\lambda)$, whose interest is strictly theoretical. Such light-cone slices have the advantage of being defined regardless of any specific model for the space-time metric. The analysis of \citet{kibble2005average} was indeed conducted on $\Sigma(\lambda)$.
%eventually assuming that $\tilde{\mu}(\lambda)=\tilde{\mu}(z)$.
We shall examine the time (or equivalently radial) shift at fixed affine parameter, $\delta\eta(\lambda)$, in the next sub-section. Before that, we propose to first consider the converse shift~$\delta\lambda(\eta)$ for pedagogical reasons.

In the background FLRW model, the $0$th component of the geodesic equation~\eqref{eq:geodesic_equation1} is integrated as
\begin{equation}
\bar{\lambda}(\eta) = \int_{\eta}^{\eta_0} \dd\eta' \; a^2(\eta') \ .
\end{equation}
In the presence of perturbations, this becomes
\begin{equation}
\lambda(\eta)
= \int_{\eta}^{\eta_0}
    \dd\eta' \; a^2(\eta')
    \left[
            1+2(\phi-\phi_0) 
            - 2\int_0^{\eta'} \dd\eta'' \; \frac{\partial\phi}{\partial\eta''}
    \right]. 
\label{eq:lambda_cte_t_full}
\end{equation}
at first order (higher-order terms are negligible here). We note that since $\ev{\phi}=0$, the mean correction to the affine parameter is simply $\ev{\delta\lambda(\eta)}/\bar{\lambda}(\eta) = -2\phi_0$. However, we may account for the fact that, in our simulation just as in observations, the gravitational potential at the observer is fixed, and estimate $\ev{\delta\lambda(\eta)}\e{d}$ with the following constrained ensemble average
\begin{equation}
\label{eq:mean_lambda_cte_t_contrained}
\ev{\delta\lambda(\eta)|\phi_0}
= -2\phi_0\int_0^{\bar{r}(\eta)} \dd r \;
    a^2(\eta_0-r)\left[1-\frac{\xi_\phi(r)}{\xi_\phi(0)}\right]
\not= 0\ ,
\end{equation}
where we chose to neglect the small ISW term of \cref{eq:lambda_cte_t_full}. In the high-$z$ limit, this result may be approximated as
\begin{equation}
\label{eq:delta_lambda_eta}
\frac{\ev{\delta\lambda(\eta)|\phi_0}}{\bar{\lambda}(\eta) }
\approx -2(1-\Xi_\infty)\phi_0 = \cst \ ,
\end{equation}
with
\begin{equation}
\Xi_\infty \equiv
\lim_{R\rightarrow\infty}
\frac{1}{\bar{\lambda}(R)} \int_0^R \dd r \; a^2(\eta_0-r) \, \frac{\xi_\phi(r)}{\xi_\phi(0)} \approx 0.15 \ .
\end{equation}

\begin{figure}
\includegraphics[width=\columnwidth]{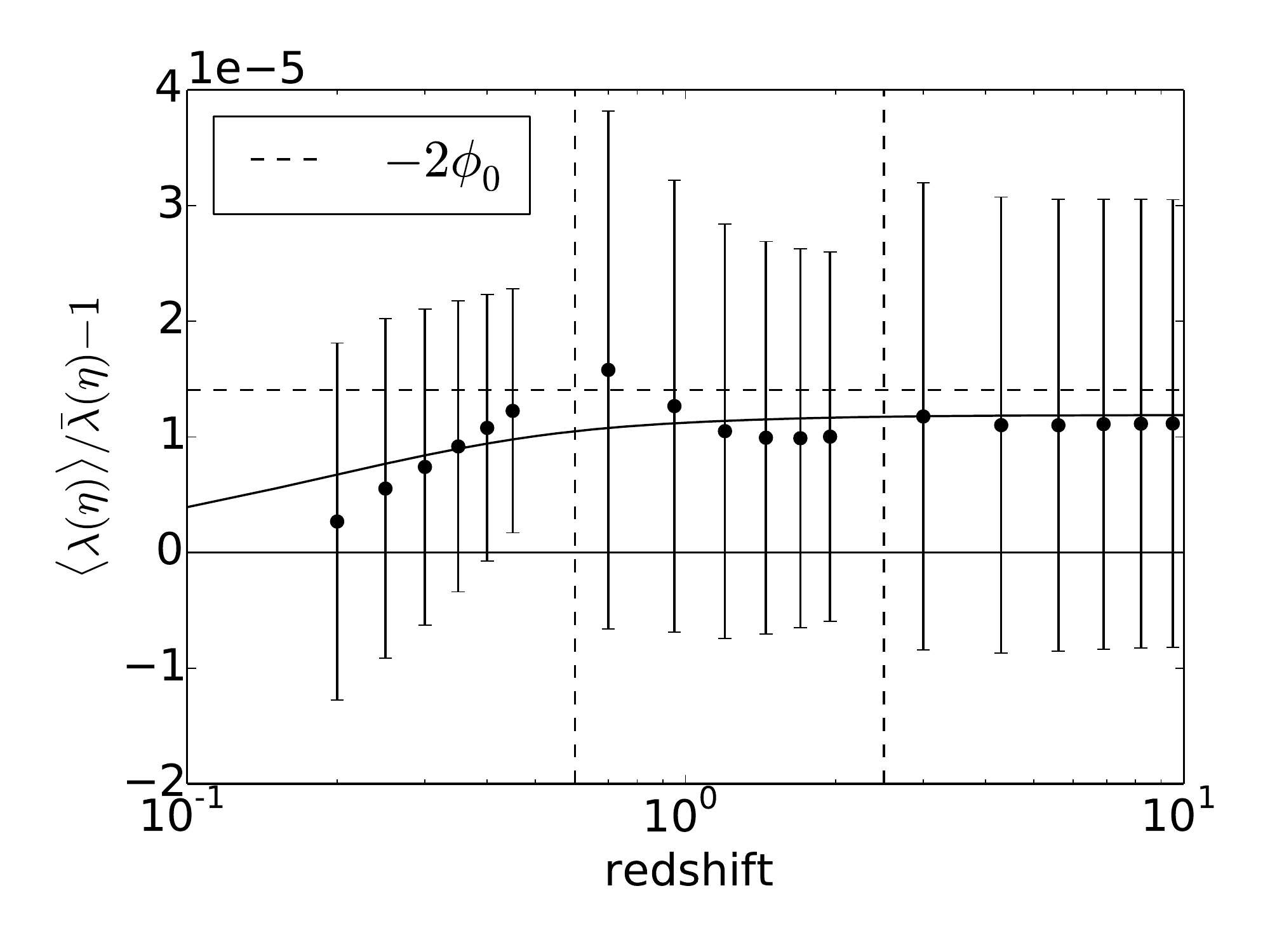} 
    \caption{Perturbation of the affine parameter at fixed time~$\eta$. Dots are numerical results obtained by directional averaging; the solid line indicates the theoretical prediction~\eqref{eq:mean_lambda_cte_t_contrained}; the dashed line shows $-2\phi_0$. Hence, the fractional difference between the dashed line and the asymptote of the solid line indicates $\Xi_\infty$. Error bars account for Poisson variance and constrained super-sample variance; see \cref{sec:appendix_variance_constrained} for details.}
    \label{fig:mean_affine_at_cte_time}
\end{figure}

As shown in \cref{fig:mean_affine_at_cte_time}, the numerical results do follow the theoretical prediction of \cref{eq:mean_lambda_cte_t_contrained} within the error bars, although the unconstrained prediction $\ev{\delta\lambda(\eta)}/\bar{\lambda}=-2\phi_0$ at first order would also be validated by the data. In particular, $\ev{\delta\lambda}/\bar{\lambda}$ is indeed observed to converge to a constant at high $z$. For the deep cone ($z>2$) the error bars are dominated by super-sample variance, but it seems that by chance the mean gravitational potential in these $400~\mathrm{deg}^2$ is very close to zero.

We stress that error bars of \cref{fig:mean_affine_at_cte_time} were computed in configuration space to account for the constraint $\phi(0)=\phi_0$ on the variance (see \cref{sec:appendix_variance_constrained}). For comparison, we also computed the variance in harmonic space using \cref{eq:correlated_variance}. Results were similar for the intermediate and deep cones, while for the full-sky cone the unconstrained error bars were twice larger than the constrained one. Had we disposed of full-sky data up to $z=10$, the error bars at high-$z$ would have been smaller than the bias of $\ev{\lambda(\eta)}$. In other words, $\ev{\delta\lambda(\eta)}=0$ would have been excluded.

\subsubsection{Spatio-temporal shift at constant affine parameter}
\label{subsubsec:shift_cst_lambda}

We now consider the time and radial shifts~$\delta\eta(\lambda)=-\delta r(\lambda)$ at fixed affine parameter, which is the converse of the above. For that operation, one would typically adopt a linearised approach and write $\delta\eta=(\dd\bar{\eta}/\dd\lambda)\delta\lambda$. However, since $\dd\bar{\eta}/\dd\lambda=1/a^2=(1+\bar{z})^2$, this derivative can become quite large as the redshift increases, so that $\delta\lambda$ may actually get out of the linear behaviour of $\bar{\eta}(\lambda)$. We may thus adopt a more accurate inversion that could be extrapolated up to the LSS. Specifically, we use
\begin{align}
\eta(\lambda)
&= \bar{\eta}\left\{\bar\lambda[\eta(\lambda)]\right\} \ ,
\\
&= \bar{\eta}\left\{\lambda[\eta(\lambda)]-\delta\lambda[\eta(\lambda)]\right\} \ ,
\\
&\approx \bar{\eta}\left\{\lambda-\delta\lambda[\bar\eta(\lambda)]\right\} \ ,
\end{align}
which we shall refrain from further expanding.

We have checked numerically that the fluctuations of $\lambda(\eta)$ about its mean are much smaller than its bias $\ev{\delta\lambda(\eta)}$. This allows us to express the average emission time at $\lambda$ as
\begin{equation}
\label{eq:mean_eta_lambda}
\ev{\eta(\lambda)}
\approx \bar{\eta}\left\{\lambda-\ev{\delta\lambda[\bar\eta(\lambda)]}\right\}
\equiv \bar\eta(\lambda)+\ev{\delta\eta(\lambda)}\ ,
\end{equation}
where $\ev{\delta\lambda}$ is given by \cref{eq:mean_lambda_cte_t_contrained}.

\begin{figure}
\includegraphics[width=\columnwidth]{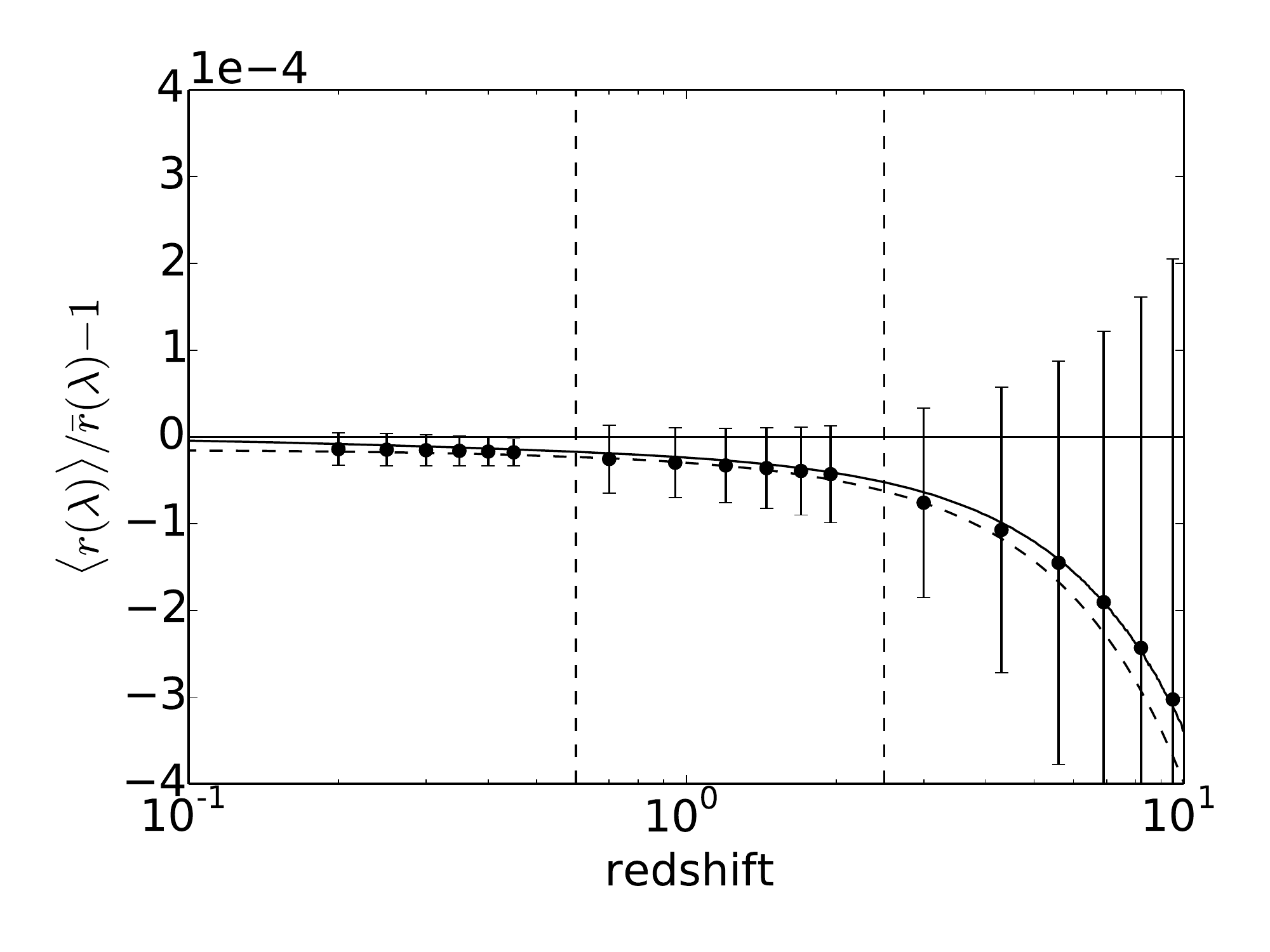} 
    \caption{Radial (or equivalently temporal) shift~$\ev{\delta r(\lambda)}=-\ev{\delta \eta(\lambda)}$ at fixed affine parameter~$\lambda$. Dots indicate numercal results from directional averaging; the solid line shows the theoretical prediction~\eqref{eq:mean_eta_lambda}, while the dashed line indicates $\bar{r}[(1-2\phi_0)\lambda]-\Bar{r}(\lambda)$ for comparison. Error bars account for Poisson and super-sample uncertainties; see \cref{sec:appendix_variance_constrained}.}
    \label{fig:mean_time_at_cte_affine}
\end{figure}

As shown in \cref{fig:mean_time_at_cte_affine}, the prediction~\eqref{eq:mean_eta_lambda} interpreted as a radial shift $\ev{\delta r(\lambda)}=-\ev{\delta\eta(\lambda)}$ is in excellent agreement with the simulation data $\ev{\delta r(\lambda)}\e{d}$, although again $0$ remains within the error bars dominated by super-sample variance at high-$z$. As expected, the amplitude of the shift blows up as $\lambda$ increases.

In fact, a linear expansion of \cref{eq:mean_lambda_cte_t_contrained} would also match our numerical results, whose redshift range is not large enough for the non-linear prescription to be critical. However, it does matter for the surface $\lambda=\lambda_*$, that is the surface of constant affine parameter dictated by the background affine parameter of the LSS. In that case the linearised version of \cref{eq:mean_lambda_cte_t_contrained} would highly over-estimate the mean shift, which is already very large, $\ev{\delta r(\lambda_*)}/\bar{r}(\lambda_*)\sim 10\%$ with $\phi_0=-2~\mathrm{km/s}$ of our simulation.

We now need to investigate the consequences of such a large shift on the area~$A(\lambda_*)$ of the surface of affine parameter~$\lambda_*$. Just as for iso-$z$ surfaces, the time and radial shifts of iso-$\lambda$ surfaces both contribute to the perturbation of their area, or equivalently to $\ev[1]{\tilde{\mu}^{-1}(\lambda)}$. Accounting for that shift only (the tilt being second-order, it would be sub-dominant), we have
\begin{align}
\frac{\delta A(\lambda_*)}{\bar{A}(\lambda_*)}
%&= \ev{\tilde{\mu}^{-1}(\lambda)}\e{d} - 1 \\
= \ev{
        \frac{\delta\left\{a^2[\eta(\lambda_*)] r^2(\lambda_*)\right\}}
            {a^2[\bar{\eta}(\lambda_*)]\bar{r}^2(\lambda_*)}
        }
\approx \frac{a^2[\ev{\eta(\lambda_*)}]-a^2[\bar{\eta}(\lambda_*)]}{a^2[\bar{\eta}(\lambda_*)]} \ ,
\label{eq:area_iso_lambda}
\end{align}
which is dominated by the difference in scale factor.\footnote{This is because the surface $\lambda=\lambda_*$ is very far from the observer. For closer surfaces the radial shift would dominate.} Again, we shall refrain from linearising \cref{eq:area_iso_lambda} because the time shift is large enough to invalidate a first-order Taylor expansion of $a(\eta)$.

\begin{figure}
\includegraphics[width=\columnwidth]{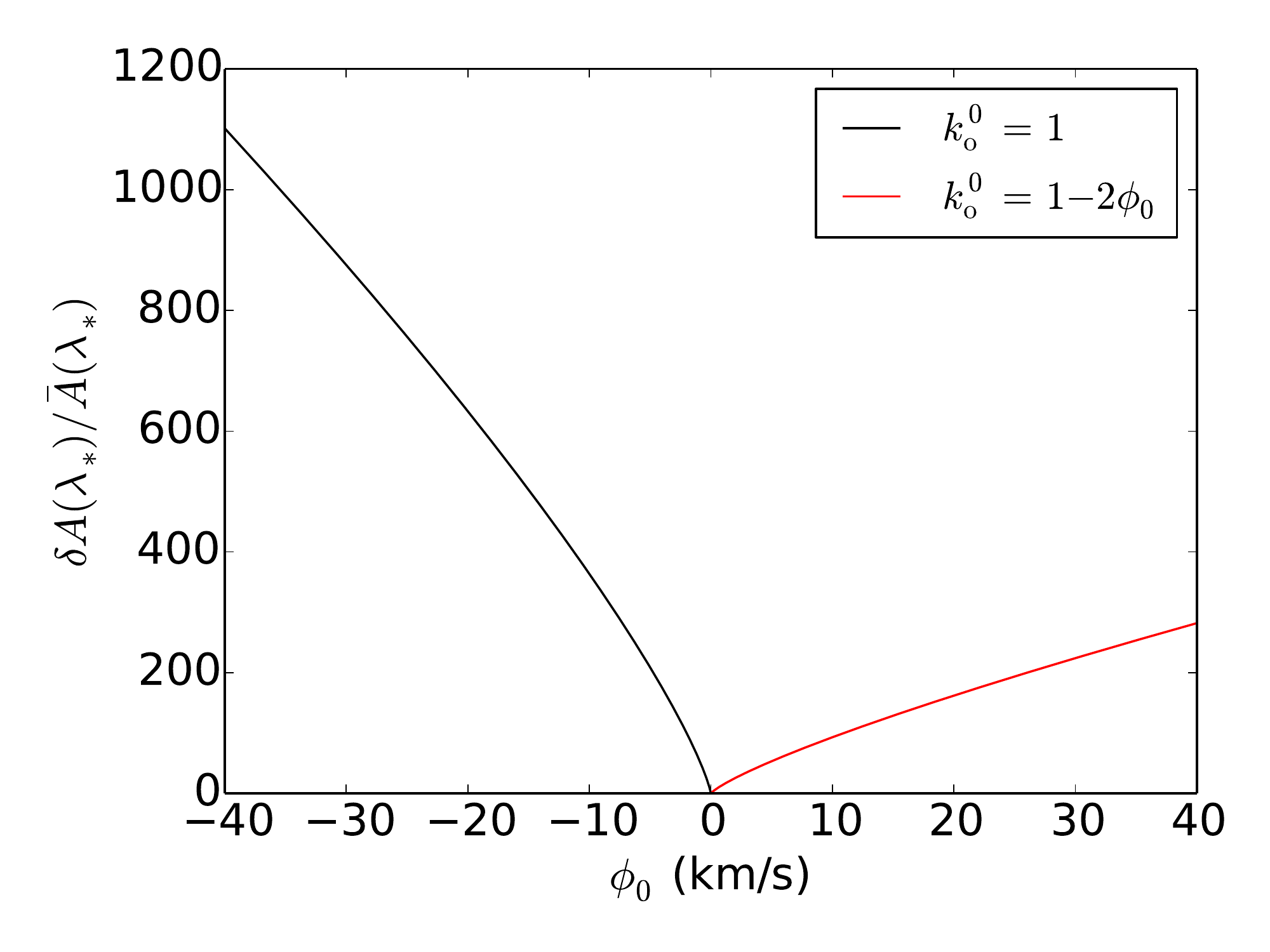} 
    \caption{Fractional change of the area~$A(\lambda_*)$ of the surface of constant affine parameter~$\lambda=\lambda_*$, as a function of the gravitational potential~$\phi_0$ at the observer. The black and red lines correspond to two different normalisations for the affine parameter.}
    \label{fig:area_bias_lambda_star}
\end{figure}

We illustrate the behaviour of \cref{eq:area_iso_lambda} in \cref{fig:area_bias_lambda_star} (black solid line) as a function of the observer's potential. Even for reasonable values of that quantity, such as $\phi_0\sim -10~\mathrm{km/s}$, which is the order of magnitude expected within galaxy clusters \citep{wojtak2011gravitational, wojtak2015local}, we find an astonishing $\delta A(\lambda_*)/\bar{A}(\lambda_*) \sim 3000\%$. We note that the black solid line of \cref{fig:area_bias_lambda_star} is only plotted for negative values of $\phi_0$. This is because for $\phi_0>0$, $\ev{\delta\eta(\lambda_*)}<0$ quickly diverges as $\phi_0$ increases; since $\lim_{\eta\rightarrow-\infty}\bar{\lambda}(\eta)<\infty$, it may not even be defined. 

At that point, we may note that the affine parameter is defined up to an arbitrary normalisation. Here, this normalisation has been chosen so that $k^0\e{o}=1$ at the observer, hence $\lambda$ coincides with cosmic time at that point. Another sensible choice consists in setting the observed frequency to unity, $\omega\e{o}=(1+2\phi_0) k^0\e{o}=1$. This corresponds to a re-normalisation $\lambda\mapsto\lambda'=(1+2\phi_0)\lambda$, where now $\lambda'$ represents proper time at the observer. This alternative normalisation eliminates the $\phi_0$ term in \cref{eq:lambda_cte_t_full} and yields $\ev[1]{\delta\lambda'(\eta)}/\bar{\lambda}(\eta)=2\Xi_\infty\phi_0$ instead of \cref{eq:delta_lambda_eta}. However, the bias on the area~$A(\lambda'_*)$ remains huge, as seen with the red line of \cref{fig:area_bias_lambda_star},\footnote{This red line is only plotted for $\phi_0\geq 0$ for the same reason as the black line is only plotted for $\phi_0\leq 0$. Namely, $\ev{\delta\eta(\lambda_*')}<0$ quickly diverges and may not even exist when $\phi_0 < 0$.} which corresponds to that alternative normalisation for the affine parameter. The only way to avoid such a huge bias would be to define $\lambda''$ with the quite un-natural normalisation $k^0\e{o}=1-2(1-\Xi_\infty)\phi_0$.

The perhaps surprising results reported in this section must be attributed to the fact that $\dd\bar{\lambda}/\dd\eta=a^2(\eta)\rightarrow 0$ when $\eta\rightarrow -\infty$, so that $\bar{\lambda}(\eta)$ flattens out and saturates in the early Universe. This makes $\eta$ extremely sensitive to even tiny changes in $\lambda$. All in all, this indicates that the affine parameter is a rather dangerous quantity to be used in theoretical analyses of light propagation down to the early Universe. The huge bias on $A(\lambda)$ shown in \cref{fig:area_bias_lambda_star} must be considered a theoretical hiccup with no observational consequences.

\subsection{Focus on the tilt correction}
\label{subsec:results_tilt}

We may close this section on numerical results by evaluating the tilt, or wrinkly-surface effect, which tends to increase the inverse amplification~$1/\tilde{\mu}$ and the area of light-cone slices. As outlined in \cref{subsubsec:shift/tilt} and illustrated in \cref{fig:shift_tilt_area}, the tilt is defined as the angle~$\iota$ between the radial direction~$\vect{\beta}$ of a point and the local direction of light propagation at that point.

The contribution of the tilt to the area~$A(p)$ of an iso-$p$ surface, or equivalently to the directional average of the inverse amplification~$\ev[1]{\tilde{\mu}^{-1}(p)}\e{d}$, where $p$ is any relevant parameter, goes as $\iota^2(p)$; it is always a second-order quantity. As such, it does not matter much which parameter~$p$ is actually considered here. Indeed, for any two parameters $p,q$ (which can stand for $z, \eta, \lambda, \ldots$) we have $\iota(p)/\iota(q)-1\sim |\text{shift}(p)-\text{shift}(q)|\ll 1$. In other words, the tilt can be considered universal in first approximation.

In light of the above discussion, we shall consider the tilt over surfaces of constant time, keeping in mind that the result would equally apply to other ways to slice the light cone. The choice of iso-$\eta$ surfaces is also motivated by the fact that, in this case,\footnote{This property would apply to other surfaces in their natural frame, which is not necessarily the comoving frame.} $\iota$ coincides with the angle formed by the normal~$\vect{n}$ to the surface and the radial direction~$\vect{\beta}$, as seen in \cref{subsubsec:LSS}. Thus,
\begin{equation}
\cos\iota(\eta,\vect{\beta})
= \vect{\beta}\cdot\vect{n}(\eta,\vect{\beta})
= \left[
    1 + 
    \left|
        \frac{1}{r(\eta,\vect{\beta})}
        \frac{\partial r}{\partial\vect{\beta}}
    \right|^2
    \right]^{-1/2} \ .
\end{equation}
At second order, the area increase due to the tilt is predicted to be (see \cref{subsec:appendix_tilt} for details),
\begin{equation}
\frac{\delta A\e{tilt}(\eta)}{\bar{A}(\eta)}
= - \frac{1}{2}\ev{\iota^2(\eta)}
= \frac{1}{\bar{r}^2(\eta)}\int_0^{\bar{r}(\eta)}\dd r \; r^2 \, J(r) \ ,
    \label{eq:area_tilt_perturbation_comoving_cte_time}
\end{equation}
with $J(r)$ given in \cref{eq:definition_J}.

\begin{figure}
\includegraphics[width=\columnwidth]{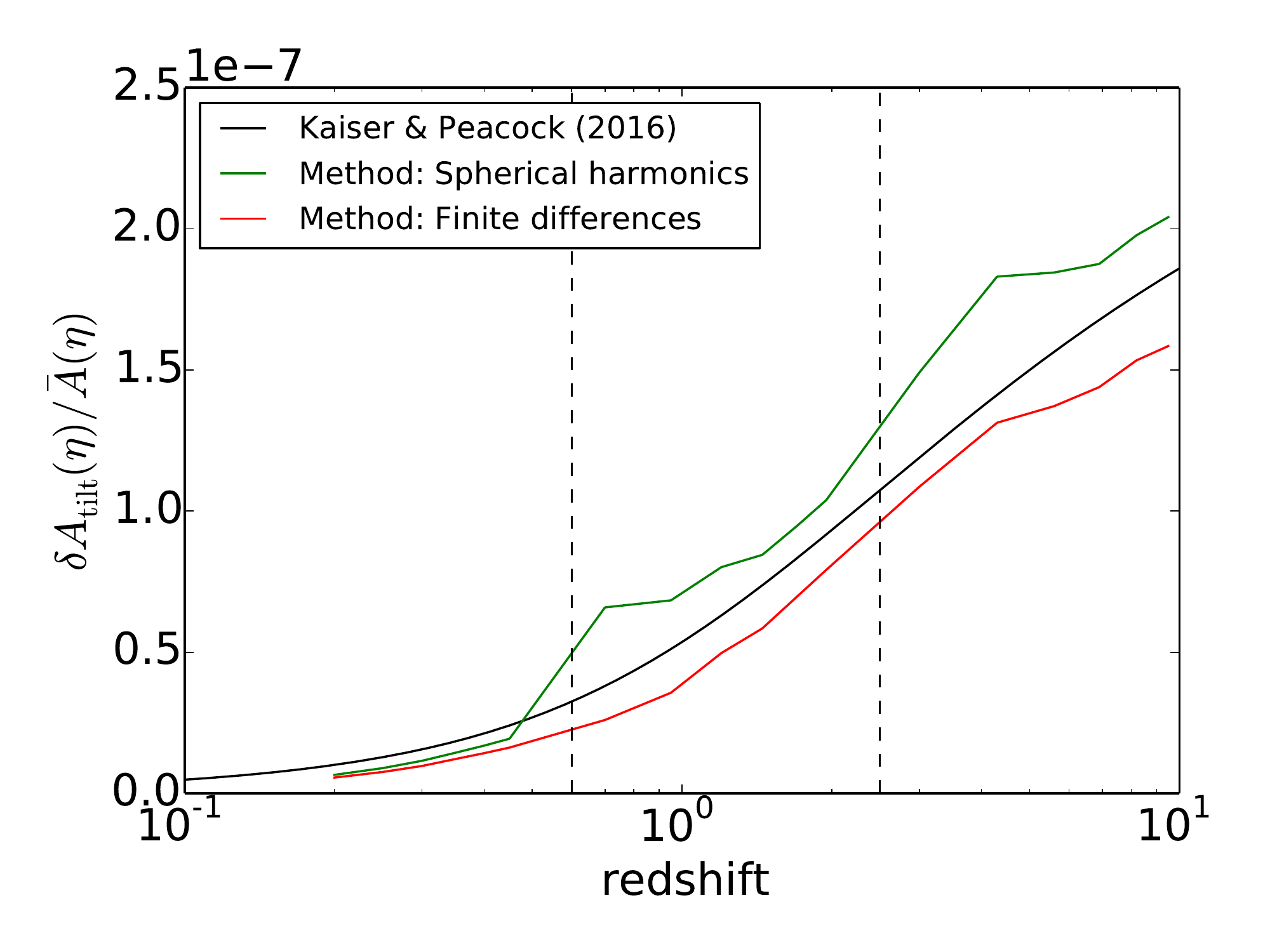} 
    \caption{Wrinkly-surface effect: relative increase of the area of surfaces of constant time due to their non-sphericity. The black solid line indicates the theoretical prediction~\eqref{eq:area_tilt_perturbation_comoving_cte_time} as originally found by KP16. The green and red lines indicate numerical results obtained from the two different methods outlined in \cref{subsec:wrinkly_surface}.}
    \label{fig:area_comoving_at_cte_time}
\end{figure}

This prediction is successfully confronted with numerical results following \cref{subsec:wrinkly_surface} in \cref{fig:area_comoving_at_cte_time}. This ends our series of numerical checks of the remarkably accurate predictions of KP16.

%______________________________________________________________

\section{Conclusion}
\label{sec:conclusion}

The general topic of this article was the effect of inhomogeneities on cosmological observables, notably the average result of cosmic distance measurements.
In the theoretical part (\cref{sec:theory}), we started by emphasising the subtle difference between the notions of geometric magnification~$\mu$ and observable magnification~$\tilde{\mu}$. We interpreted that difference in terms of shifts and tilts associated with light propagation in the inhomogeneous Universe. We then reviewed and compared various notions of averaging involved in cosmology, in particular directional averaging $\ev{\cdots}\e{d}$ and source-averaging~$\ev{\cdots}\e{s}$.

We argued that, because of the exact identity $\ev[1]{\mu^{-1}}\e{d}=1$, one may expect the approximate relations~$\ev[1]{\tilde{\mu}^{-1}}\e{d}\approx \ev{\tilde{\mu}}\e{s}\approx 1$. Departures from the exact $\ev[1]{\tilde{\mu}^{-1}}\e{d}=1$ are due to the shift and tilt effects. We rigorously showed that such statements may be reformulated in terms of the area~$A$ of slices of the light cone; namely $A$ should be almost unaffected by cosmological inhomogeneities, which is the conjecture of \citet{weinberg1976apparent}. Finally, we have reviewed how $\ev[1]{\tilde{\mu}^{-1}}\e{d}\approx \ev{\tilde{\mu}}\e{s}\approx 1$ allows one to predict the statistical bias of any distance measure, such as the magnitude in SN surveys.

Most of the above theoretical considerations had been investigated in the past, notably in KP16, within the framework of cosmological perturbation theory at second order. The main addition of this article is their thorough analysis via ray tracing in a high-resolution $N$-body simulation up to $z = 10$. This tool (\cref{sec:numerical_methods}) allowed us to account for the inhomogeneity of the Universe down to very small scales, where the perturbation theory fails. We produced \textsc{Healpix} maps and halo catalogues to generate direction-averaged and source-averaged mock observations respectively. Our main results are the following:

\begin{enumerate}[(i)]
\item At all redshifts, we confirmed that $\ev[1]{\mu^{-1}(z)}\e{d} \approx \ev[1]{\tilde{\mu}^{-1}(z)}\e{d} \approx \ev{\mu(z)}\e{s}\approx \ev{\tilde{\mu}(z)}\e{s} \approx 1$ within our error bars, which account for both Poisson and super-sample variance. Source-averaged quantities were found to be more biased due to real-space clustering.
\item Still within error bars, the bias on the direction-averaged distance is $\ev{\delta d(z)}\e{d} = -\ev[1]{\kappa^2}/2$, while the source-averaged distance and distance modulus are biased as $\ev{\delta d(z)}\e{s} = 3\ev[1]{\kappa^2}/2$ and $\ev{\Delta m(z)}\e{s} = 5\ev[1]{\kappa^2}/\ln 10$. All these numerical results thus agree very well (within error bars) with theoretical predictions. A large scatter is observed at low redshift due to peculiar velocities.
\item In order to further test the accuracy of Weinberg's conjecture, we investigated in detail the discrepancy between geometric and observable magnifications $\mu, \tilde{\mu}$, that is, the effects of shift and tilt. We found that the fractional area perturbations of $\Sigma(\eta)$ are well recovered by the predictions of KP16, of the order of order $10^{-7}$. However, this bias turned out to be much smaller than the super-sample variance on time delays, which is on the order of $10^{-5}$. Furthermore, we checked that propagating light rays on a coarse grid (rather than the AMR grid) does not impact the discrepancy between $\mu$ and $\tilde{\mu}$, which shows that the latter is relatively insensitive to very small scales.
\end{enumerate}

Summarising, our results show no unexpectedly large bias for source and direction-averaged observables. They also confirm Weinberg's conjecture that the area of surfaces of constant redshift, or constant time, are almost unaffected by inhomogeneities, with corrections remaining below a part in a million for the latter (while the former might be slightly larger at low $z$ due to peculiar velocities, but so is the associated variance).

As a theoretical curiosity, we also considered the area bias of surfaces of constant affine parameter, $\Sigma(\lambda)$. We found that at very high redshift, the bias grows quickly and eventually diverges, because $\eta\mapsto\lambda(\eta)$ reaches a flat asymptote for $\eta\rightarrow-\infty$. In practice, this leads to absurdly large corrections to $A(\lambda)$, and shows that one should avoid the use of the affine parameter to describe light rays in the early Universe.

For the present analysis we used finite ray bundles to compute the lensing distortion matrix~$\vect{\mathcal{A}}$. This allowed us, as a side product, to test the predictions of the finite-beam formalism developed by \citet{fleury2017weak,fleury2019cosmic,fleury2019weak}. Specifically, we checked that the convergence and shear power spectrum for finite ray bundles were suppressed for scales smaller than the bundle's width, in excellent agreement with \citet{fleury2019cosmic}. Last, we found that treating the gravitational potential field~$\phi$ as a constrained field near the observer, improves the agreement between numerical data and theoretical predictions.

Several extensions are possibles for this work: at very large scale it could be interesting to see the impact of a fully general relativistic treatment, either by correcting the results from a Newtonian $N$-body code \citep{chisari2011connection,fidler2015general} or by directly using GR simulations \citep{adamek2016gevolution, barrera-hinojosa2020gramses}. At smaller scales, one could investigate the effect of strong lensing, allowing for multiple images for a single source. Also, we studied the bias on the distance-redshift relation within the $\Lambda$CDM framework. One could perform a similar analysis using different cosmologies, for example by changing the nature of the dark sector or departing from GR, to see how such new physics may be degenerate with observational biases.

\begin{acknowledgements}
This paper is the continuation of a work that started during Vincent Reverdy's PhD thesis \citep{reverdy2014propagation}. MAB thanks Sylvain de la Torre for pointing out the paper of \citet{desjacques2020statistics}. We thank the referee John Peacock for many relevant comments which significantly improved the quality of this manuscript, especially in \cref{sec:theory}. This work was granted access to HPC resources of TGCC through allocations made by GENCI (Grand Equipement National de Calcul Intensif) under the allocations A0050402287 and A0070402287. PF received the support of a fellowship from ``la Caixa'' Foundation (ID 100010434). The fellowship code is LCF/BQ/PI19/11690018.
\end{acknowledgements}

%-------------------------------------------------------------------
\bibliographystyle{aa}
\bibliography{biblio} % if your bibtex file is called example.bib

\appendix

\section{Alternative approach to the shift and tilt corrections}
\label{sec:shift/tilt_spherical}

\begin{figure}[t]
    \centering
    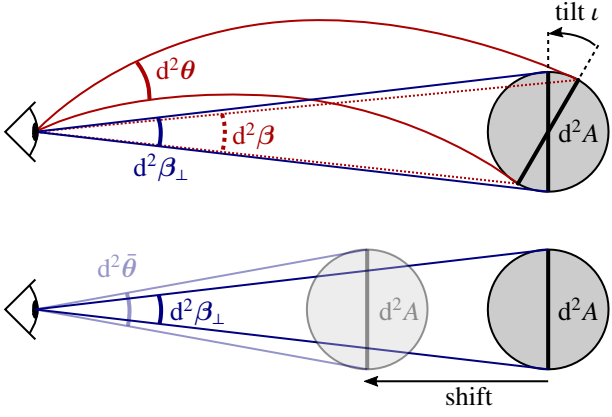
    \caption{Illustration of the shift and tilt effects, implying that the observable magnification differs from the geometric magnification, in the case of an isotropic spherical source.}
    \label{fig:shift_tilt_spherical}
\end{figure}

In \cref{subsubsec:shift/tilt} we have presented the shift and tilt corrections to the magnification by assuming that the source was an infinitesimal element of the iso-$z$ surface. This choice of formulation was justified by its tight connection with the subsequent considerations on the area of light-cone slices. Having said that, the curious reader may wonder about the generality of the shift and tilt corrections as defined therein. We may now consider an individual source instead of an element of iso-$z$ surface. This appendix aims to show that the results of \cref{subsubsec:shift/tilt} equally applies to this case.

We assume, for simplicity, that the source at redshift $z$ is spherical with isotropic emission.\footnote{Anisotropic emission would bring additional effects that go beyond the problematic of this article.} Recall that the geometric and observational magnifications are respectively defined as
\begin{align}
\mu &= \pm
\frac{\dd^2\vect{\theta}}
{\dd^2\vect{\beta}} \ ,
\\
\tilde{\mu} &= \pm
\frac{\dd^2\vect{\theta}}
{\dd^2\bar{\vect{\theta}}} \ ,
\end{align}
where $\dd^2\vect{\theta}$ is the angular size of an image, $\dd^2\vect{\beta}$ the coordinate solid angle covered by the source, and $\dd^2\bar{\vect{\theta}}$ the apparent size of that source if it were placed at the same redshift in an FLRW universe (see \cref{fig:shift_tilt_spherical}). The difference between $\mu$ and $\tilde{\mu}$ lies in the subtle difference between $\dd^2\vect{\beta}$ and $\dd^2\bar{\vect{\theta}}$.

Since the source is spherical, it is equivalent to a disk of area $\dd^2 A$ orthogonal to the direction of light propagation at emission. However, due to light deflection, this disk is tilted with respect to the radial direction. As a consequence, the disk covers a smaller coordinate solid angle~$\dd^2\vect{\beta}$ than its non-tilted counterpart~$\dd^2\vect{\beta}_\perp$ (top of \cref{fig:shift_tilt_spherical}). Both are related by $\dd^2\vect{\beta}=\dd^2\vect{\beta}_\perp\,\cos\iota$.

Besides, due to redshift corrections in the inhomogeneous Universe (peculiar velocities, Sachs-Wolfe effect, etc.) the time and radial coordinates of the source are shifted with respect to their background counterparts. As a result, a source with a given size is seen under a different solid angle in both cases (bottom of \cref{fig:shift_tilt_spherical}), namely
\begin{equation}
\frac{\dd^2\bar{\vect{\theta}}}{\dd^2\vect{\beta}_\perp}
= \frac{a^2[\eta(z)]\, r^2(z)}
        {a^2[\bar{\eta}(z)] \, \bar{r}^2(z)} \ .
\end{equation}

Summarising, the geometric and observational magnifications are related by
\begin{equation}
\frac
{\tilde{\mu}(z, \vect{\theta})}
{\mu(z, \vect{\theta})}
= \frac{\dd^2\vect{\beta}}{\dd^2\bar{\vect{\theta}}}
=
\underbrace{
\frac
{\dd^2\vect{\beta}_\perp}
{\dd^2\bar{\vect{\theta}}}
}_{\text{shift}}
\underbrace{
\frac
{\dd^2\vect{\beta}}
{\dd^2\vect{\beta}_\perp}
}_{\text{tilt}}
=
\frac
{a^2[\bar{\eta}(z)] \, \bar{r}^2(z)}
{a^2[\eta(z)]\, r^2(z)}
\, \cos\iota \ ,
\end{equation}
which is indeed equivalent to \cref{eq:geometric_vs_observable_magnification}.

\section{Finite-beam corrections}
\label{sec:appendix_finite_beams_calculations}

In this appendix we derive the finite-beam corrections~\eqref{eq:finite_beam_convergence_powerspectrum}, \eqref{eq:finite_beam_shear_powerspectrum} to the power spectra of convergence and shear. The computation will follow the general philosophy of \cite{fleury2017weak, fleury2019cosmic, fleury2019weak}, but it will differ in the details, due to the specific four-ray set-up used to compute $\kappa$ and $\gamma$ in this article.

\subsection{Estimators of convergence and shear}

As shown by \cite{fleury2019cosmic}, finite-beam corrections to cosmic convergence and shear occur on very small scales. Thus, we can safely work in the flat-sky approximation in the following. In that context, the lens equation reads
\begin{equation}
\vect{\beta}
= \vect{\theta} - \vect{\alpha}(\vect{\theta}) \ ,
\end{equation}
where $\vect{\beta}$ is the position of a point source, $\vect{\theta}$ the position of its image, and $\vect{\alpha}$ the displacement angle. An infinitesimal image is a collection of points $\vect{\theta}$ whose separation is much smaller than the typical angular scale over which $\vect{\alpha}(\vect{\theta})$ varies appreciably. In that case, one Taylor-expands $\vect{\alpha}(\vect{\theta})$ at first order and gets
\begin{equation}
\vect{\alpha}(\vect{\theta})
=
\vect{\alpha}(\vect{0})+
\begin{pmatrix}
\kappa+\gamma_1 & \gamma_2 \\
\gamma_2 & \kappa-\gamma_1
\end{pmatrix}
\vect{\theta} \ ,
\end{equation}
which defines the convergence $\kappa$ and shear $\gamma$. We note that we neglected the rotation~$\omega\sim |\gamma|^2$ [see \cite{Fleury:2015hgz}] for simplicity. Since $\vect{\alpha}(\vect{0})$ could be absorbed in a re-definition of the origin of the source plane, we set it to zero for convenience, $\vect{\alpha}(\vect{0})=\vect{0}$.

In the following calculation, it will be very convenient to associate a complex number~$\cplx{\theta}=\theta_x + \ii\theta_y$ with any 2-dimensional vector~$\vect{\theta}=\theta_x\vect{e}_x+\theta_y\vect{e}_y$. The lens equation then reads $\cplx{\beta}=\cplx{\theta}-\cplx{\alpha}(\cplx{\theta})$, and for infinitesimal images
\begin{equation}
\cplx{\alpha}(\cplx{\theta})
= \kappa\,\cplx{\theta} + \gamma\,\cplx{\theta}^* \ ,
\end{equation}
where a star denotes complex conjugation, and $\gamma=\gamma_1+\ii\gamma_2$ is the complex shear.

In this article, as explained in \S~\ref{sec:ray_bundles}, the distortion matrix, and hence convergence and shear, are estimated using a four-ray-bundle method. About a direction $\vect{\theta}$, four rays are shot in the directions $\vect{\theta}\pm \eps\vect{e}_x$, $\vect{\theta}\pm \eps\vect{e}_y$, and traced to get the associated four source positions. With complex notations, the corresponding estimators of convergence and shear are found to read
\begin{align}
\label{eq:kappa_finite_beam}
\kappa(\cplx{\theta};\eps)
&= \frac{1}{4\eps^2}\sum_{p=0}^3 \Re\left[\cplx{\eps}_p^* \cplx{\alpha}(\cplx{\theta}+\cplx{\eps}_p) \right],\\
\gamma(\cplx{\theta};\eps)
&= \frac{1}{4\eps^2}\sum_{p=0}^3 \cplx{\eps}_p \cplx{\alpha}(\cplx{\theta}+\cplx{\eps}_p) ,
\end{align}
where
\begin{equation}
\cplx{\eps}_p \equiv \eps \ex{\ii p \frac{\pi}{2}}
= \eps \, \ii^p \qquad
p = 0,1,2,3 \ ,
\end{equation}
are the four shifts with respect to the central ray $\cplx{\theta}$ used here. 

We note that the infinitesimal-beam case is recovered as $\eps\rightarrow 0$,
\begin{align}
\lim_{\eps\rightarrow 0}
\kappa(\cplx{\theta};\eps)
&= \Re(\cplx{\partial}\cplx{\alpha})
= \cplx{\partial}\cplx{\alpha}
= \kappa(\cplx{\theta};0),\\
\lim_{\eps\rightarrow 0}
\gamma(\cplx{\theta};\eps)
&= \cplx{\partial}^*\cplx{\alpha}
=\gamma(\cplx{\theta};0) \ ,
\end{align}
where we introduced the complex derivative
\begin{equation}
\cplx{\partial}
\equiv \frac{\partial}{\partial\cplx{\theta}}
=\frac{1}{2}\left(\frac{\partial}{\partial\theta_x}-\ii\frac{\partial}{\partial\theta_y}\right) .
\end{equation}

\subsection{Fourier transform}

Before moving to the actual computation of the power spectra if $\kappa, \gamma$, it is useful to express their Fourier transforms, as a function of their infinitesimal-beam counterparts. We use the convention
\begin{align}
\tilde{f}(\vect{\ell})
&= \int\dd^2\vect{\theta}\;
\ex{-\ii\vect{\ell}\cdot\vect{\theta}}\,
f(\vect{\theta}), \\\
f(\vect{\theta})
&= \int\frac{\dd^2\vect{\ell}}{(2\pi)^2}\;
\ex{\ii\vect{\ell}\cdot\vect{\theta}}\,
\tilde{f}(\vect{\ell}) \ .
\end{align}
In the infinitesimal beam case, this implies
\begin{align}
\tilde{\kappa}(\vect{\ell};0)
&= \frac{1}{2} \, \ii\cplx{\ell}^* \, \tilde{\cplx{\alpha}}(\vect{\ell}), \\
\tilde{\gamma}(\vect{\ell};0)
&= \frac{1}{2} \, \ii\cplx{\ell} \,
\tilde{\cplx{\alpha}}(\vect{\ell}) \ .
\end{align}
We start with convergence. Taking the Fourier transform of \cref{eq:kappa_finite_beam}, and using $\tilde{\kappa}(-\vect{\ell};0)=[\tilde{\kappa}(\vect{\ell};0)]^*$, we find
\begin{equation}
\tilde{\kappa}(\vect{\ell};\eps)
= C(\vect{\ell};\eps) \, \tilde{\kappa}(\vect{\ell};0) \ ,
\end{equation}
with the finite-beam filter
\begin{equation}
C(\vect{\ell};\eps)
\equiv
\frac{1}{2\eps^2}
\sum_{p=0}^3
    \Im\left[ \frac{\cplx{\eps}_p}{\cplx{\ell}} \right]
    \ex{\ii\vect{\eps}_p\cdot\vect{\ell}} \ .
\end{equation}
Similarly, the Fourier transform of shear reads
\begin{equation}
\tilde{\gamma}(\vect{\ell};\eps)
= S(\vect{\ell};\eps) \, \tilde{\gamma}(\vect{\ell};0) \ ,
\end{equation}
with the finite-beam filter
\begin{equation}
S(\vect{\ell};\eps)
= \frac{1}{2\ii\eps^2} \sum_{p=0}^3
    \frac{\cplx{\eps}_p}{\cplx{\ell}} \, 
    \ex{\ii\vect{\eps}_p\cdot\vect{\ell}} \ .
\end{equation}

\subsection{Power spectra}

\begin{figure*}
\includegraphics[width=1.02\columnwidth]{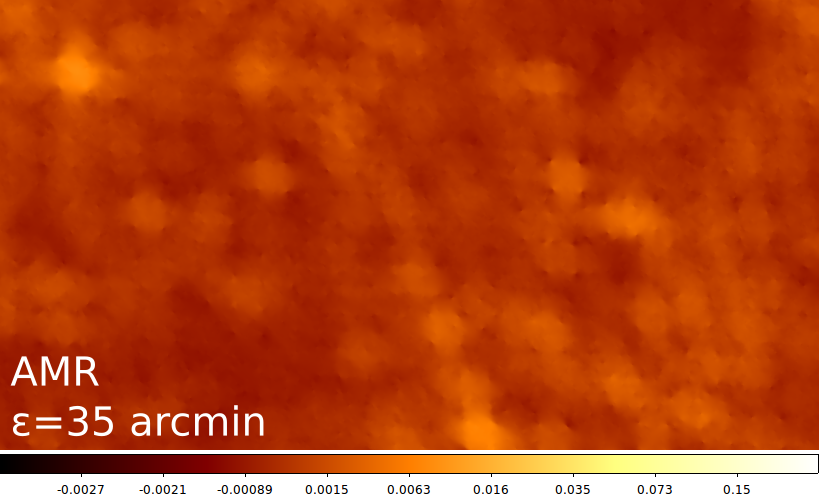}
\includegraphics[width=1.02\columnwidth]{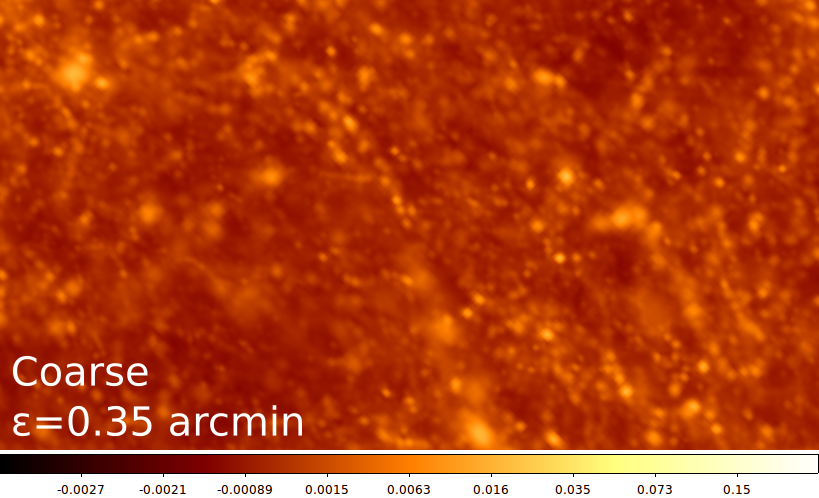}
\includegraphics[width=1.02\columnwidth]{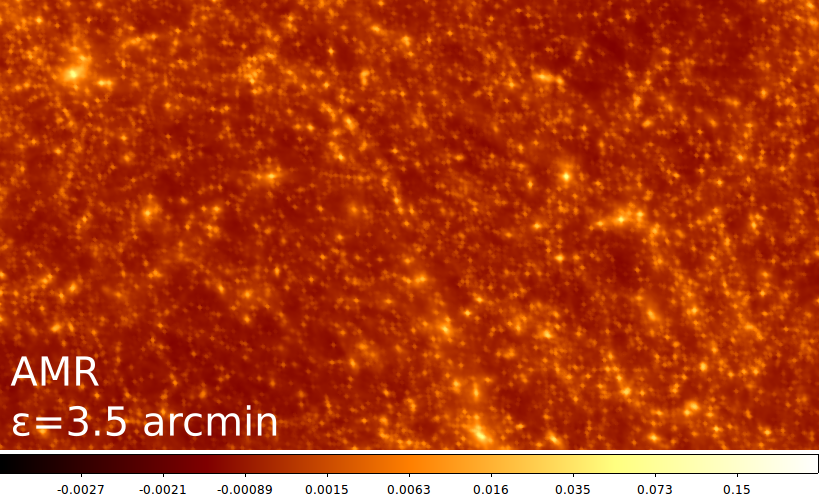} 
\includegraphics[width=1.02\columnwidth]{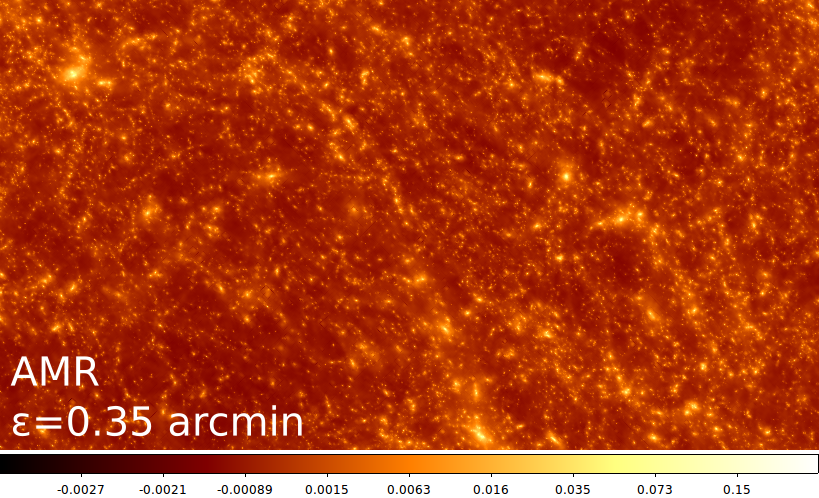} 
    \caption{Convergence maps at $z = 0.2$ using \textsc{Healpix} with nside = 2048. \emph{Top right panel}: coarse grid and $\eps =0.35$ arcmin. \emph{Top left panel}: AMR grid and  $\eps = 35$ arcmin. \emph{Bottom left panel}: AMR grid and $\eps = 3.5$ arcmin. \emph{Bottom right panel}: AMR grid and $\eps = 0.35$ arcmin.}
  \label{fig:ds9_convergence_maps}
\end{figure*}

We are now ready to compute the power spectra of $\kappa(\vect{\theta};\eps), \gamma(\vect{\theta};\eps)$, and in particular to evaluate how the results of the four-ray set-up may differ from the theoretical predictions with infinitesimal beams. The convergence power spectrum can be defined via
\begin{equation}
\ev{
\tilde{\kappa}(\vect{\ell}_1;\eps)\, \tilde{\kappa}(\vect{\ell}_2;\eps)
}
= (2\pi)^2\delta\e{D}(\vect{\ell}_1+\vect{\ell}_2) P_\kappa(\vect{\ell}_1;\eps) \ ,
\end{equation}
from which we deduce that
\begin{equation}
P_\kappa(\vect{\ell};\eps)
= C(\vect{\ell};\eps)C(-\vect{\ell};\eps) P_\kappa(\ell;0)\ .
\end{equation}
We note that, contrary to $P_\kappa(\ell;0)$, $P_\kappa(\vect{\ell};\eps)$ depends on the orientation of $\vect{\ell}$. This is due to the anisotropic square-like geometry of the four-beam set-up. However, in practice we effectively calculate the isotropic part from ray tracing,
\begin{equation}
P_\kappa(\ell;\eps)
\equiv
\int_0^{2\pi} \frac{\dd\psi}{2\pi}
P_\kappa(\vect{\ell};\eps) \ .
\end{equation}
where $\psi$ denotes the polar angle of $\vect{\ell}=\ell(\cos\psi,\sin\psi)$. After a tedious but straightforward calculation, we finally get
\begin{align}
\frac{P_\kappa(\ell;\eps)}{P_\kappa(\ell;0)}
&= \int_0^{2\pi} \frac{\dd\psi}{2\pi} \;
    C(\vect{\ell};\eps)C(-\vect{\ell};\eps), \\
&= \frac{1+J_2(2\eps\ell)+2J_2(\sqrt{2}\eps\ell)-J_0(2\eps\ell)}{2(\eps\ell)^2}.
\end{align}
The calculation for shear is slightly different but technically simpler. The power spectrum can be defined via
\begin{equation}
\ev{\tilde{\gamma}(\vect{\ell}_1;\eps)
    \tilde{\gamma}^*(-\vect{\ell}_2);\eps)
    }
= (2\pi)^2\delta\e{D}(\vect{\ell}_1+\vect{\ell}_2)\,
    P_\gamma(\vect{\ell}_1;\eps) \ ,
\end{equation}
so that
\begin{equation}
P_\gamma(\vect{\ell};\eps)
= \left| S(\vect{\ell};\eps) \right|^2 P_\gamma(\ell;0) \ .
\end{equation}
Taking the isotropic part then yields
\begin{align}
\frac{P_\gamma(\ell;\eps)}{P_\gamma(\ell;0)}
&= \int_0^{2\pi} \frac{\dd\psi}{2\pi} \;
    \left| S(\vect{\ell};\eps) \right|^2, \\
&= \frac{1-J_0(2\eps\ell)^2}{(\eps\ell)^2} \ .
\end{align}
As we can see in \cref{fig:ds9_convergence_maps}, similarly to resolution effects on simulations \citep{lepori2020weak}, the finite-beam effect acts as a smoothing on weak lensing maps.

\section{Details on the area of constant-time surfaces: shift and tilt}
\label{appendix:LSS}

In this appendix, we derive the theoretical predictions for the perturbation of the area of surfaces of constant time. As shown in \cref{subsubsec:LSS}, at second order, we have
\begin{equation}
\label{eq:delta_A_eta_appendix}
\frac{\delta A(\eta)}{\bar{A}(\eta)}
= \int_{\mathbb{S}^2} \frac{\dd^2\vect{\beta}}{4\pi}
    \left[
    \frac{2\delta r(\eta,\vect{\beta})}{\bar{r}(\eta)}
    + \frac{\delta r^2(\eta,\vect{\beta})}{\bar{r}^2(\eta)}
    + \frac{1}{2} \iota^2(\eta,\vect{\beta})
    \right] \ ,
\end{equation}
where $\delta r(\eta,\vect{\beta})=r(\eta,\vect{\beta})-\bar{r}(\eta)$ is the radial perturbation of the iso-$\eta$ surface, which may be decomposed into a geometrical and a time-delay (Shapiro) contribution as $\delta r=\delta r\e{geo}+\delta s$; $\iota$ denotes the tilt between the normal iso-$\eta$ surface and the radial direction.

We compute the ensemble average of $\delta r\e{geo}, \delta s, \iota$ in the spirit of KP16. Although our final result~\eqref{eq:perturbation_area_LSS_KP16} agrees with KP16, our derivations slightly differ. We also derive the angular correlation functions for $\delta r\e{geo}$ and $\delta s$, which is necessary to estimate their super-sample variance, and hence to allow a consistent confrontation of theory and numerical results.

\subsection{Comoving radius reached at fixed comoving distance travelled}
\label{subsec:distance_reached_vs_distance_travelled}

Due to gravitational lensing, light rays are wiggly, and hence the comoving radius that they reach is smaller than their comoving distance travelled. For a given comoving distance travelled $s$, the radius thus reads~$r=\bar{r}+\delta r\e{geo}(s)$ with $\bar{r}=s$. Here we carefully derive the expression of $\delta r\e{geo}<0$ and its statistical properties.

\subsubsection{Expression of $\delta r\e{geo}$}

We consider a photon observed in the direction~$\vect{\theta}$ from a source in the direction~$\vect{\beta}$; we may write its trajectory as $\vect{x}(r)=r\vect{\beta}+\vect{\xi}(r)$, where $\vect{\xi}\perp\vect{\beta}$ is the transverse displacement with respect to the axis spanned by $\vect{\beta}$. If the source is located at $r\e{s}$, then by definition $\vect{\xi}(0)=\vect{\xi}(r\e{s})=\vect{0}$ and $\dot{\vect{\xi}}(0)=\vect{\theta}-\vect{\beta}\equiv\vect{\alpha}$ in the limit of small angles, where a dot denotes a derivative with respect to $r$.

With such conditions, the equation of motion $\ddot{\vect{\xi}} = -2\nabla_{\perp}\phi$ is solved as
\begin{align}
\label{eq:xi_dot}
\dot{\vect{\xi}}(r)
&= \vect{\alpha} - 2\int^r_0\dd r' \; \nabla_{\perp}\phi(r') \ ,
\\
\label{eq:xi}
\vect{\xi}(r)
&= r\vect{\alpha} - 2\int^r_0\dd r' \; (r-r')\nabla_{\perp}\phi(r') \ .
\end{align}
The condition $\vect{\xi}(r\e{s})=\vect{0}$ then implies the familiar
\begin{equation}
\vect{\alpha}
= 2 \int_0^{r\e{s}}\dd r \; \frac{r\e{s}-r}{r\e{s}} \, \nabla_{\perp}\phi(r) \ .
\end{equation}
We note that the presence of $\vect{\alpha}$ in \cref{eq:xi_dot,eq:xi} comes from our definition of $\vect{\xi}$ as the transverse displacement with respect to the $\vect{\beta}$-axis; had we considered, as KP16, transverse displacement with respect to the $\vect{\theta}$-axis, $\vect{\alpha}$ would not be present. Our convention is more adapted (i) to the geometry of the problem, and (ii) to the fact that we will eventually consider ensemble averages where $\vect{\beta}$ is fixed.

The total path length depends on the local inclination~$\iota(r)=|\dot{\vect\xi}(r)|$ of the ray with respect to the axis spanned by $\vect{\beta}$. Specifically, we have, at second order,
\begin{equation}
s(r\e{s})
= \int_0^{r\e{s}} \frac{\dd r}{\cos\iota(r)}
= r\e{s}
    + \underbrace{
                \frac{1}{2} \int_0^{r\e{s}} \dd r \; |\dot{\vect{\xi}}(r)|^2
               }_{-\delta r\e{geo}(r\e{s})} \ ,
\end{equation}
which defines the wiggly-ray correction~$\delta r\e{geo}<0$ to the radius reached after travelling $s$.\footnote{The cautious reader may note that our definition of $\delta r\e{geo}$ differs from (A21) of KP16. Specifically, our approach involves the angle~$\iota$ between the photon direction and $\vect{\beta}$, while KP16 consider instead the angle between photon propagation and the instantaneous photon position $\vect{\beta}+\vect{\xi}(r)/r$. Our methods yield the same final result for $\delta r\e{geo}$.} Substituting the expression~\eqref{eq:xi_dot} of $\dot{\vect{\xi}}$ then yields
\begin{align}
\delta r\e{geo} 
&\equiv - \frac{1}{2} \int_0^{r\e{s}}\dd r \; |\dot{\vect{\xi}}(r)|^2 \ ,
\\
&= -\frac{1}{2} r\e{s} |\vect{\alpha}|^2
    + 2\vect{\alpha}\cdot\int_0^{r\e{s}} \dd r \int_0^{r} \dd r' \; \nabla_\perp\phi(r') \ ,
\nonumber\\&\qquad 
- 2 \int_0^{r\e{s}} \dd r \int_0^r \dd r' \int_0^r \dd r'' \;
    \nabla_\perp\phi(r')\cdot\nabla_\perp\phi(r'') \ .
\label{eq:calculation_delta_r_geo}
\end{align}

We may now simplify this expression by playing with the order of integration. We recommend the reader to draw the various integration domains in order to make sense of the following operations. The integrals in the second term of \cref{eq:calculation_delta_r_geo} read
\begin{align}
\int_0^{r\e{s}} \dd r \int_0^{r} \dd r' \; \nabla_\perp\phi(r')
&= \int_0^{r\e{s}} \dd r' \int_{r'}^{r\e{s}} \dd r \; \nabla_\perp\phi(r') \ ,
\\
&= \int_0^{r\e{s}} \dd r' \; (r\e{s}-r') \nabla_\perp\phi(r') \ ,
\\
&= \frac{1}{2} \, r\e{s} \vect{\alpha} \ ,
\end{align}
so that the whole second term is simply $r\e{s}|\vect{\alpha}|^2$.
As for the third term of \cref{eq:calculation_delta_r_geo}, we may perform the following operations, aiming at moving the integration over $r$ to the right. The first step is identical as above, and reads
\begin{equation}
\int_0^{r\e{s}} \dd r \int_0^r \dd r' \int_0^r \dd r''
= \int_0^{r\e{s}} \dd r' \int_{r'}^{r\e{s}} \dd r \int_0^r \dd r'' \ .
\end{equation}
The second inversion is more involved because it features an integration region made of right-angled triangle and a rectangle. A possible operation is
\begin{equation}
\int_{r'}^{r\e{s}} \dd r \int_0^r \dd r''
=
\int_{0}^{r'} \dd r'' \int_{r'}^{r\e{s}} \dd r
+
\int_{r'}^{r\e{s}} \dd r'' \int_{r''}^{r\e{s}} \dd r \ ,
\end{equation}
so that after a couple of additional manipulations we have
\begin{multline}
\int_0^{r\e{s}} \dd r \int_0^r \dd r' \int_0^r \dd r'' \;
    \nabla_\perp\phi(r')\cdot\nabla_\perp\phi(r'') \\
= 2 \int_0^{r\e{s}} \dd r' \int_0^{r'} \dd r'' \; (r\e{s}-r') \nabla_\perp\phi(r')\cdot\nabla_\perp\phi(r'')
\ .
\end{multline}

Gathering all three terms of \cref{eq:calculation_delta_r_geo} and using the symmetry of the double integration in $|\vect{\alpha}|^2$, we finally obtain
\begin{equation}
\label{eq:delta_r_geo_appendix}
\delta r\e{geo}
=
-4
\int_0^{r\e{s}} \dd r
\int_0^r \dd r' \;
\frac{(r\e{s}-r)r'}{r\e{s}} \,
\nabla_\perp\phi(r)\cdot\nabla_\perp\phi(r') \ ,
\end{equation}
in agreement with the third line of (A21) in KP16.

\subsubsection{Ensemble average of $\delta r\e{geo}$}
\label{appendix:ensemble_average_rgeo}
For $\delta A(\eta)$ we need to evaluate the average of $\delta r\e{geo}$ over $\vect{\beta}$. Following \cref{subsubsec:ensemble_average}, we apply the ergodicity principle to translate this into an ensemble average,
\begin{equation}
\int_{\mathbb{S}^2} \frac{\dd^2\vect{\beta}}{4\pi} \, \delta r\e{geo}(\eta,\vect{\beta})
=
\ev{\delta r\e{geo}(\eta,\vect{\beta})} ,
\end{equation}
which holds up to super-sample variance.

Introducing the Fourier transform~$\tilde{\phi}$ of $\phi$, we have
\begin{equation}
\nabla_\perp\phi
=\int \frac{\dd^3\vect{k}}{(2\pi)^3} \;
\ex{\ii\vect{k}\cdot\vect{x}} \,
(\ii\vect{k}_\perp)
\,\tilde{\phi}(\vect{k}) \ ,
\end{equation}
and using Limber's approximation we may proceed as
\begin{align}
\ev{\nabla_\perp\phi\cdot\nabla_\perp\phi'}
&= \int\frac{\dd^3\vect{k}}{(2\pi)^3}
    \frac{\dd^3\vect{k}'}{(2\pi)^3} \;
    \ex{\ii(\vect{k}\cdot\vect{x}
        +\vect{k}'\cdot\vect{x}')}\nonumber\\
&\hspace{2cm}\times
    (-\vect{k}_\perp\cdot\vect{k}_\perp')
    \ev{\tilde{\phi}(\vect{k})\tilde{\phi}(\vect{k}')} \ ,
\\
&= \int\frac{\dd^3\vect{k}}{(2\pi)^3} \;
    \ex{\ii\vect{k}\cdot(\vect{x}-\vect{x}')} \,
    \vect{k}_\perp^2
    P_{\phi}(\eta, \eta', \vect{k}) \ ,
\\
&\approx \delta\e{D}(r-r')
    \underbrace{
    \int\frac{\dd^2\vect{\ell}}{(2\pi r)^2} \;
    \left(\frac{\ell}{r}\right)^2
    P_\phi\left(\eta,\frac{\ell}{r}\right) 
    }_{\equiv J(r)/2} \ ,
\end{align}
where we introduced the power spectrum~$P_\phi$ of the gravitational potential, which is evaluated down the background light cone with $\eta=\eta_0-r$. In the last line we adopted the notation of KP16 and identified the integral
\begin{align}
J(r)
&\equiv
\frac{2}{r^4}\int_0^{\infty} \frac{\dd\ell}{2\pi} \; \ell^3
    P_\phi\left(\eta_0-r, \frac{\ell}{r}\right) \ , \\
&= 
2\int_0^{\infty} \frac{\dd k}{2\pi} \; k^3
    P_\phi\left(\eta_0-r, k\right) \ , \\
    &= 
2\pi\int_0^{\infty} k\,
\mathcal{P}_\phi\left(\eta_0-r, k\right)
\; \dd\ln k \ ,
\end{align}
where $\mathcal{P}_\ph$ denotes the dimensionless power spectrum of $\phi$, which is related to the standard one by
$2\pi^2 k^{-3} \mathcal{P}_\phi(k) = P_\phi(k)$.

Substituting the above in the expression of $\delta r\e{geo}$, we find
\begin{equation}
\label{eq:delta_r_geo_average}
\ev{\delta r\e{geo}} = -\int_0^{r\e{s}}
\dd r \; \frac{(r\e{s}-r)r}{r\e{s}} \, J(r) \ .
\end{equation}
Importantly, during that last step a factor $1/2$ appears, due to the integration of a Dirac delta on half of its domain:
\begin{equation}
\int_0^{r} \dd r' \;
\delta\e{D}(r'-r) f(r')
= \frac{1}{2} \, f(r) \ .
\end{equation}

\subsection{Distance travelled at fixed time}

We now turn to the effect of time delays, which implies that during a time $\eta_0-\eta$, a photon travels a comoving distance $s=\eta_0-\eta+\delta s$, with
\begin{equation}
\delta s(\eta) = \int_{\eta}^{\eta_0} \dd\eta \; 2\phi[\eta, \vect{x}(\eta)] \ .
\end{equation}
Combined with the wiggly-ray effect, this implies that the radius reached at fixed time reads $r(\eta)=\bar{r}(\eta)+\delta r\e{geo}(\eta)+\delta s(\eta)$.

\subsubsection{Post-Born expansion of $\delta s$}

We note that $\delta s$ is a first-order quantity evaluated on the perturbed trajectory~$\vect{x}$ of the photon. Since it is involved in $\delta r$ together with $\delta r\e{geo}$ which is a second-order quantity, we must account for post-Born corrections for consistency. These will turn out to be exactly minus twice $\delta r\e{geo}$.

Just as in \cref{subsec:distance_reached_vs_distance_travelled}, we write the perturbed photon path as $\vect{x}(r)=r\vect{\beta}+\vect{\xi}(r)$, so that $\delta s$ becomes
\begin{equation}
\label{eq:calculation_delta_s}
\delta s
= \int_0^{r\e{s}} \dd r \; 2\phi(r) 
    + \underbrace{\int_0^{r\e{s}} \dd r \; 2\vect{\xi}(r)\cdot\nabla_\perp\phi(r) }_{\equiv \delta s\e{pB}}\ ,
\end{equation}
where we changed to an integration over comoving radius without any loss of generality. The second term, $\delta s\e{pB}$, in \cref{eq:calculation_delta_s} encodes post-Born corrections. Substituting the expression~\eqref{eq:xi} of $\vect{\xi}$ we may rewrite it as
\begin{align}
\delta s\e{pB}
&= 4 \int_0^{r\e{s}} \dd r \int_0^{r\e{s}} \dd r' \; 
    \frac{r(r\e{s}-r')}{r\e{s}} \, \nabla_\perp\phi(r')\cdot\nabla_\perp\phi(r)
\nonumber\\&\quad
- 4 \int_0^{r\e{s}} \dd r \int_0^{r} \dd r' \; 
    (r-r') \, \nabla_\perp\phi(r')\cdot\nabla_\perp\phi(r) \ ,
\\
&= 4 \int_0^{r\e{s}} \dd r \int_r^{r\e{s}} \dd r' \; 
   \frac{r(r\e{s}-r')}{r\e{s}} \, \nabla_\perp\phi(r')\cdot\nabla_\perp\phi(r)
\nonumber\\&\quad
   + 4 \int_0^{r\e{s}} \dd r \int_0^{r} \dd r' \; 
    \frac{r'(r\e{s}-r)}{r\e{s}} \, \nabla_\perp\phi(r')\cdot\nabla_\perp\phi(r) \ ,
\\
&= 8  \int_0^{r\e{s}} \dd r \int_0^{r} \dd r' \; 
    \frac{r'(r\e{s}-r)}{r\e{s}} \, \nabla_\perp\phi(r')\cdot\nabla_\perp\phi(r) \ ,
\\
\delta s\e{pB} &= -2\delta r\e{geo} \ .
\label{eq:area_average_shapiro}
\end{align}

\subsubsection{Ensemble average of $\delta s$}
\label{subsubsec:extracting_post_Born}

The ensemble average of $\phi$ being zero, the only contribution from the ensemble average of $\delta s$ comes from the post-Born term,
\begin{equation}\label{eq:delta_s_average_result}
\ev{\delta s} = \ev{\delta s\e{pB}} = - 2\ev{\delta r\e{geo}} > 0\ .
\end{equation}
The expression of $\ev[1]{\delta r\e{geo}}$ is given in \cref{eq:delta_r_geo_average}.

In \cref{subsubsec:delta_s_results} we find that numerical estimates of $\ev{\delta s}$ are overwhelmed by the super-sample variance of its first-order contribution. In order to check \cref{eq:delta_s_average_result} numerically, it would be convenient to extract its post-Born term. This can actually be done with the following estimator for the average of $\delta s\e{pB}$:
\begin{align}
\widehat{\delta s\e{pB}}
\equiv 
\frac{\ev[1]{\mu^{-1}\delta s}\e{d}}{\ev[1]{\mu^{-1}}\e{d}}
-\ev{ \delta s}\e{d}
= \ev{\delta s(\vect{\beta})} - \ev{\delta s(\vect{\theta})} \ .
\label{eq:estimator_area_vs_direction_shapiro}
\end{align}
A calculation along the same lines as before indeed shows that the post-Born contribution drops from $\ev{\delta s(\vect{\theta})}$, that is when ensemble averaging is taken whilst $\vect{\theta}$ is kept fixed. Since, however, the Born contribution to both $\ev{\delta s(\vect{\beta})}, \ev{\delta s(\vect{\theta})}$ is identical, their difference eliminates it from the final result.

\subsubsection{Ensemble average of $\delta s^2$}

Since $\delta s$ is a first-order quantity, we must also evaluate its mean square as a contribution to the term $\ev[1]{(\delta r/\bar{r})^2}$ in \cref{eq:delta_A_eta_appendix}. The computation is straightforward and yields
\begin{equation}
\ev{\delta s^2}
\approx 4\int_0^{r\e{s}} \dd r \; \xi_\phi(r) \ ,
\end{equation}
in Limber's approximation, where $\xi_\phi$ is the two-point correlation function of $\phi$.

\subsection{Total effect of the shift}

Gathering the geometric and time-delay effects, we conclude that, at second order
\begin{align}
\frac{\delta A\e{shift}(\eta)}{\bar{A}(\eta)}
&\equiv \ev{\frac{2\delta r(\eta,\vect{\beta})}{\bar{r}(\eta)}
    + \frac{\delta r^2(\eta,\vect{\beta})}{\bar{r}^2(\eta)}} \ ,
\\
&= \ev{\frac{2\delta r\e{geo}}{r\e{s}}} 
        + \ev{\frac{2\delta s}{r\e{s}}}
        + \ev{\frac{\delta s^2}{r\e{s}^2}} \ ,
\\
&=
2\int_0^{r\e{s}}
\dd r \; \frac{(r\e{s}-r)r}{r\e{s}^2} \, J(r)
+ 4 \int_0^{r\e{s}} \dd r \; \xi_\phi(r) \ .
\end{align}
In practice, the second term is negligible with respect to the first one and it can be discarded.

\subsection{Tilt or wrinkly-surface effect}
\label{subsec:appendix_tilt}

We now turn to the increase of the area due to its wrinkles. As seen in \cref{eq:delta_A_eta_appendix}, this effect is controlled by the angle~$\iota(\eta,\vect{\beta})=|\dot{\vect{\xi}}(\eta,\vect{\beta})|$ formed by the normal to the iso-$\eta$ surface and the direction spanned by $\vect{\beta}$. Specifically,
\begin{equation}
\frac{\delta A\e{tilt}(\eta)}{\bar{A}(\eta)}
\equiv \frac{1}{2}\int_{\mathbb{S}^2} \frac{\dd\vect{\beta}}{4\pi} \; \iota^2(\eta,\vect{\beta})
= \frac{1}{2}\ev{ |\dot{\vect{\xi}}(\eta,\vect{\beta})|^2 } \ ,
\end{equation}
From the expression~\eqref{eq:xi_dot} of $\dot{\vect{\xi}}$, we find
\begin{equation}
|\dot{\vect{\xi}}|^2 = 4\int_0^{r\e{s}} \dd r \int_0^{r\e{s}} \dd r' \;
            \frac{r r'}{r\e{s}^2} \, \nabla_\perp\phi(r)\cdot\nabla_\perp\phi(r') \ ,
\end{equation}
and hence, in Limber's approximation,
\begin{equation}
\frac{1}{2}\ev{|\dot{\vect{\xi}}(\eta,\vect{\beta})|^2}
= \int_0^{r\e{s}} \dd r \; \frac{r^2}{r\e{s}^2} \, J(r) \ ,
\end{equation}
in agreement with Eq.~(A41) of KP16.

Combining this wrinkly surface contribution with the total contribution of the shift then yields the final result
\begin{equation}
\frac{\delta A(\eta)}{\bar{A}(\eta)}
= \int_0^{r\e{s}} \dd r \; \frac{(2r\e{s}-r)r}{r\e{s}^2} \, J(r) \ ,
\end{equation}
up to the negligible $\delta s^2$ term.

\section{Variance calculations}

For any statistical quantity $X$, its variance is given by \cref{eq:uncertainty_average_X}, which depends on the angular power spectra of $X$. In what follows, we give $C_\ell^X$ for the quantities in \cref{sec:results}.

\subsection{Variance for source and directional averages}
\label{sec:appendix_variance_source_direction_averages}

The quantities of interest for source and directional averages are functions of $\mu$ and $\tilde{\mu}$, which can be rewritten in terms of $\kappa$ and $\tilde{\kappa}$. Their angular power spectra can be rewritten as
\begin{eqnarray}
 C_\ell^\mu &=& C_\ell^{1/\mu} = 4C_\ell^\kappa \ , \\
 C_\ell^{\tilde{\mu}} &=& C_\ell^{1/{\tilde{\mu}}} = 4C_\ell^{\tilde{\kappa}} \ , \\
 C_\ell^d &=& C_\ell^{\tilde{\kappa}} \ , \\
 C_\ell^{\Delta m} &=& \left[5/\ln 10\right]^2 C_\ell^{\tilde{\kappa}} \ .
\end{eqnarray}
Therefore, to estimate the variance we need to compute the angular power spectra of $\kappa$ and $\tilde{\kappa}$.

As discussed in \cref{eq:Doppler_magnification}, $\tilde{\kappa}$ may be approximated as
\begin{equation}
    \tilde{\kappa} = \kappa + \tilde{\kappa}_{v} \ ,
\end{equation}
where the contribution from redshift perturbations $\tilde{\kappa}_{v}$ only contain the effect of peculiar velocities. For simplicity, we only account for the auto-correlation of $\kappa$ and $\tilde{\kappa}_{v}$, so that
\begin{equation}
    C_\ell^{\tilde{\kappa}} = C_\ell^\kappa + C_\ell^{\tilde{\kappa}_{v}} \ .
\end{equation}
We generate $C_\ell^{\kappa}$ with \textsc{Nicaea} \citep{kilbinger2017precision} which uses the non-linear prescription from \textsc{Halofit} \citep{smith2003stable, takahashi2012revising}. 

The angular power spectrum of $\tilde{\kappa}_{v}$ is simply
\begin{equation}
    C_{\ell}^{\tilde{\kappa}_{v}} = 4\pi  \left( 1 - \frac{1}{\mathcal{H}r}\right)^2 \int \frac{\dd k}{k} \; j_\ell^2(kr) \, \frac{k^3P_v(k)}{2\pi^2} \ ,
    \label{eq:angular2PCF_kappatilde}
\end{equation}
with $P_v(k)$ the peculiar-velocity power spectrum.

We only consider the variance from $\kappa$ and $\tilde{\kappa}_{v}$ for simplicity. In principle, one should also account for cross-terms as well as additional variance terms, for example due to real-space clustering for source-averaged quantities, see \citep{fleury2017how}.

We note that we do not account for the constraint at the observer (see \cref{subsec:constrained_grf_method}), because this effect is only relevant when the field of interest is very correlated at large scales, such as the gravitational potential whose power spectrum scales as $P(k)/k^4$. This is not the case for the quantities investigated here, hence we do not need to account for the constraint at the observer.

\subsection{Angular power spectrum of $\delta r\e{geo}$}
\label{sec:appendix_variance_rgeo}

The geometrical shift~$\delta r\e{geo}$ at fixed distance travelled is a second-order quantity, which depends on the power spectrum of $\phi$. Hence its own angular power spectrum will depend on the four-point function of $\phi$. In that context, a full spherical-harmonic computation would be rather involved; furthermore, expect most of the power to be held by small scales. Hence, we choose to compute a flat-sky power spectrum~$P\e{geo}(\ell)\approx C_\ell[\delta r\e{geo}]$.

We start from the definition of the two-point correlation function as
\begin{align}
\label{eq:definition_correlation_delta_r_geo}
\xi\e{geo}(\theta)
\equiv
\ev{\delta r\e{geo}(\vect{\theta}_1)\, \delta r\e{geo}(\vect{\theta}_2)}
- \ev{\delta r\e{geo}(\vect{\theta}_1)}\ev{\delta r\e{geo}(\vect{\theta}_2)} \ ,
\end{align}
with $\theta\equiv|\vect{\theta}_1-\vect{\theta}_2|$.
We note that since $\delta r\e{geo}$ is second-order, it does not really matter whether we are considering observed or `true' angular positions here.

Substituting the expression~\eqref{eq:delta_r_geo_appendix} of $\delta r\e{geo}$ in the first term of \cref{eq:definition_correlation_delta_r_geo} and using its Fourier transform yields
\begin{multline}
\ev{\delta r\e{geo}(\vect{\theta}_1)\, \delta r\e{geo}(\vect{\theta}_2)}
= 16 \int_0^{r\e{s}} \dd r_1 \int_0^{r_1} \dd r'_1
    \int_0^{r\e{s}} \dd r_2 \int_0^{r_2} \dd r'_2 \\
    \times
    \left[ \frac{(r\e{s}-r_1)r_1'}{r\e{s}} \right]
    \left[ \frac{(r\e{s}-r_2)r_2'}{r\e{s}} \right]
    \int\frac{\dd^3\vect{k}_1}{(2\pi)^3}
        \frac{\dd^3\vect{k}_1'}{(2\pi)^3}
        \frac{\dd^3\vect{k}_2}{(2\pi)^3}
        \frac{\dd^3\vect{k}_2'}{(2\pi)^3} \\
        \times
        \exp\ii\left[
                (r_1\vect{k}_1 + r'_1\vect{k}'_1)\cdot\vect{\theta}_1
                + (r_2\vect{k}_2 + r'_2\vect{k}'_2)\cdot\vect{\theta}_2
            \right] \\
        \times
        (\vect{k}_{\perp 1}\cdot\vect{k}_{\perp 1}')
        (\vect{k}_{\perp 2}\cdot\vect{k}_{\perp 2}')\,
        \ev{\tilde{\phi}(\vect{k}_1) \tilde{\phi}(\vect{k}'_1)
                \tilde{\phi}(\vect{k}_2) \tilde{\phi}(\vect{k}'_2)} \ ,
\end{multline}
which raises the difficulty of computing the four-point correlation $\ev[1]{\tilde{\phi}(\vect{k}_1) \tilde{\phi}(\vect{k}'_1) \tilde{\phi}(\vect{k}_2) \tilde{\phi}(\vect{k}'_2)}$. Assuming that $\phi$ is reasonably modelled by a Gaussian random field, we may neglect the tri-spectrum and apply Wick's theorem as
\begin{multline}
\ev{\tilde{\phi}(\vect{k}_1) \tilde{\phi}(\vect{k}'_1)
                \tilde{\phi}(\vect{k}_2) \tilde{\phi}(\vect{k}'_2)}
=
\ev{\tilde{\phi}(\vect{k}_1) \tilde{\phi}(\vect{k}_1')}
\ev{\tilde{\phi}(\vect{k}_2) \tilde{\phi}(\vect{k}_2')}
\\
+
\ev{\tilde{\phi}(\vect{k}_1) \tilde{\phi}(\vect{k}_2)}
\ev{\tilde{\phi}(\vect{k}_1') \tilde{\phi}(\vect{k}_2')}
+
\ev{\tilde{\phi}(\vect{k}_1) \tilde{\phi}(\vect{k}_2')}
\ev{\tilde{\phi}(\vect{k}_1') \tilde{\phi}(\vect{k}_2)} \ .
\end{multline}
Each of the three terms leads to a different contribution. The first one exactly compensates $\ev[1]{\delta r\e{geo}}^2$ in \cref{eq:definition_correlation_delta_r_geo}; the third one vanishes in Limber's approximation; the second one holds the interesting correlation and we eventually get
\begin{multline}
\label{eq:intermediate_C}
\xi\e{geo}(\theta)
= 16\int_0^{r\e{s}} \dd r \int_0^{r} \dd r'
\left[ \frac{(r\e{s}-r)r'}{r\e{s}} \right]^2
\frac{1}{(rr')^4} \\
\times
\int\frac{\dd^2\vect{\ell}}{(2\pi)^2}
    \frac{\dd^2\vect{\ell}'}{(2\pi)^2} \;
    (\vect{\ell}\cdot\vect{\ell}')^2
    \ex{\ii(\vect{\ell}+\vect{\ell}')\cdot(\vect{\theta}_1-\vect{\theta}_2)}
    P_\phi\left(\frac{\ell}{r}\right)
     P_\phi\left(\frac{\ell'}{r'}\right) \ .
\end{multline}

The power spectrum $P\e{geo}(L)$ must satisfy
\begin{equation}
C\e{geo}(\theta)
= \int\frac{\dd^2\vect{\ell}}{(2\pi)^2} \; \ex{\ii\vect{\ell}\cdot\vect{\theta}} \, P\e{geo}(\ell) \ .
\end{equation}
Thus, introducing the variable $\vect{L}\equiv \vect{\ell}+\vect{\ell}'$ and performing the change of variable $(\vect{\ell}, \vect{\ell}')\mapsto(\vect{\ell}, \vect{L})$ in \cref{eq:intermediate_C}, we immediately identify
\begin{multline}
P\e{geo}(L)
=
16\int_0^{r\e{s}} \dd r
    \int_0^{r} \dd r'
        \left[ \frac{(r\e{s}-r)r'}{r\e{s}} \right]^2
    \frac{1}{(rr')^4} 
    \int\frac{\dd^2\vect{\ell}}{(2\pi)^2} \\
    \times [(\vect{L}-\vect{\ell})\cdot\vect{\ell}]^2 \,
     P_\phi\left(\eta_0-r,\frac{\ell}{r}\right)
     P_\phi\left(\eta_0-r',
        \frac{|\vect{L}-\vect{\ell}|}{r'}\right) \ .
\end{multline}
In the last integral, the direction of the vector $\vect{L}$ does not matter, because integration over $\vect{\ell}$ makes everything isotropic. In practice, one may take $\vect{L}$ to be aligned with $\vect{e}_x$. With that convention, the two-dimensional integral over $\vect{\ell}$ becomes
\begin{multline}
\int\frac{\dd^2\vect{\ell}}{(2\pi)^2} \;
     [(\vect{L}-\vect{\ell})\cdot\vect{\ell}]^2 \,
     P_\phi\left(\eta_0-r,\frac{\ell}{r}\right)
     P_\phi\left(\eta_0-r',
        \frac{|\vect{L}-\vect{\ell}|}{r'}\right)
\\
= \int_0^\infty\frac{\ell\dd\ell}{2\pi} \;
    P_\phi\left(\eta_0-r,\frac{\ell}{r}\right)
    \int_0^{2\pi}\frac{\dd\psi}{2\pi} \;
    [\ell(L\cos\psi-\ell)]^2 \\
    \times P_\phi\left(\eta_0-r',         
        \frac{\sqrt{L^2+\ell^2-2L\ell\cos\psi}}
                {r'}\right) \ .
\end{multline}
Introducing the integration variable $k\equiv \ell/r$, and then making the change $L\rightarrow \ell$, we get the final result
\begin{multline}
\label{eq:pkl_deltar_cte_s}
P\e{geo}(\ell)
=
16 \int_0^{r\e{s}}\dd r\int_0^{r}\dd r'
\left[ \frac{(r_0-r)}{r_0 r'} \right]^2
\\ \times
\int_0^\infty \frac{k^3\dd k}{2\pi}
\int_0^{2\pi} \frac{\dd\psi}{2\pi} \;
(\ell\cos\psi-kr)^2
\\ \times
P_\phi(\eta_0-r, k) \,
P_\phi\left(\eta_0-r', \frac{\sqrt{\ell^2+(kr)^2-2\ell kr\cos\psi}}{r'} \right) .
\end{multline}

\subsection{Angular power spectrum of $\delta r(z)$}
\label{sec:appendix_variance_redshift}

Regarding the relative perturbations on the comoving distance at constant observed redshift, since we only account for the Doppler effect, the prediction is proportional to the velocity field. We therefore compute its angular power spectrum similarly to \cref{eq:angular2PCF_kappatilde} but with the pre-factor $1/\mathcal{H}r$ instead of $(1-1/\mathcal{H}r)$.

\subsection{Constrained variance}
\label{sec:appendix_variance_constrained}

All the other quantities that we investigate in \cref{subsec:results_shift}, that is $\delta r(\eta)$, $\delta\lambda(\eta)$ and $\delta r(\lambda)$,
are obtained by line-of-sight integrations of the gravitational potential. As such, these are particularly impacted by the constraint at the observer (see \cref{subsec:constrained_grf_method}). Here we show how to compute the variance for such quantities.

We consider a scalar quantity $X$ that is a line-of-sight projection of the potential $\phi$, with
\begin{equation}
X = \int_{0}^{r\e{s}} \dd r \; \mathcal{K}_X(r) \, \phi(\eta_0-r, r) \ ,
\end{equation}
where $\mathcal{K}_X$ is the kernel associated with $X$. To estimate the variance given the constrained field $\phi(\bm{r})$ it is easier to work in configuration space and compute the variance using (see \cref{subsec:variance_theory})
\begin{equation}
 \label{eq:correlated_variance_configurationspace}
 \sigma^2_{\rm ss} = \int \dd^2\vect{\theta} \, \dd^2\vect{\theta}' \;
    W(\vect{\theta}) \, W(\vect{\theta}')
   \ev{X(\vect{\theta}) X(\vect{\theta}')} \ .
\end{equation}
In \cref{eq:configuration_space_sample_variance} we used a window function for a cone-shaped geometry with circular base for simplicity. Actually, our narrow cones are pyramid-shaped (we expect the difference to be negligible compared to the circular case). To compute \cref{eq:correlated_variance_configurationspace}, we set the integration boundaries to $\varphi = [-\Delta/2, \Delta/2]$, $\vartheta=[\pi/2 + \Delta/2, \pi/2 - \Delta/2]$, where ($\varphi$, $\vartheta$) are the angles in spherical coordinates. For the intermediate and deep cones, $\Delta = 50$ and 20 degrees respectively.

Assuming that $\ev{X(\vect{\theta}) X(\vect{\theta}')} =  \omega_X(|\bm{\theta}-\bm{\theta}'|)$ is statistically isotropic, we find
\begin{equation}
\label{eq:angular_2PCF}
    \omega_X(|\bm{\theta}-\bm{\theta}'|) =\int_0^{r_s}\dd r\int_0^{r\e{s}} \dd r' \; \mathcal{K}_X(r) \mathcal{K}_X(r') \, H_X(r\bm{\theta},r'\bm{\theta}').
\end{equation} 
Then, to compute the relevant variance for \cref{subsec:results_shift}, we used the functions shown in \cref{tab:constrained_functions}.
\begin{table}[h!]
	\centering
	\caption{Expressions used in \cref{eq:angular_2PCF} for various quantities studied in \cref{subsec:results_shift}, with $\mathcal{K}_X \equiv \mathcal{K}_X(r)$ and $H_X \equiv H_X(r\bm{\theta},r'\bm{\theta}')$. For $H_X$ we use the relations in \cref{subsec:constrained_grf_method}.}	
	\begin{tabular}{c|c|c} % four columns, alignment for each
	$X$ & $\mathcal{K}_X$ & $H_X$  \\
	\hline
    $\delta r(\eta)/\bar{r}(\eta)$ & $ 2/\bar{r}(\eta)$ & $\ev{\phi(r\bm{\theta})\phi(r'\bm{\theta}')|\phi_0}$  \\
    $\delta\lambda(\eta)/\bar{\lambda}(\eta)$  & $ 2a^2(\eta)/\bar{\lambda}(\eta)$ & $\ev{(\phi(r\bm{\theta})-\phi_0)(\phi(r'\bm{\theta}')-\phi_0)|\phi_0}$  \\
    $\delta r(\lambda)/\bar{r}(\lambda)$  & $2a^2(\lambda)/(\bar{r}(\lambda)\bar{a}^2(\lambda))$ & $\ev{(\phi(r\bm{\theta})-\phi_0)(\phi(r'\bm{\theta}')-\phi_0)|\phi_0}$  \\
	\end{tabular}
	\label{tab:constrained_functions}
\end{table}

\end{document}